\pdfoutput=1

\documentclass[longtitle,final,3p,times,twocolumn]{elsarticle}

\usepackage{graphicx}

\usepackage{amssymb}

\usepackage{lineno}

\newcommand*\patchAmsMathEnvironmentForLineno[1]{%
  \expandafter\let\csname old#1\expandafter\endcsname\csname #1\endcsname
  \expandafter\let\csname oldend#1\expandafter\endcsname\csname end#1\endcsname
  \renewenvironment{#1}%
     {\linenomath\csname old#1\endcsname}%
     {\csname oldend#1\endcsname\endlinenomath}}%
\newcommand*\patchBothAmsMathEnvironmentsForLineno[1]{%
  \patchAmsMathEnvironmentForLineno{#1}%
  \patchAmsMathEnvironmentForLineno{#1*}}%
\AtBeginDocument{%
  \patchBothAmsMathEnvironmentsForLineno{equation}%
  \patchBothAmsMathEnvironmentsForLineno{align}%
  \patchBothAmsMathEnvironmentsForLineno{flalign}%
  \patchBothAmsMathEnvironmentsForLineno{alignat}%
  \patchBothAmsMathEnvironmentsForLineno{gather}%
  \patchBothAmsMathEnvironmentsForLineno{multline}%
}
\usepackage[english]{babel}
\usepackage[utf8]{inputenc}
\usepackage[T1]{fontenc}
\usepackage{amsmath}
\usepackage{amsfonts}
\usepackage{url}
\usepackage{color}
\usepackage[pdftex, unicode={true}, bookmarks={false}]{hyperref}
\hypersetup{
  colorlinks={true},
  linkcolor={black},
  pdfauthor={
    Johan Nyberg, 
    Department of Physics and Astronomy, 
    Uppsala University},
  pdfcreator={pdflatex},
  pdfkeywords={},
  pdfsubject={},
  pdftitle={},
  urlcolor={blue}
}
\usepackage{multirow}
\usepackage{threeparttable}
\usepackage[alsoload=binary]{siunitx}
\newcommand{\geant}{\ensuremath{\textsc{Geant4}}}
\newcommand{\ROOT}{\ensuremath{\textsc{ROOT}}}

\usepackage{relsize}
\newcommand\CC{C\nolinebreak[4]\hspace{-.05em}\raisebox{.4ex}{\relsize{-3}{\bf ++}}}
\biboptions{sort&compress}

\journal{Nuclear Instruments and Methods in Physics Research A}

\begin{document}

\begin{frontmatter}




\title{AGATA -- Advanced Gamma Tracking Array}


\author[ankara]{S.~Akkoyun}
\author[valencia]{A.~Algora}
\author[darmstadt]{B.~Alikhani}
\author[gsi]{F.~Ameil}
\author[lnl]{G.~de Angelis}
\author[univ-strasbourg,cnrs-strasbourg]{L.~Arnold}
\author[csnsm]{A.~Astier}
\author[ankara,uppsala,stockholm]{A.~Ata\c{c}}
\author[ipn-orsay]{Y.~Aubert}
\author[ipn-lyon]{C.~Aufranc}
% ...

\address[ankara]{Department of Physics, Faculty of Science, Ankara University, 06100 Tando\u{g}an, Ankara, Turkey}
\address[valencia]{IFIC, CSIC-Universitat de Val\'{e}ncia, E-46980 Paterna, Spain}
\address[darmstadt]{IKP, TU Darmstadt, Schlossgartenstra{\ss}e 9, D-64289 Darmstadt, Germany}
\address[gsi]{GSI Helmholtzzentrum f\"{u}r Schwerionenforschung GmbH, D-64291 Darmstadt, Germany}
\address[lnl]{INFN Laboratori Nazionali di Legnaro, IT-35020 Padova, Italy}
\address[univ-strasbourg]{Universit\'{e} de Strasbourg, IPHC, 23 rue du Loess, 67037 Strasbourg, France}
\address[cnrs-strasbourg]{CNRS, UMR 7178, 67037 Strasbourg, France}
\address[csnsm]{CSNSM, CNRS, IN2P3, Universit\'{e} Paris-Sud, F-91405 Orsay, France}
\address[uppsala]{Department of Physics and Astronomy, Uppsala University, Uppsala, Sweden}
\address[stockholm]{The Royal Institute of Technology, SE-10691 Stockholm, Sweden}
\address[ipn-orsay]{IPNO, CNRS/IN2P3, Universit\'{e} Paris-Sud, F-91406 Orsay, France}
\address[ipn-lyon]{Universit\'{e} de Lyon, Universit\'{e} Lyon 1, CNRS-IN2P3, Institut de Physique Nucl\'{e}aire de Lyon, F-69622 Villeurbanne, France}
% ...
\address[univ-florence]{Universit\`{a} di Firenze, Dipartimento di Fisica e Astronomia, IT-50019 Firenze, Italy}
% ...
\address[valencia-ee]{Department of Electronic Engineering, University of Valencia, Burjassot (Valencia) Spain}
\address[univ-padova]{Dipartimento di Fisica, Universit\`{a} di Padova, IT-35131 Padova, Italy}
% ...
\address[univ-milano]{Dipartimento di Fisica, Universit\`{a} di Milano, IT-20133 Milano, Italy}
% ...
\address[astro-univ-padova]{Dipartimento di Astronomia, Universit\`{a} di Padova, IT-35131 Padova, Italy}
% ...
\address[univ-istanbul]{Istanbul University, Istanbul, Turkey}
% ...
\address[univ-warsaw]{Faculty of Physics, Warsaw University of Technology, ul. Koszykowa 75, 00-662 Warszawa, Poland}
% ...
\address[univ-sofia]{Faculty of Physics, St. Kliment Ohridski University of Sofia, Bulgaria}
% ...

\cortext[cor1]{Corresponding authors}
\cortext[cor2]{Principal corresponding author}
\fntext[dead]{Deceased}
\fntext[pa_lbl]{Present address: Lawrence Berkeley National Laboratory,
  Berkeley, CA, USA}
\fntext[pa_nscl]{Present address: National Superconducting Cyclotron Laboratory
  Michigan State University, East Lansing, Michigan 48824-1321, USA}

{\small{
\begin{abstract} 
  The Advanced GAmma Tracking Array (AGATA) is a European project to
  develop and operate the next generation $\gamma$-ray
  spectrometer. AGATA is based on the technique of $\gamma$-ray energy
  tracking in electrically segmented high-purity germanium
  crystals. This technique requires the accurate determination of the
  energy, time and position of every interaction as a $\gamma$ ray
  deposits its energy within the detector volume. Reconstruction of
  the full interaction path results in a detector with very high
  efficiency and excellent spectral response. The realization of
  $\gamma$-ray tracking and AGATA is a result of many technical
  advances. These include the development of encapsulated
  highly-segmented germanium detectors assembled in a triple cluster
  detector cryostat, an electronics system with fast digital sampling
  and a data acquisition system to process the data at a high
  rate. The full characterization of the crystals was measured and
  compared with detector-response simulations. This enabled
  pulse-shape analysis algorithms, to extract energy, time and
  position, to be employed. In addition, tracking algorithms for event
  reconstruction were developed. The first phase of AGATA is now
  complete and operational in its first physics campaign. In the
  future AGATA will be moved between laboratories in Europe and
  operated in a series of campaigns to take advantage of the different
  beams and facilities available to maximize its science output. The
  paper reviews all the achievements made in the AGATA project
  including all the necessary infrastructure to operate and support
  the spectrometer.
\end{abstract}
}}



\end{frontmatter}


\section{Introduction} \label{s:intro}
Contemporary nuclear physics research aims at understanding the
microscopic and mesoscopic features of the nuclear many-body system,
determined by the effective interactions and underlying symmetries.
These aims are often addressed by studying the nuclear system under
extreme values of isospin, mass, angular momentum or temperature. In
particular many facets of the nuclear system can be probed and
understood by studying nuclei far from stability. With the inception
of the new generation of Radioactive Ion Beam (RIB) facilities, in the
case of Europe FAIR (Darmstadt, Germany), HIE-ISOLDE (CERN, Geneva,
Switzerland), SPIRAL2 (Caen, France) and SPES (Legnaro, Italy), where
a much wider range of unstable proton- and neutron-rich nuclei will
become accessible, a new era is being opened for nuclear physics
experiments.

For five decades, high resolution $\gamma$-ray spectroscopy, in
particular with the large germanium detector arrays, has become a
cornerstone in nuclear structure studies. Intense research and
development efforts during the 1980's and the 1990's both in Europe
and in the USA led to the construction of efficient $4\pi$
escape-suppressed $\gamma$-ray spectrometers
\cite{SharpeySchafer1988293,Beausang1996}. Although the
escape-suppression technique significantly improves the peak-to-total
ratio in $\gamma$-ray spectra, it limits the solid angle occupied by
the germanium detectors and therefore the efficiency of the
$\gamma$-ray detection system. This detection technique culminated
with the design and construction of the EUROBALL
\cite{Beck1992443,Simpson1997} and GAMMASPHERE \cite{Lee1990}
spectrometers in Europe and in the USA, respectively. This contributed
in a significant way to the impressive progress made in nuclear
structure research since then.

In recent years, new important technical advances, namely that of
position sensitive Ge crystals and tracking array technology, were
developed to cope with Doppler effects due to large source velocities
and the experimental conditions at the future facilities for intense
radioactive and high-intensity stable ion beams. These conditions are
expected to be extremely challenging, requiring unprecedented levels
of sensitivity and count-rate capabilities. The required performance
figures are beyond reach with conventional escape-suppressed arrays.
The use of electrically segmented Ge crystals enables the
identification of the individual points of interaction of the $\gamma$
rays within the volume of the Ge crystals as well as the determination
of the deposited energy with high resolution.  Besides the
highly-segmented Ge detectors, the realisation of such an array
requires digital sampling electronics to extract energy, time, and
position information from the detectors' output signals using
pulse-shape analysis methods.  The path of the $\gamma$ rays in the Ge
crystals can then be reconstructed, making use of "tracking"
algorithms on the position and energy information of the individual
interactions, and the full energy of the original $\gamma$ ray can be
determined. With the $\gamma$-ray tracking technique
\cite{Deleplanque1999292,Lee2003}, the Compton-suppression shields
become unnecessary resulting in a large gain in efficiency while
maintaining spectral quality.  Furthermore, the direction of emission
of each individual $\gamma$ ray can be determined with high precision,
which is crucial for a good Doppler energy correction and hence to
achieve a good energy resolution even when $\gamma$ rays are emitted
from a fast moving nucleus, as is the case in most nuclear reactions.

This radically new concept constitutes a dramatic advance in
$\gamma$-ray detection that will have wide-ranging applications also
in medical imaging, astrophysics, nuclear safeguards and
radioactive-waste monitoring, as well as establish a new level of
detection capability for nuclear-structure studies. Given the
importance of this development and its far-reaching implications, a
European collaboration currently consisting of over 40 institutions
from 12 countries has been established to develop and construct a
European $4\pi$ tracking spectrometer called AGATA (Advanced GAmma
Tracking Array). A similar project, GRETINA/GRETA, is also ongoing in
the USA \cite{Lee2003,Lee2004}. The development of a tracking
spectrometer in Europe is based on progress made within many previous
projects, e.g. MINIBALL \cite{Eberth2001389,Habs1997111}, MARS
\cite{Kroll2001227}, some of which were co-ordinated and supported by
the EU TMR programme (project title: Development of Gamma-Ray Tracking
Detectors for $4\pi$ Gamma-Ray Arrays)
\cite{Lieder2001279,Lieder2001399}.

AGATA is a mobile instrument that will move between major laboratories
in Europe take advantage of the range of different beams and equipment
at each laboratory and to optimize the use of beam time at these
facilities. AGATA will therefore be operated in a series of science
campaigns at specific European facilities. It is now fully operational
in its first physics campaign at INFN Laboratori Nazionali di Legnaro
(LNL) in Italy, utilizing the wide range of stable beams
available. Subsequently it will operate at the GSI facility in Germany
and the GANIL laboratory in France and later at new radioactive beams
facilities such as FAIR, SPIRAL2, SPES and HIE-ISOLDE. The
spectrometer will be expanded over time, in phases, towards the full
$4\pi$ 180 detector system.

This paper describes concisely the AGATA spectrometer and summarizes
all the necessary developments that have been performed by the AGATA
collaboration for its design, construction and operation. These
developments range from advances in Ge detector technology, digital
data acquisition systems, signal decomposition and $\gamma$-ray
interaction reconstruction, and in many areas of the infrastructure
needed to support and operate such a complex device.

\section{Conceptual design} \label{s:condes}
The conceptual design of AGATA explored the possible configurations of
a $\gamma$-ray tracking array and compared their performance in a
consistent way.  This process was described extensively in
\cite{Farnea2010331}. Here only the basic ideas and the most important
results will be summarised.

It is evident that, in order to maximise the detection efficiency of
AGATA, the solid angle coverage should be maximised. In addition, to
minimise the development and maintenance costs, the solid angle should
be covered with only a few elementary shapes.  The passive parts of
the array should be minimised by using composite detectors, implying
grouping (clustering) more crystals within the same
cryostat. Moreover, to simplify the handling and maintenance of such
complex objects, the detectors should rely on the encapsulation
technique originally developed for the EUROBALL Cluster detectors
\cite{Eberth1996135}.  An additional requirement in the conceptual
design of AGATA was to keep a sufficiently large inner space inside
the array in order to host the foreseen complementary instrumentation,
which often is indispensable in the physics programme of AGATA.

The most elegant way to achieve a large solid angle coverage with a
few elementary shapes relies on a decomposition of the icosahedron,
namely of the platonic polyhedron having the largest number of faces.
Such a decomposition will always result in \num{12} regular pentagons
and a variable number of irregular hexagons once projected onto the
spherical surface.  As discussed in more detail in
\cite{Farnea2010331}, the configurations having 120 or 180 hexagons
were soon identified as the most attractive ones for AGATA, given the
possibility to cover the solid angle with a few crystal shapes (2 and
3 for the case of 120 and 180 hexagons, respectively) and to easily
form clusters of crystals (one kind of cluster with 4 and 3 crystals
for the case of 120 and 180 hexagons, respectively). The contribution
to the overall detection efficiency provided by the 12 pentagonal
detectors was considered too limited to justify the extra costs for
development. Therefore, the geometry of the array was optimised by
minimizing the size of the pentagons.  Indeed, the pentagonal holes
can be utilised for mechanical support, insertion of complementary
detectors and for the beam entry and exit pipes.

The possible configurations for AGATA were evaluated through detailed
Monte Carlo simulations of the full array.  The simulation code for
AGATA is based on the \CC\ classes of
\geant\ \cite{Agostinelli2003250}, which provide a full description of
the microscopic interactions of radiation with matter, as well as
tools to implement the geometry of complex detector arrangements and
to process and extract the relevant information. The \geant\ geometry
libraries were complemented with a specific class capable of
describing irregular convex polyhedral shapes as the AGATA elementary
shapes.

Establishing the optimal geometrical configuration for AGATA was a
complex problem where the ingredients considered went beyond the mere
overall performance figures of the array and other factors such as the
reliability, simplicity, symmetry and cost of the adopted solution
were taken into account.

\begin{figure}[ht]
  \begin{center}
    \includegraphics[width=0.99\columnwidth]
                    {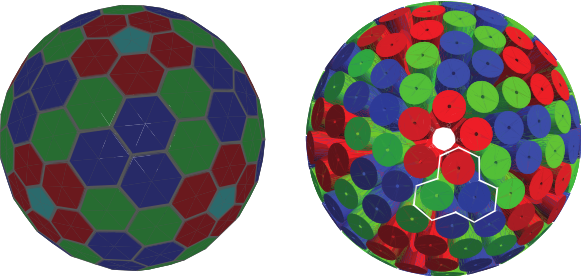}
    \caption{(Colour online) Computer aided design images of the
      tiling of the sphere (left) and the 180 crystal configuration
      (right). The cryostats and the detector encapsulation are not
      shown.}
    \label{fig:cd01}
  \end {center}
\end{figure}

\begin{figure}[ht]
  \begin{center}
    \includegraphics[width=0.99\columnwidth]
                    {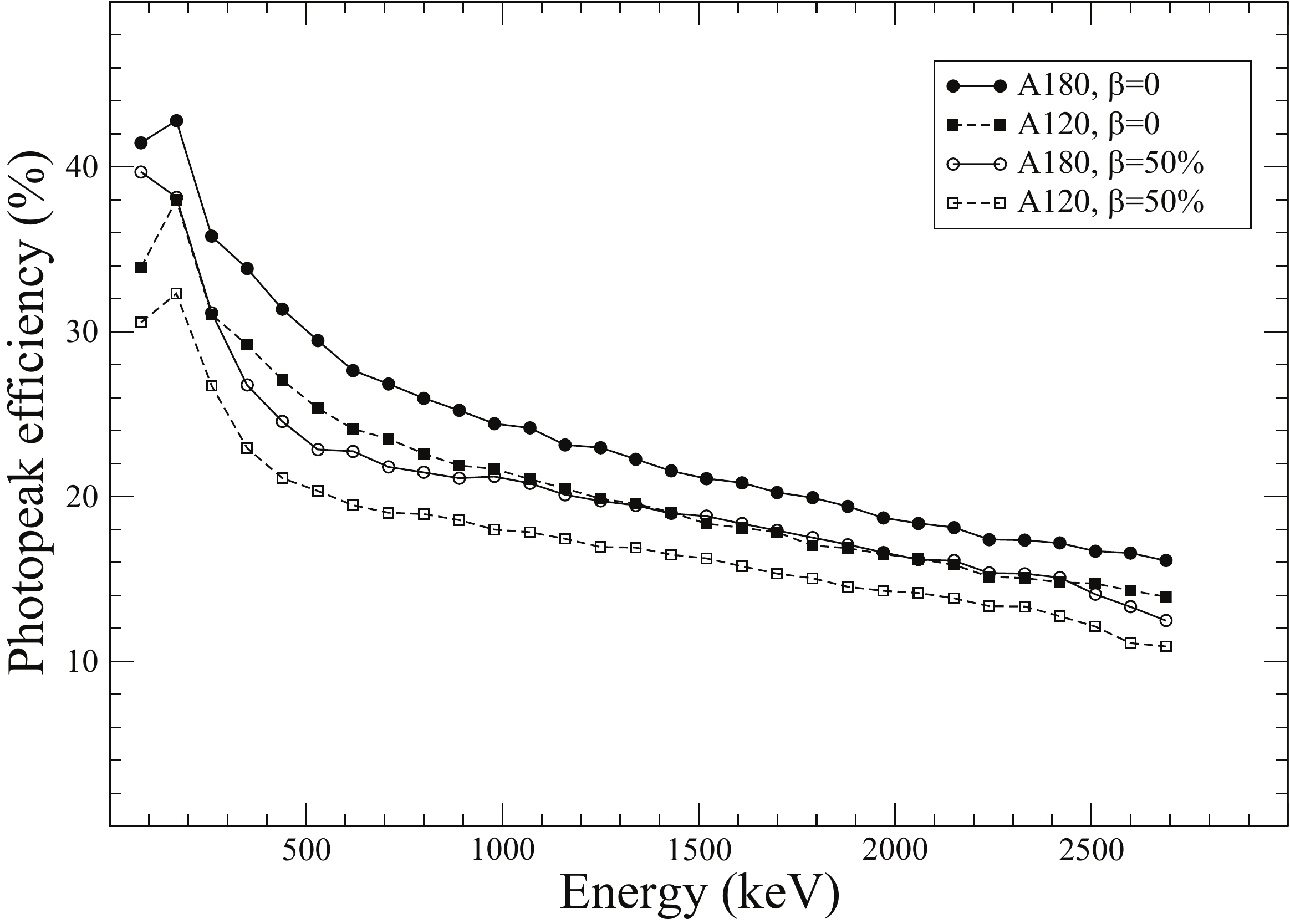}
    \caption{Simulated full-energy efficiency for the 120 crystals and
      the 180 crystals (AGATA) arrays as a function of the
      $\gamma$-ray energy and at multiplicity $M_{\gamma}=30$, emitted
      by a point source recoiling along the $z$ axis, with velocity
      $v/c=0$, and $v/c=0.5$.  }
    \label{fig:cd03}
  \end {center}
\end{figure}

\begin{figure}[ht]
  \begin{center}
    \includegraphics[width=0.99\columnwidth]{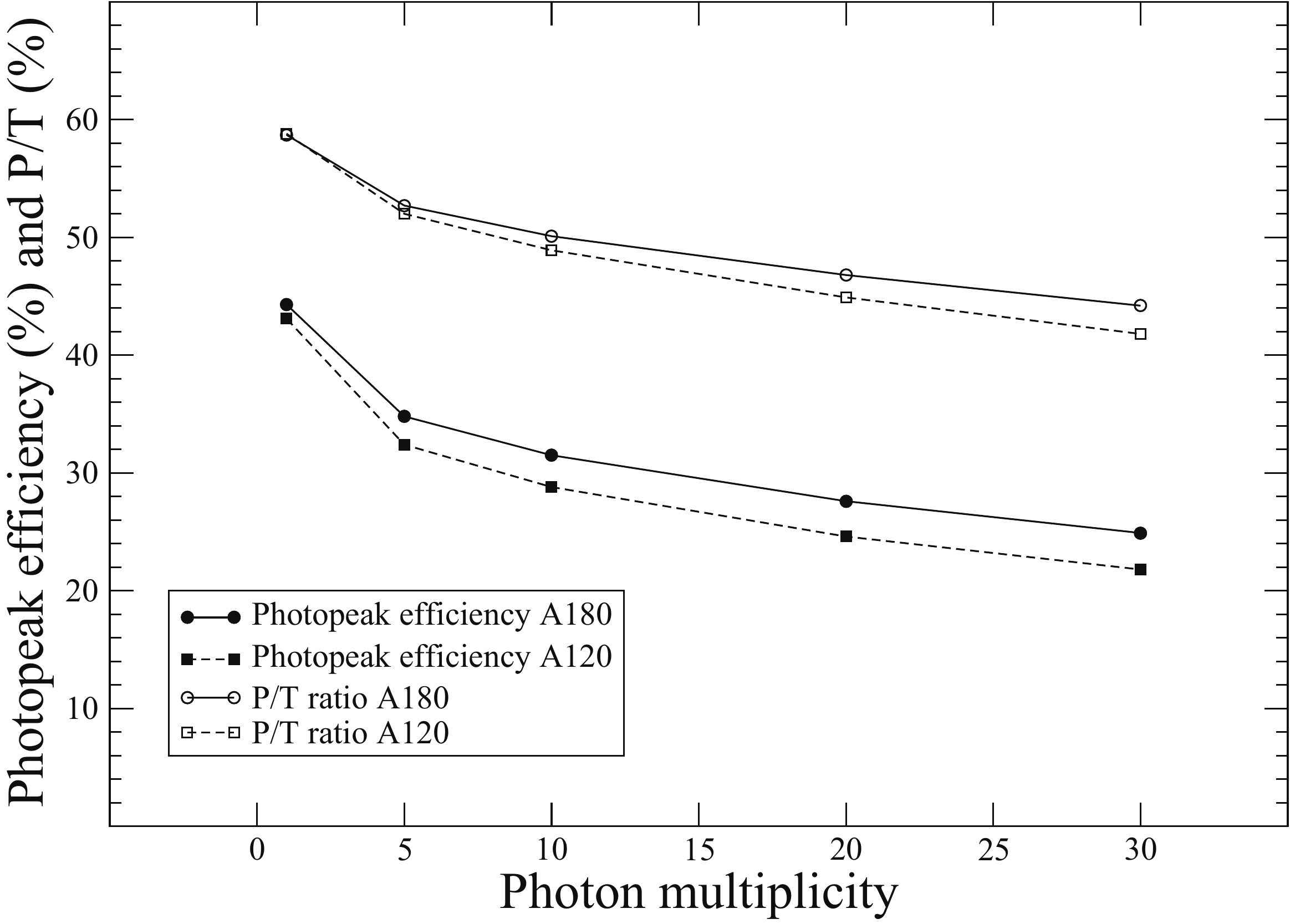}
    \caption{Simulated full-energy efficiency and peak-to-total (P/T)
      ratio for the 120 crystals and the 180 crystals (AGATA) array
      for \SI{1}{\MeV} photons as a function of the $\gamma$-ray
      multiplicity. The $\gamma$ rays have been emitted by a stopped
      point-like source.  }
    \label{fig:cd02}
  \end {center}
\end{figure}

From the results of the Monte Carlo simulations described in
\cite{Farnea2010331}, it can be concluded that the configuration with
180 hexagonal crystals (Fig.~\ref{fig:cd01}) has better energy
resolution, full-energy efficiency (Fig.~\ref{fig:cd03}) and
peak-to-total (P/T) ratio than any configuration based on 120
detectors in a broad range of experimental conditions. The difference
in performance is particularly evident at high $\gamma$-ray
multiplicity, see Fig.~\ref{fig:cd02}. The performance figures for
\SI{1}{\MeV} photons of the final optimised geometry with 180
detectors are \SI{82}{\percent} solid angle coverage,
\SI{43}{\percent} (\SI{28}{\percent}) full-energy efficiency and
\SI{59}{\percent} (\SI{43}{\percent}) P/T ratio at a photon
multiplicity $M_\gamma=1$ ($M_\gamma=30$).  The final configuration
for AGATA was therefore chosen to be based on 180 segmented hexagonal
crystals.

With hexagonally shaped crystals the azimuthal segmentation is quite
naturally based on 6 sectors, each of them centred on the crystal
corners.  The pattern of the longitudinal segmentation was optimised
on the basis of detailed electric field simulations.  On one hand, the
effective volume of the segments should be balanced. On the other
hand, the number of longitudinal segments should be large enough to
achieve the required position resolution after the pulse-shape
analysis process, namely \SI{5}{\mm} FWHM as calculated in the Monte
Carlo simulations.  A longitudinal segmentation based on 6 ``rings''
was considered the best compromise between the required performance
and the cost of the associated electronics.  For further details
regarding the crystal segmentation scheme, see
subsection~\ref{ss:crystals}.

\section{Detectors} \label{s:det}
The AGATA detectors are based on encapsulated and electrically
segmented closed-end coaxial $n$-type high-purity germanium (HPGe)
crystals. The crystals have a tapered hexagonal geometry with an
asymmetric shape to fit into the $4\pi$ \num{180} detector
geometry (see Fig.~\ref{fig:cd01}). This geometry is realised with
three different shapes, with a triplet of crystals arranged in
identical triple cryostats, such that the full $4\pi$ array has 60
cryostats. With this configuration, having a \SI{9}{\cm} thick
germanium shell, a solid angle coverage of up to \SI{82}{\percent} is
realised. Its rather large inner radius of \SI{22.5}{\cm} (to the
endcap face of the cryostat) allows the use of most ancillary
detectors.

The AGATA triple cluster (ATC) detector contains three \num{36}-fold
segmented HPGe crystals. The total energy deposited in each crystal
is collected in the central contact (core) leading to 37 signals per
crystal. Therefore, the ATC detector contains \num{111}
high-resolution spectroscopy channels.  A more detailed description of
the ATC detector is given in \cite{Wiens2010223,Lersch2011133}. All
signal channels are equipped with a cold preamplifier stage operated
close to the liquid nitrogen temperature of the cryostat. The core
preamplifier is characterised by low noise and a large dynamic range
for energy detection, pulse-shape analysis and timing properties.  A
novel reset technique of the core preamplifier allows for an increased
counting rate capability of the detectors of more than \SI{50}{\kHz},
preserving an energy resolution close to the nominal one.  Moreover,
the energy range of the ATC detector is substantially
extended from \SIrange{20}{180}{\MeV}, see subsection~\ref{ss:preamp}.
These developments are documented in
\cite{Pullia2006,Zocca2009,Pascovici2008}. Despite the high electronic
integration density only small cross-talk contributions, typically
less than \num{1} in \num{d-3}, are measurable between the segments
within a crystal. This cross-talk contribution is an expected effect
caused by capacitive coupling of the signals via the bulk Ge material
and can be well described within an electronic model of the combined
crystal and preamplifier assembly
\cite{Bruyneel2009196,Bruyneel200999}.

\subsection{The AGATA crystals} \label{ss:crystals}

\begin{figure}[th]
  \centering
  \includegraphics[width=1.0\columnwidth]{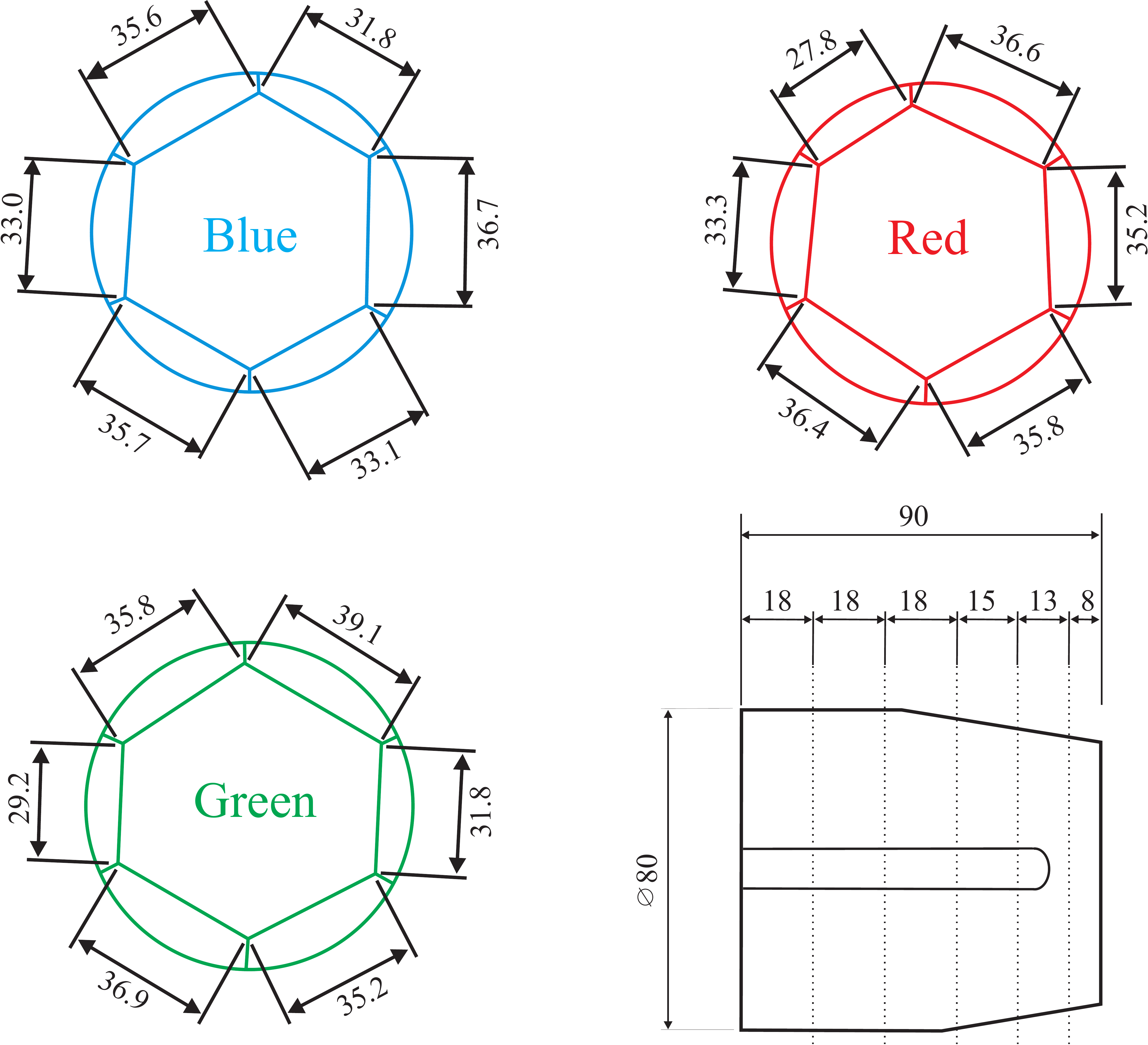}
  \caption{(Colour online) Drawing of the three AGATA crystal
    geometries. The AGATA triple cluster detector combines the three
    different crystal shapes. The side view (lower right) shows the
    position of the segmentation lines. All dimensions are given in
    \si{\mm}.}
  \label{fig:3-shapes}
\end{figure}

\begin{figure}[th]
  \centering
  \includegraphics[width=1.0\columnwidth]{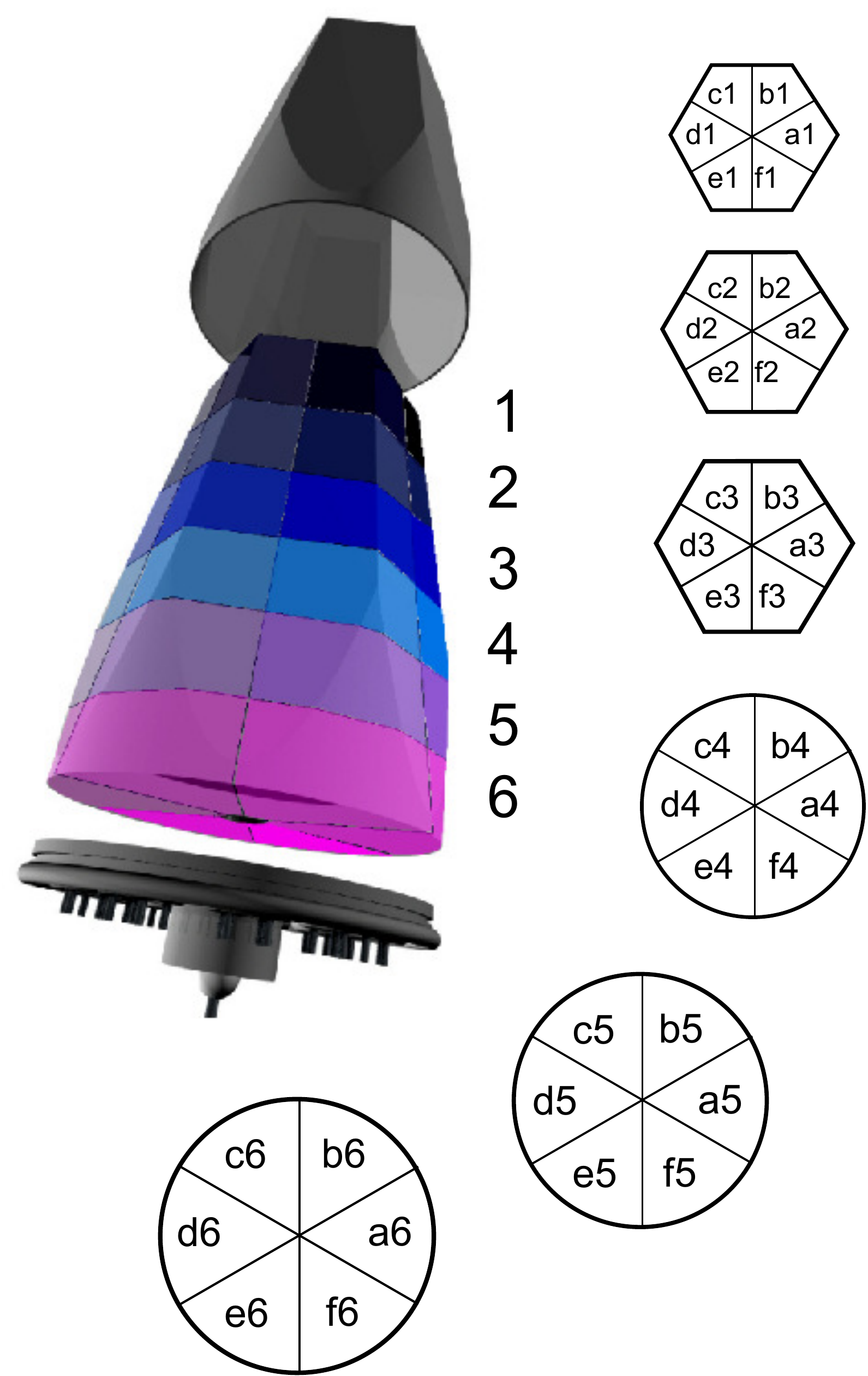}
  \caption{(Colour online) Segment labelling scheme of the AGATA HPGe
    capsules. Along the crystal axis the external contact is
    subdivided into six rings labelled 1 to 6. Each ring is subdivided
    into six sectors labelled a to f.}
  \label{fig:segment_labelling}
\end{figure}

All detectors are produced by the company Canberra, France. The three
types of detectors employed in AGATA merely differ in their irregular
hexagonal shape (see Fig.~\ref{fig:3-shapes}). The different
geometries are assigned a letter and a colour: A -- red, B -- green
and C -- blue.  A serial number is also assigned to each crystal
(A001, A002, etc.). The crystals have a length of $90{\pm}1$\;\si{\mm}
and a diameter of $80^{+0.7}_{-0.1}$\;\si{\mm} at the rear. At the
front they are tapered to a hexagonal shape with a \ang{10} tapering
angle. The crystal's central hole has a diameter of \SI{10}{\mm} and
extends to \SI{13}{\mm} from the front end. The 6-fold sector-wise
segmentation goes through the middle of each flat hexagonal side. The
6-fold longitudinal segmentation forms rings of
\num{8}, \num{13}, \num{15}, \num{18}, \num{18} and \SI{18}{\mm}
in thickness starting at the hexagonal
front face of the crystal (see Fig.~\ref{fig:3-shapes}). The
thicknesses of the rings have been optimised for a uniform
distribution of the $\gamma$-ray interactions and optimal pulse-shape
sensitivity \cite{Kroll2001227}. The segment labelling scheme of the
AGATA crystals is shown in Fig. \ref{fig:segment_labelling}.

The weight of a bare AGATA Ge crystal is about \SI{2}{\kg}. All
crystals are made of $n$-type HPGe material with an impurity
concentration specified to be between \num{0.4} and
\SI{1.8e10}{\per\cubic\cm}. The surfaces of these crystals are very
delicate and therefore each crystal is encapsulated into a
hermetically sealed aluminium canister with a \SI{0.8}{\mm} wall
thickness (see Fig. \ref{fig:segment_labelling}). The encapsulation
technology was developed for the EUROBALL cluster detectors
\cite{Eberth2008283} and extended to segmented detectors within the
frame of the MINIBALL project \cite{Eberth2001389}.  The distance
between capsule walls and crystal side faces is
\SIrange{0.4}{0.7}{\mm}. The $6{\times}7$ connector feed-throughs
provide access to each of the \num{36} segmented outer contacts. The core
contact, which is used for applying the high voltage and to obtain the
core energy signal, is isolated with ceramic material. Efficient
$\gamma$-ray tracking requires extreme care with the close packing of
the crystals. The capsules are accurately mounted in the ATC detector
with a \SI{0.5}{\mm} spacing between the flat surfaces.

\subsection{The AGATA cryostats}

The cryostats of the ATC detectors were assembled and successfully
commissioned in a common effort by the company CTT Montabaur, together
with the AGATA collaboration.

The preamplifiers of segment and core contacts are divided in two
spatially separated parts. The cold input stages of the preamplifiers
are operated close to the Ge crystals.  Cooling and mounting in close
proximity to the detector is required to optimise noise performance.
In addition a good electronic shielding between the input stages is
required in order to minimise cross-talk effects. The AGATA cryostats
employ a separated cooling scheme for the encapsulated Ge detector and
the cold part of the preamplifier electronics. While the Ge detectors
are cooled to \SI{90}{\kelvin}, the FETs (Field Effect Transistors) are
operated at temperatures near \SI{130}{\kelvin} where their noise
contribution is minimal. The other adjacent parts of the preamplifier
electronics contribute less to the noise performance and are therefore
situated outside the vacuum, where they are readily accessible. The
electric connection between the two parts is made by several hundreds
of individual thin wires with low thermal conductivity.

The thermal isolation is established by a vacuum with pressure values
below \SI{1e-6}{\milli\BAR}. This pressure is maintained over long
periods by the active getter materials built into the cryostat.

Although each individual FET has only an electric power consumption of
$\sim$\SI{20}{\milli\watt}, the total consumption of the \num{111} FETs in one
ATC detector adds up to \SI{2.3}{\watt}. Together with the enhanced
thermal connection by the wiring inside the cryostat and the radiative
heat absorption, a considerable cooling capacity is demanded. The
Dewar of the triple cryostat contains up to \SI{4.5}{\litre} of liquid
nitrogen. It has a length of \SI{38}{\cm} and an outer diameter of
\SI{25}{\cm}. A full Dewar is sufficient for about \SI{8}{\hour} of
continuous operation.  An electronic measurement of the liquid
nitrogen filling level, which is based on a capacitance measurement
between a metallic cylindrical tube inside the Dewar and the inner
wall of the cryostat~\cite{Lersch2011133}, is incorporated in the
Dewar.  The temperature is monitored by two platinum resistance
thermometers of the type PT100, one attached to the copper cooling
finger close to the Dewar and the other one positioned close to the
crystals.

\begin{figure}[ht]
  \centering
  \includegraphics[width=1.0\columnwidth]{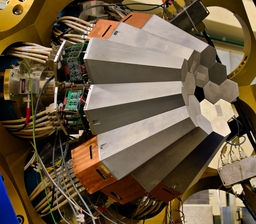}
  \caption{(Colour online) Photograph of the setup with five AGATA
    triple cluster detectors installed at LNL in Italy.}
  \label{fig:ad_photo}
\end{figure}

The triple cryostats have a length of \SI{92}{\cm} and a weight of
\SI{48}{\kg} including the crystals and liquid nitrogen. Very tight
tolerances are demanded for the manufacturing of the cryostat endcaps
such that the final spacing between the endcap side faces of
neighbouring triple cryostats is
\SI{0.5}{\mm}. Fig.~\ref{fig:ad_photo} shows \num{5} ATC detectors
mounted into the support structure at LNL and demonstrates the
challenges in the design, assembly and on-site installation of such
cryostats.

\subsection{Preamplifiers} \label{ss:preamp}

The preamplifiers for the AGATA detectors require, besides the
traditional good energy and timing properties, also fast and clean
transfer functions to register unperturbed signal traces for
pulse-shape analysis. In addition, a high count-rate capability was
demanded in order to exploit fully the high geometrical
efficiency. New preamplifiers have been developed by the AGATA
collaboration which fulfill these requirements
\cite{Pascovici2008,Zocca2009,Pullia2006}.  The segment and core
signals of the AGATA detectors are read out through advanced
charge-sensitive resistive feed-back preamplifiers, employing a new
fast-reset technique for dead time and dynamic range optimisation as
well as an original circuit structure for maximising the open-loop
gain of the charge-sensing stage.

The preamplifiers have a cold and a warm part.  The cold part consists
of a low-noise silicon FET, model BF862, a \SI{1.0}{\pico\farad} feedback
capacitance and a \SI{1}{\giga\ohm} feedback resistance.  A dedicated
shielding was developed for the cold preamplifier board for minimising
the inter-channel cross talk. The warm part, operated at room
temperature, is separated by \SI{15}{\cm} cabling inside the vacuum
from the cold part of the cryostat and comprises a low-noise
transimpedance amplifier, a pole-zero stage, a differential output
buffer, and fast-reset circuitry.

The core preamplifier \cite{Pascovici2008} is characterised by low
noise and a large dynamic range for energy detection, pulse-shape
analysis and timing properties. Transient signals are not deformed due
to the large bandwidth. The core preamplifier possesses high
count-rate capabilities and an embedded precision pulser.

The fast reset is provided by the desaturation circuitry, which is
capable of detecting saturated signals. In such a situation a current
source is connected, discharging the capacitance in the pole-zero
network and achieving a fast restoration of the output level.

This innovative time-over-threshold (TOT) technique \cite{Zocca2009}
for high-resolution spectroscopy extends the $\gamma$-ray energy range
up to about \SI{180}{\MeV}. In this technique the energy is obtained
through a precise determination of the reset time, required by the
desaturation circuitry, which is measured as a time difference.  The
time measurement is started when the electronic pulse exceeds the
saturation threshold and is stopped when the pulse amplitude becomes
smaller than this threshold.  The time difference is strictly related
to the pulse height of the saturating energy signal and allows an
energy determination with a good energy resolution (FWHM) of about
\SI{0.2}{\percent}. This TOT energy resolution value can be compared
for energies around \SI{10}{\MeV} with values of about
\SI{0.15}{\percent} achieved by the standard pulse height mode. Beyond
10~MeV the FWHM values are even comparable in both modes.

A custom programmable high-precision pulser is located on the
core-preamplifier board. Its applications are: testing, calibration,
time alignment and dead-time corrections, which are relevant for
efficiency measurements of the detector. The pulser is used to inject
calibration pulses to the core electrode itself as well as to all
segment electrodes through the detector bulk capacitance.  The output
signal of the pulser is DC coupled to the source pin of the core input
FET through a resistor divider consisting of a \SI{48.5}{\ohm}
resistor and a grounded \SI{1.8}{\ohm} resistor. Thereafter the signal
reaches each of the \num{36} detector segments via the capacitive
coupling of the core to the segments.

A detailed description of the newly developed segment preamplifiers is
given in \cite{Pascovici2008}. Three segment preamplifier channels are
integrated on one printed circuit board. The power consumption per
segment channel is limited to \SI{350}{\milli\watt} allowing the 108 closely
packed spectroscopic channels to be operated close to the vacuum
feed-throughs in air.

Differential signal outputs of the 111 spectroscopic channels are
transmitted through \num{21} MDR (Mini D Ribbon) high-speed digital
data-transmission cables.  The segment MDR connectors merge the
\num{6} segment signals of each sector. The core preamplifier has an
individual MDR connector for the preamplifier output signal and the
pulser control signals. Table 3 of ref. \cite{Wiens2010223} summarises
the most relevant specifications and their values.

\subsection{AGATA detector specifications}

The individual AGATA detectors have to meet the detector
specifications given by the collaboration. The core energy resolution
(FWHM) is specified to be better or equal to \SI{2.35}{\keV}
(\SI{1.35}{\keV}) at \SI{1.33}{\MeV} (\SI{122}{\keV}) and the peak
shape FWTM/FWHM (FWTM = full width at tenth maximum) is smaller than
\num{2.00}. The segment FWHM at \SI{1.33}{\MeV} (\SI{60}{\keV}) is
specified to be better or equal to \SI{2.30}{\keV} (\SI{1.30}{\keV})
with a mean value of the 36 segments values better or equal to
\SI{2.10}{\keV} (\SI{1.20}{\keV}). The cross talk between channels has
to be smaller than \num{1e-3}.  Upon delivery, the specifications are
verified during a customer acceptance test, which is performed by the
AGATA collaboration at three sites, IKP Cologne, University of
Liverpool and CEA Saclay. For these measurements single test cryostats
are used, which are equipped with \num{37} cold input
stages. Typically standard analogue commercial electronics is used for
energy-resolution measurements.  The cross-talk properties are
determined with a \num{37} channel coincidence electronics based on
high-speed digital sampling electronics. They are extracted from a
$^{60}$Co measurement after adding the coincident signals of any pair
of segments as the variation of the \SI{1332.5}{\keV} full-energy peak
position. The peak shift should not exceed \SI{0.65}{\keV}. The most
relevant parts of the specifications are summarised in detail in Table
1 of ref. \cite{Wiens2010223}.

\subsection{Performance of the AGATA triple cluster detector}

\begin{figure}[th]
  \centering
  \includegraphics[width=0.85\columnwidth]{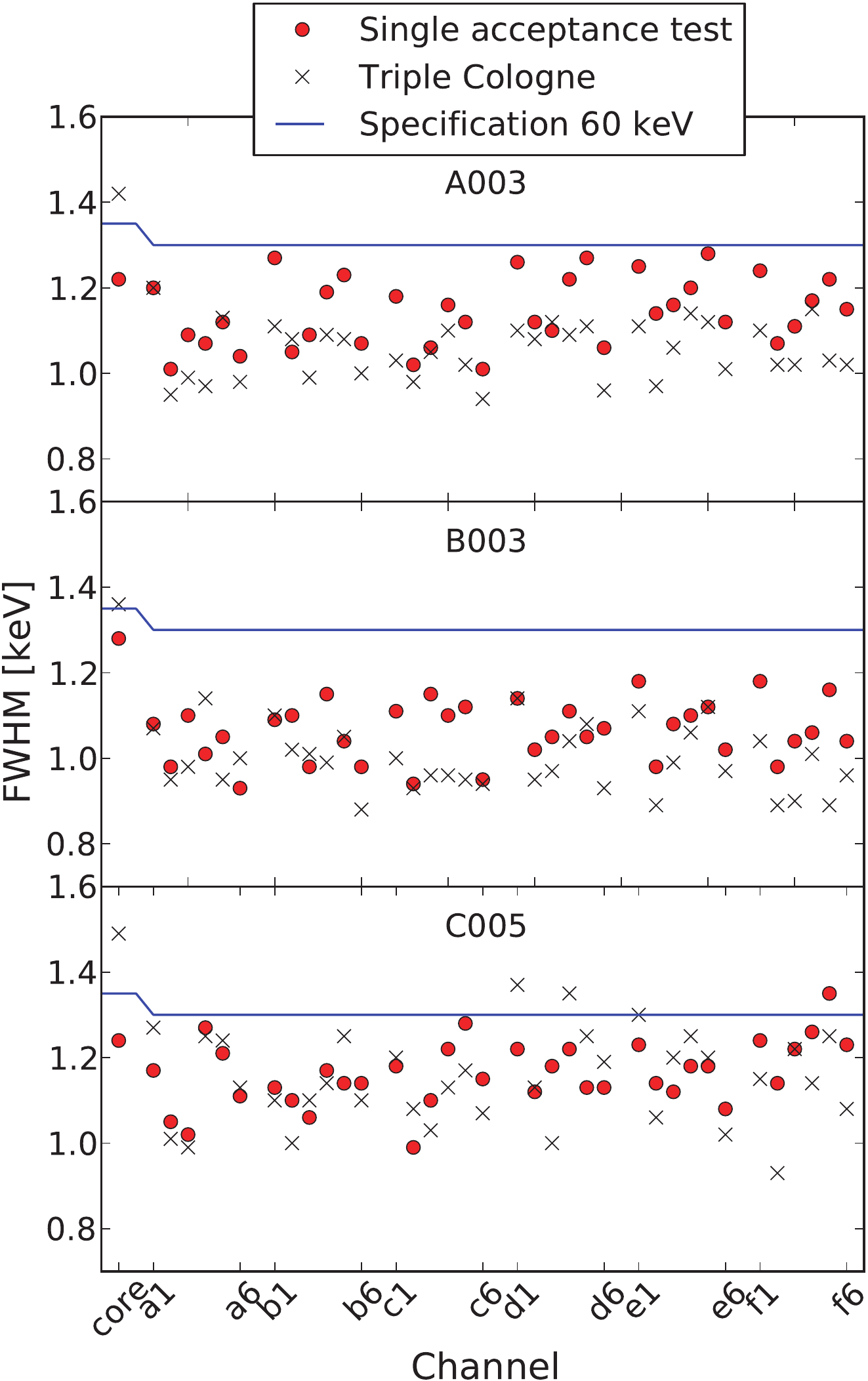}
  \caption{(Colour online) Energy resolution values of the crystals
    A003, B003 and C005 for the core signals at
    $E_{\gamma}=122$\;\si{\keV} and for the segment signals at
    $E_{\gamma}=60$\;\si{\keV}. The measurements were done at IKP
    Cologne using standard analogue electronics. The filled circles
    are the results of measurements performed with the crystals
    mounted in a single test cryostat while the crosses show
    measurements performed with the crystals mounted in a triple
    cluster detector (ATC2). The specification limits at
    \SI{122}{\keV} and \SI{60}{\keV} are shown as solid lines.}
  \label{fig:atc2_low}
\end{figure}

\begin{figure}[ht]
  \centering
  \includegraphics[width=0.85\columnwidth]{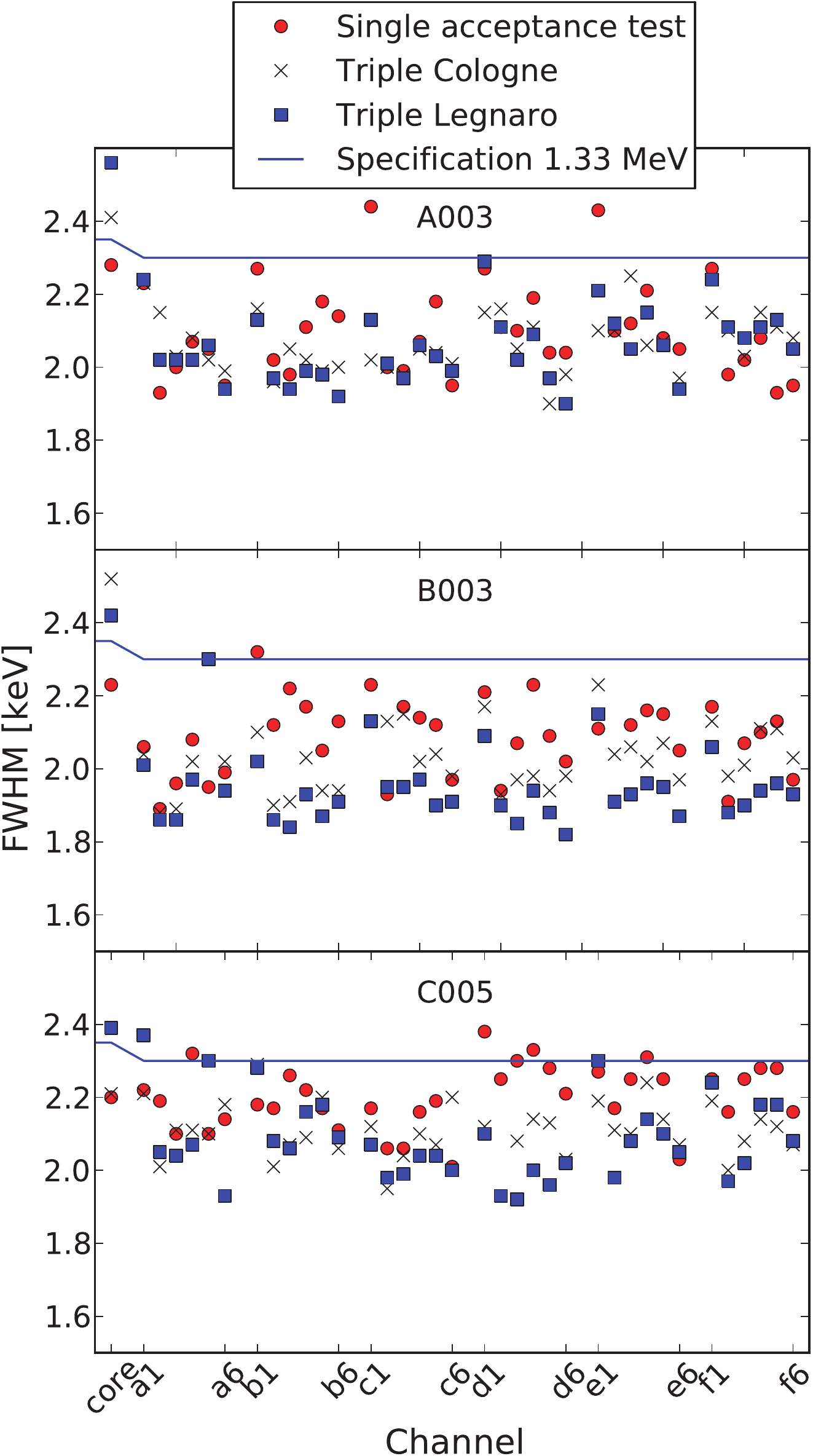}
  \caption{(Colour online) Energy resolution values of the crystals
    A003, B003 and C005 for the core and segment signals at
    $E_{\gamma}=1332.5$\;\si{\keV}. The filled circles and crosses
    show results of measurements performed at IKP Cologne using
    standard analogue electronics and with the crystals mounted in a
    single test cryostat and in a triple cryostat (ATC2),
    respectively. The filled squares show results of measurements
    performed at LNL with ATC2 mounted on the frame and by using the
    AGATA digital electronics and data acquisition system.  The energy
    specification limit at \SI{1332.5}{\keV} is shown as a solid line.
  }
\label{fig:atc2_high}
\end{figure}

The ATC detector is equipped with three single core preamplifier
boards and $3{\times}12$ triple segment preamplifier boards. Energy
resolutions at \SI{60}{\keV} for all segments and at \SI{122}{\keV}
for the cores, are measured with analogue electronics. At higher
energies (\SI{1332.5}{\keV}), the measurements are performed also with
digital electronics.  The results obtained with a triple cryostat are
compared in Fig.~\ref{fig:atc2_low} at low energy and in
Fig.~\ref{fig:atc2_high} at high energy with the measured performance
of the same crystals in a single test cryostat.  Average values of the
energy resolution measurements for the first five ATC detectors are
summarised in Tables~\ref{tab:ATC1_performance} and
~\ref{tab:atcs_fwhm_high}.

\begin{table*}		
  \centering
  \caption{Energy resolution values for the first five triple cluster
    detectors measured with analogue electronics at IKP Cologne. The
    core FWHM values were measured at \SI{122}{\keV}. The segment's
    average values and their standard deviations were measured at
    \SI{60}{\keV}. Not all the measurements were performed for ATC5.
    \vspace*{0.5\baselineskip}
  }
  \label{tab:ATC1_performance}
  \begin{tabular}{|c|c|cc|cc|}
    \hline
    \multirow{3}{*}{Detector} & \multirow{3}{*}{Crystal} &
    \multicolumn{2}{c}{Core FWHM [keV]} &
    \multicolumn{2}{|c|}{Segment average FWHM [keV]} \\
    \cline{3-6}
    & & Single   & Triple    & Single   & Triple \\
    & & cryostat & cryostat  & cryostat & cryostat \\
    \hline
    \multirow{3}{*}{ATC1}
    & A001 &1.34 &1.44 &1.08 $\pm$ 0.07 &1.01 $\pm$ 0.05\\
    & B002 &1.29 &1.41 &1.09 $\pm$ 0.09 &1.04 $\pm$ 0.07\\
    & C002 &1.28 &1.21 &1.03 $\pm$ 0.08 &0.97 $\pm$ 0.06\\
    \hline
    \multirow{3}{*}{ATC2}
    & A003 &1.22 &1.42 &1.14 $\pm$ 0.08 &1.05 $\pm$ 0.07\\
    & B003 &1.28 &1.36 &1.06 $\pm$ 0.07 &1.00 $\pm$ 0.07\\
    & C005 &1.24 &1.49 &1.16 $\pm$ 0.07 &1.14 $\pm$ 0.11\\
    \hline
    \multirow{3}{*}{ATC3}
    & A002 &1.26 &1.44 &1.03 $\pm$ 0.08 &0.93 $\pm$ 0.11\\
    & B005 &1.08 &1.43 &1.05 $\pm$ 0.08 &1.05 $\pm$ 0.08\\
    & C006 &1.09 &1.42 &1.15 $\pm$ 0.10 &1.14 $\pm$ 0.11\\
    \hline
    \multirow{3}{*}{ATC4}
    & A005 &1.23 &1.28 &1.04 $\pm$ 0.09 &1.11 $\pm$ 0.17\\
    & B001 &1.29 &1.27 &1.02 $\pm$ 0.07 &1.03 $\pm$ 0.10\\
    & C003 &1.16 &1.33 &1.00 $\pm$ 0.09 &1.11 $\pm$ 0.38\\
    \hline
    \multirow{3}{*}{ATC5}
    & A004 &1.27 &1.21 &1.17 $\pm$ 0.08 &               \\
    & B009 &1.36 &1.54 &1.11 $\pm$ 0.07 &1.05 $\pm$ 0.11\\
    & C004 &1.30 &     &1.11 $\pm$ 0.08 &               \\
    \hline
  \end{tabular}
\end{table*}

\begin{table*}		
  \centering
  \caption{Energy resolution values (FWHM) at \SI{1332.5}{\keV} for
    the first five AGATA triple cluster detectors. The single cryostat
    measurements were performed at IKP Cologne with analogue
    electronics as was the measurements marked ATC Cologne. The
    measurements at LNL were performed with the ATC detectors mounted
    on the frame using the AGATA digital electronics and data
    acquisition system. Average values and their standard deviations
    are given for the segments. Not all the measurements were
    performed for ATC5 at Cologne.
    \vspace*{0.5\baselineskip}
  }
  \label{tab:atcs_fwhm_high}
  \begin{tabular}{|c|c|ccc|ccc|}
    \hline
    \multirow{3}{*}{Detector} & \multirow{3}{*}{Crystal} &
    \multicolumn{3}{c}{Core FWHM [keV]} &
    \multicolumn{3}{|c|}{Segment average FWHM [keV]} \\
    \cline{3-8}
    & & Single   & ATC at  & ATC at  & Single   & ATC at  & ATC at\\
    & & cryostat & Cologne & LNL     & cryostat & Cologne & LNL\\
    \hline
    \multirow{3}{*}{ATC1}
    &A001 &2.33 &2.46 &2.50 &$2.09 \pm 0.16$ &$2.19 \pm 0.10$ &$2.01 \pm 0.13$ \\
    &B002 &2.27 &2.46 &2.43 &$2.13 \pm 0.11$ &$2.10 \pm 0.14$ &$1.99 \pm 0.09$\\
    &C002 &2.25 &2.33 &2.42 &$2.03 \pm 0.12$ &$2.11 \pm 0.12$ &$1.94 \pm 0.11$\\
    \hline
    \multirow{3}{*}{ATC2}
    &A003 &2.28 &2.41 &2.56 &$2.10 \pm 0.13$ &$2.06 \pm 0.08$ &$2.06 \pm 0.10$\\
    &B003 &2.23 &2.52 &2.42 &2.08 $\pm$	0.11 &2.02 $\pm$ 0.09 &1.94 $\pm$ 0.08\\
    &C005 &2.20 &2.21 &2.39 &2.21 $\pm$	0.09 &2.21 $\pm$ 0.08 &2.08 $\pm$ 0.11\\
    \hline
    \multirow{3}{*}{ATC3}
    &A002 &2.31 &2.40 &2.52 &2.07 $\pm$ 0.11 &2.02 $\pm$ 0.09 &1.98 $\pm$ 0.09\\
    &B005 &2.29 &2.42 &2.49 &2.09 $\pm$ 0.14 &2.13 $\pm$ 0.11 &2.04 $\pm$ 0.13\\
    &C006 &2.16 &2.27 &2.58 &2.12 $\pm$ 0.09 &2.09 $\pm$ 0.09 &2.13 $\pm$ 0.15\\
    \hline
    \multirow{3}{*}{ATC4}
    &A005 &2.23 &2.40 &2.19 &2.03 $\pm$ 0.10 &2.08 $\pm$ 0.13 &1.91 $\pm$ 0.11\\
    &B001 &2.17 &2.50 &2.30 &2.06 $\pm$ 0.11 &2.04 $\pm$ 0.11 &1.91 $\pm$ 0.11\\
    &C003 &2.34 &2.35 &2.40 &$2.08 \pm 0.11$ &$2.08 \pm 0.09$ &$2.04 \pm 0.21$\\
    \hline
    \multirow{3}{*}{ATC5}
    &A004 &2.31 &2.36 &2.33 &$2.10 \pm 0.11$ &                &$2.04 \pm 0.12$\\
    &B009 &2.33 &2.49 &2.63 &$2.03 \pm 0.14$ &                &$1.96 \pm 0.14$\\
    &C004 &2.23 &     &2.26 &$2.17 \pm 0.10$ &                &$2.04 \pm 0.24$\\
    \hline
  \end{tabular}
\end{table*}

The values obtained for the segments in the triple configuration are
on average even better than in the single test cryostat. Since these
resolutions are dominated by electronic noise, it demonstrates the
successful design and integration of the new AGATA triple cluster
detector. Especially the electronic properties comprising the cold and
warm parts of the new AGATA preamplifier assembly is causing very low
noise contributions in the triple cryostat despite the high
integration density of 111 analogue channels. In addition, an improved
grounding was applied as a result of various iterations during the
project.  This reduces unwanted high-frequency and noise components
and brings the energy resolution of the ATC detector at low energy to
a value well within the specification.

\begin{figure}[ht]
  \centering
  \includegraphics[width=0.99\columnwidth]{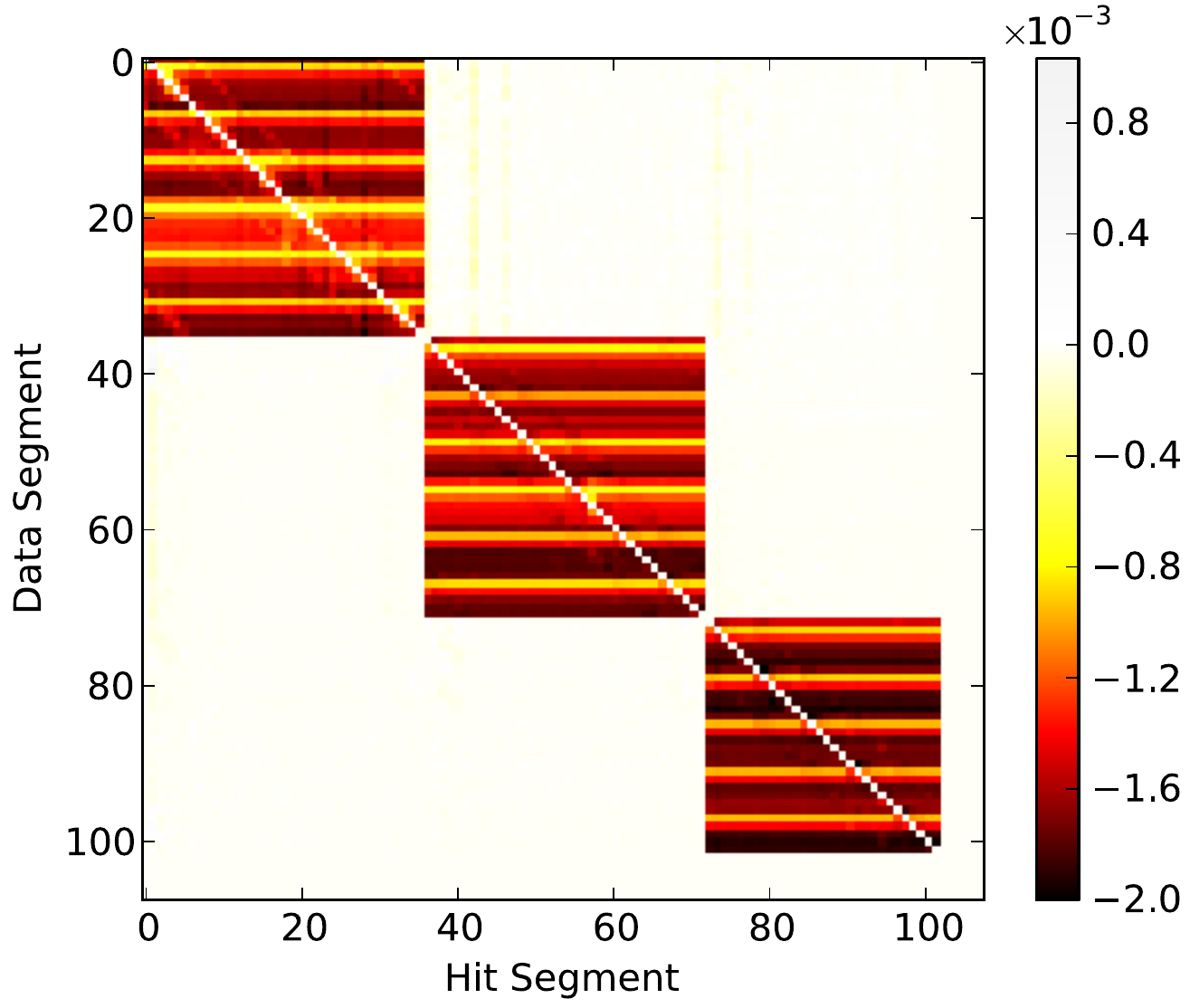}
  \caption{(Colour online) Observed relative cross-talk contributions
    of nearly all segment combinations between the three crystals of
    the ATC2 detector are plotted (only 105 channels were operational
    at the time of the measurement).  The colour scale gives the
    relative cross talk in units of \num{e-3}.  A relative energy
    shift on the $\sim$\num{e-3} level is caused by cross talk for
    segment combinations within one of the three detectors. Cross talk
    between the three different detector capsules is observed to be
    negligible. The other tested ATC detectors show similar results.
  }
  \label{fig:crosstalk}
\end{figure}

The cross-talk contributions were investigated by analysing the
coincident traces over a \SI{7}{\micro\second}-long time period using the
digital acquisition system on all three detectors. After
identification of the true energy deposition in exactly one segment
the coincident and simultaneous baseline shifts, which occur in all
remaining \num{107} nonhit segments, are recorded for these one-fold
events.  The correlation between the energy deposition in a single
detector segment and the energy shift in all other segments is
determined over an energy range given by $\gamma$-ray emission from
$^{60}$Co and $^{137}$Cs sources. The observed cross talk is within
the specifications and cross-talk contributions are only observed for
segment combinations within the same detector crystal. The regular
pattern is mainly caused by the different capacitances between the
core and segment electrodes. A subset of $105{\times}105$ combinations
is shown in Fig~\ref{fig:crosstalk}.

The method applied to quantify this result is based on all possible
combinations, shown in Fig.~\ref{fig:crosstalk} for an ATC detector
with the $110{\times}111$ possible cross-talk matrix elements within
the full triple cryostat. The cross-talk contributions between
segments of different detectors are on the \num{e-4} to \num{e-5}
level, which is well within the acceptable limits and which can be
disregarded for a standard operation of the ATC detector.

The new method to determine precise and absolute cross-talk matrix
elements was applied to all available AGATA triple detectors.  The
cross-talk pattern was measured to be at the \SI{0.1}{\percent} level,
comparable with the values shown in Fig~\ref{fig:crosstalk}. The
observed structure can be entirely attributed to the capacitive
coupling between core and segments via the bulk Ge material. This
behavior is well reproduced by a linear electronic model. This
cross-talk contribution is considered as the inherent cross-talk limit
given by the construction of the new detectors
\cite{Bruyneel200999}. The calculated values are closely approached by
the obtained results. The cross-talk contributions in all AGATA
detectors behaved very similarly, implying that the development of the
AGATA cryostats and the preamplifier electronics has progressed in
such a way that the fundamental constraints are accounted for,
detectable and understood.

\section{Detector characterisation} \label{s:detchar}
The success of the AGATA project relies on the ability to reconstruct
the trajectories of $\gamma$ rays scattered within the germanium
detectors. The deposited energy and the location of the photon
interactions can be extracted from the measurement and the analysis of
the waveforms that arise on the segments during the charge
collection. The pulse-shape analysis algorithms currently developed
use databases of calculated pulse shapes. These calculated pulses need
to be validated with real pulse shapes taken at various points within
a detector. For the development of tracking algorithms, it is also
crucial to determine experimentally the interaction position
sensitivity in three dimensions in the whole volume of a detector.

The AGATA collaboration has performed a detailed analysis of the
response function of the crystals. This work included developing a
theoretical basis data set, which describes the detector response
function and then validating this against the equivalent experimental
data. Such knowledge provides the project with the information
necessary to enable pulse-shape analysis and $\gamma$-ray tracking.

The AGATA collaboration has two operational experimental
characterisation centres, based at the University of Liverpool in the
UK and at CSNSM Orsay in France. Three new centres are being
commissioned at GSI Darmstadt in Germany, at IPHC Strasbourg in France
and at the University of Salamanca in Spain.  This would significantly
increase the number of AGATA detectors that could be experimentally
characterized.

\subsection{Liverpool scanning system} \label{ss:lss}

\begin{figure}[ht]
  \begin{center}
    \includegraphics[width=\columnwidth]{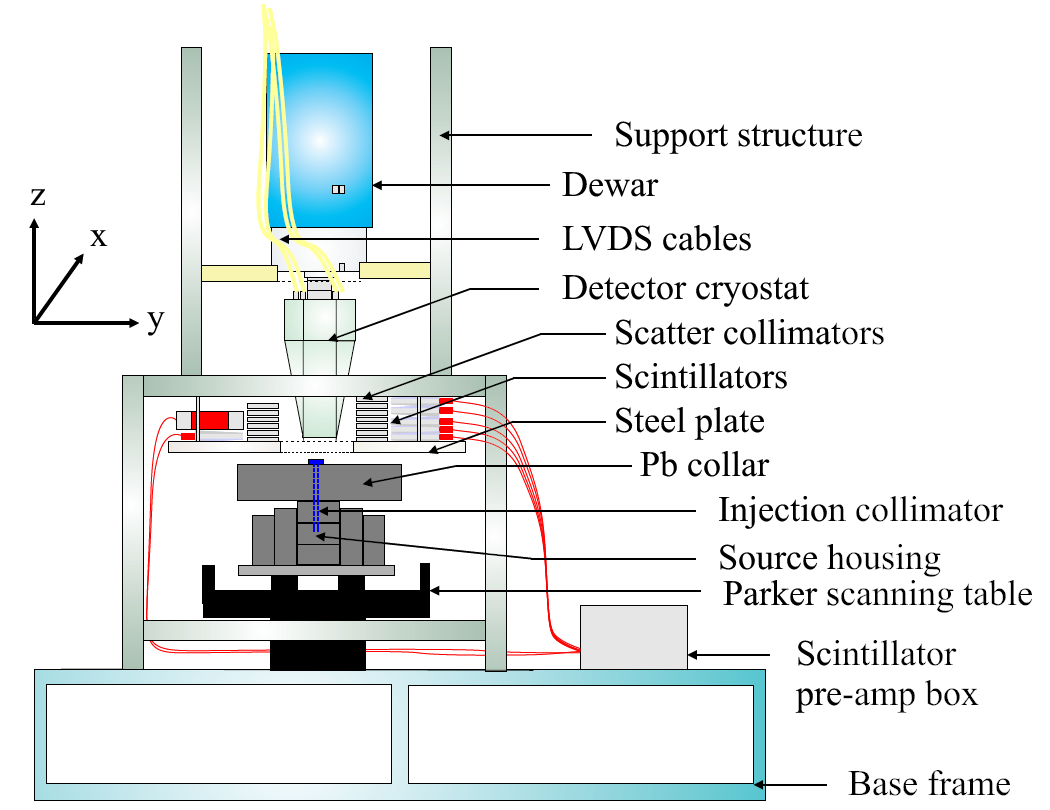}
    \caption{(Colour online) Schematic diagram of the University of
      Liverpool scanning table assembly.}
    \label{fig:liverpool_scanning_system}
  \end{center}
\end{figure}

A schematic diagram outlining the University of Liverpool AGATA
detector scanning system is shown in
Fig.~\ref{fig:liverpool_scanning_system} \cite{Descovich2005535}. The
figure displays the mechanical configuration of the system, with the
detector vertically mounted above the collimated source assembly. A
\SI{920}{\mega\becquerel} $^{137}$Cs source is mounted at the end of a
\SI{1}{\mm} diameter coaxial tungsten collimator of \SI{120}{\mm} in
length. The collimator is mounted in a lead collar and source housing
assembly, which shields the system operators from the mounted source.
The whole assembly is mounted on a precision Parker $x$-$y$
positioning table. The table is moved in precise computer-controlled
steps through the use of Pacific Scientific stepper motors and two
Parker Automation axis indexers. The system has a position accuracy of
\SI{100}{\micro\metre} and can scan over an area of the size
\SI{30x30}{\cm\square}.

A steel frame is constructed around the scanning table. From this, a
steel plate is suspended by threaded rods, on which the scatter
collimators and scatter detectors can be supported. The rods enable
the plate, and hence the collimators, to be moved in the
$z$-direction. The steel plate has a square cut from its centre and is
positioned to a height just above the lead collar. The AGATA detector
is then inserted into the frame and positioned in the centre of the
plate.

The experimental preamplifier pulse shapes observed as a function of
the interaction position of a beam of $\gamma$ rays can be
recorded. The signals are digitized by utilizing either the GRT4 VME
modules \cite{Lazarus2004}, which sample the signal with an
\SI{80}{\MHz} frequency over a $14$-bit dynamic range, or the GRETINA
VME digitizer modules, which have a \SI{100}{\MHz} sampling frequency
and the same $14$-bit dynamic range. Both digitizer systems utilize an
external trigger provided by the core-energy signal from the AGATA
detector. Data is read out and recorded simultaneously for all
\num{37} channels from a single crystal.

Scan data can be collected in singles and coincidence modes.  Singles
data yields $x$-$y$ information on the interaction position of a
$\gamma$ ray.  Such a measurement is relatively quick to perform;
however, the $z$-position information has large uncertainties, as it
is only defined by the segment size. Coincidence scanning utilizes
Compton scattering to define a single interaction position in
$x$-$y$-$z$.  Such a methodology demands that the $\gamma$ ray Compton
scatters to an angle of about \ang{90} in the germanium
detector, depositing the remaining energy in coincidence in a
collimated ring of scintillation detectors. This method is very
precise but can result in a very slow procedure due to the low
coincidence rate between the germanium detector and the scintillators.
A full characterisation of a crystal with a grid of \SI{1}{\mm} takes
2-3 months.  A detailed analysis of singles and coincidence data for
the prototype AGATA symmetric detectors can be found in
\cite{Dimmock2009a,Dimmock2009b}.

\begin{figure}[ht]
  \begin{center}
    \includegraphics[width=\columnwidth]{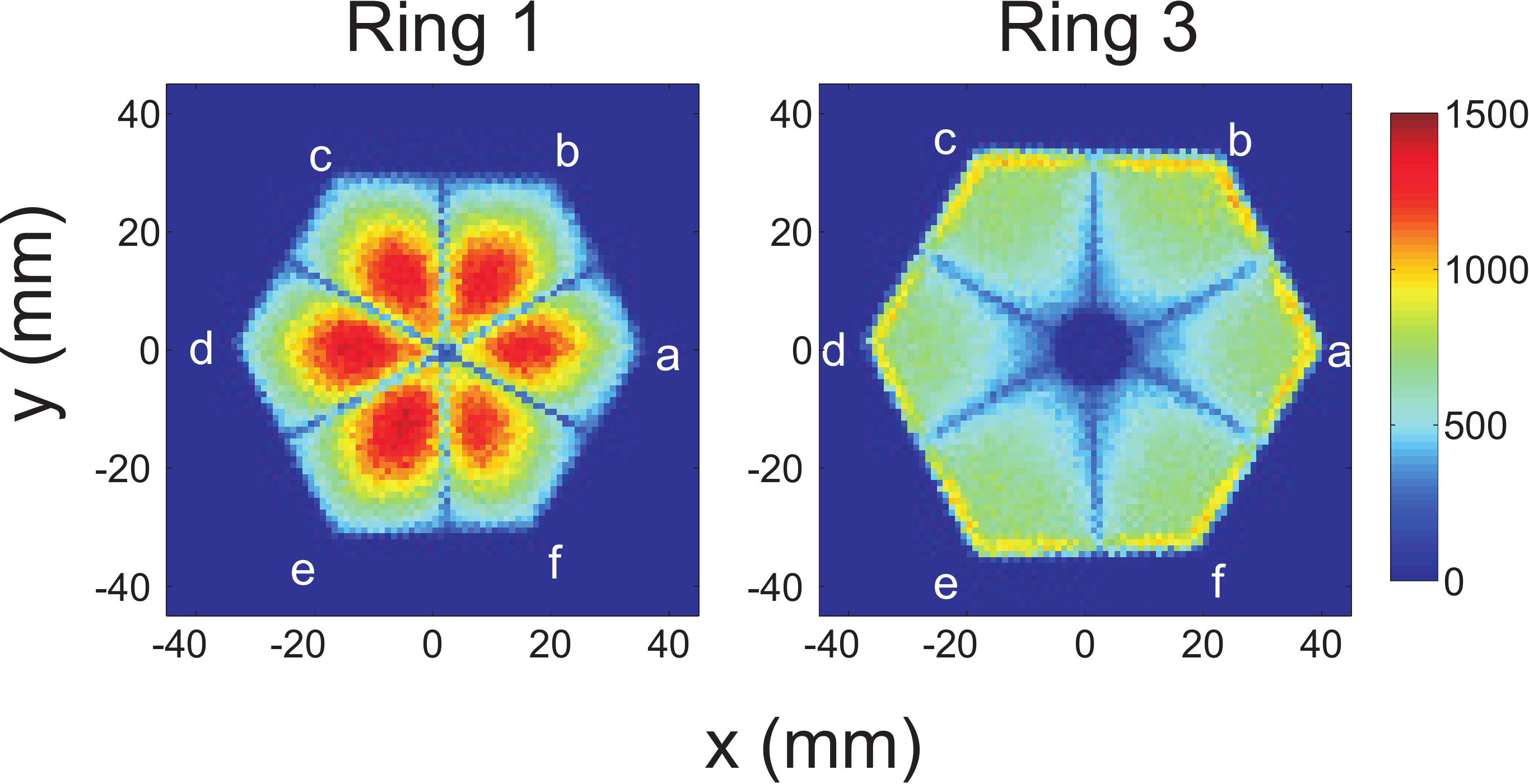}
    \caption{(Colour online) Distribution of the intensity of
      \SI{662}{\keV} full-energy depositions confined to a single
      segment in rings 1 and 3 of crystal C001. The colour scale gives
      the number of counts in the \SI{662}{\keV} peak recorded in
      \SI{1}{\minute}.  }
    \label{fig:singlesint}
  \end{center}
\end{figure}

An example of a $^{137}$Cs singles scan of the front face of crystal
C001 is shown in Fig.~\ref{fig:singlesint}. Data were recorded on a
\SI{1}{\mm} grid for \SI{1}{\minute} at each position. The system was
triggered externally with a low-energy threshold of
$\sim$\SI{650}{\keV} on the core contact in order to eliminate
unwanted Compton scattered or background events. The resulting
distributions of the intensity of \SI{662}{\keV} full-energy
depositions confined to a single segment in rings 1 and 3 are shown in
Fig.~\ref{fig:singlesint}. The $x$ and $y$ axes represent the position
of the scanning table in a range of \SI{\pm 40}{\mm}.  The plots
clearly show the segmentation pattern of the detector and the presence
of the coaxial hole in ring 3.

\begin{figure}[ht]
  \begin{center}
    \includegraphics[width=\columnwidth]{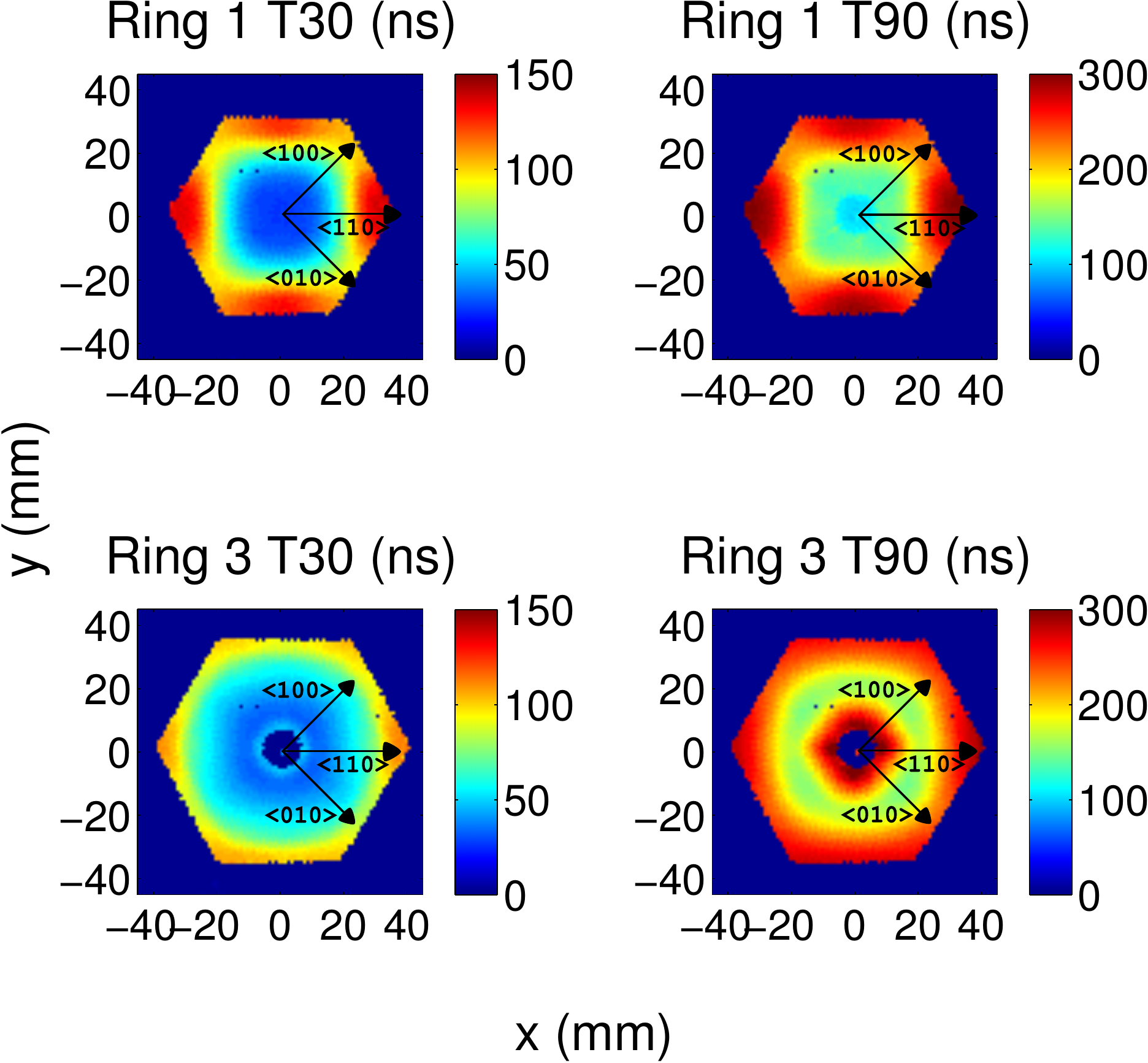}
    \caption{(Colour online) [Left] The distribution of the $T30$
      ($10$-$30\%$) rise time of the core signal for \SI{662}{\keV}
      full-energy depositions confined to a single segment in ring 1
      (top) and 3 (bottom) of crystal C001.  [Right] The corresponding
      $T90$ ($10$-$90\%$) distribution. The crystallographic axes are
      illustrated. The colour scale gives the rise time in ns.
    }
    \label{fig:corerisetime}
  \end{center}
\end{figure}

The distributions of the $T30$ (\SIrange{10}{30}{\percent} of the
maximum amplitude) and $T90$ (\SIrange{10}{90}{\percent}) rise times
of the core signal are plotted in Fig.~\ref{fig:corerisetime}.  In a
coaxial $n$-type crystal, such as those used in AGATA, the $T30$
distribution is dominated by the drift time of the electrons towards
the core contact, while the $T90$ distribution is determined by both
the electron and the hole transport.  As a result, $T30$ is expected
to increase as the interaction point is moved from the core contact to
the outer electrode, while a minimum value for $T90$ is expected at
the locations in which the electron and hole collection times are
equal. This behaviour is confirmed by the plots shown in
Fig.~\ref{fig:corerisetime}.  In ring 3, $T30$ ranges from 
$\sim$\SI{30}{\ns} for small radii to $\sim$\SI{100}{\ns} for large radii,
while $T90$ ranges from $\sim$\SI{100}{\ns} to $\sim$\SI{240}{\ns}.  For
ring 1 (front ring of the crystal) both the $T30$ and $T90$
distributions have a different character due to the more complex
electric field distribution in this region of the detector. Minimum
rise times are observed at small radii due to the quasi-planar nature
of the electric field leading to short charge collection times through
the \SI{13}{\mm} distance from the front face to the hole drill depth
in the crystal.  These plots also show the influence of the
face-centred cubic lattice orientation of the Ge crystal on the rise
times. For pulses measured at the same radius, a maximum variation of
\SI{30}{\percent} with respect to the crystal axes is observed for the
time required to collect the charge carriers.  This effect must be
taken into account in the theoretical simulation, if a reliable
validation is to be achieved.

The coincidence between an AGATA detector and an array of scintillator
crystals can be used to select interactions at a specific location
within the crystals.  In practice, several events for each location
are needed in order to average the corresponding waveforms and
eliminate, as much as possible, the effects of the noise.  The
averaging procedure, performed for each location, starts with a
scaling of each waveform by pre-calculated gain factors, derived from
the $^{152}$Eu baseline difference energy calibration, and with a
baseline subtraction. The baseline is derived for each individual
trace from an average of the initial 10 samples in each trace.

\begin{figure}[ht]
  \begin{center}
    \includegraphics[width=1.0\columnwidth]{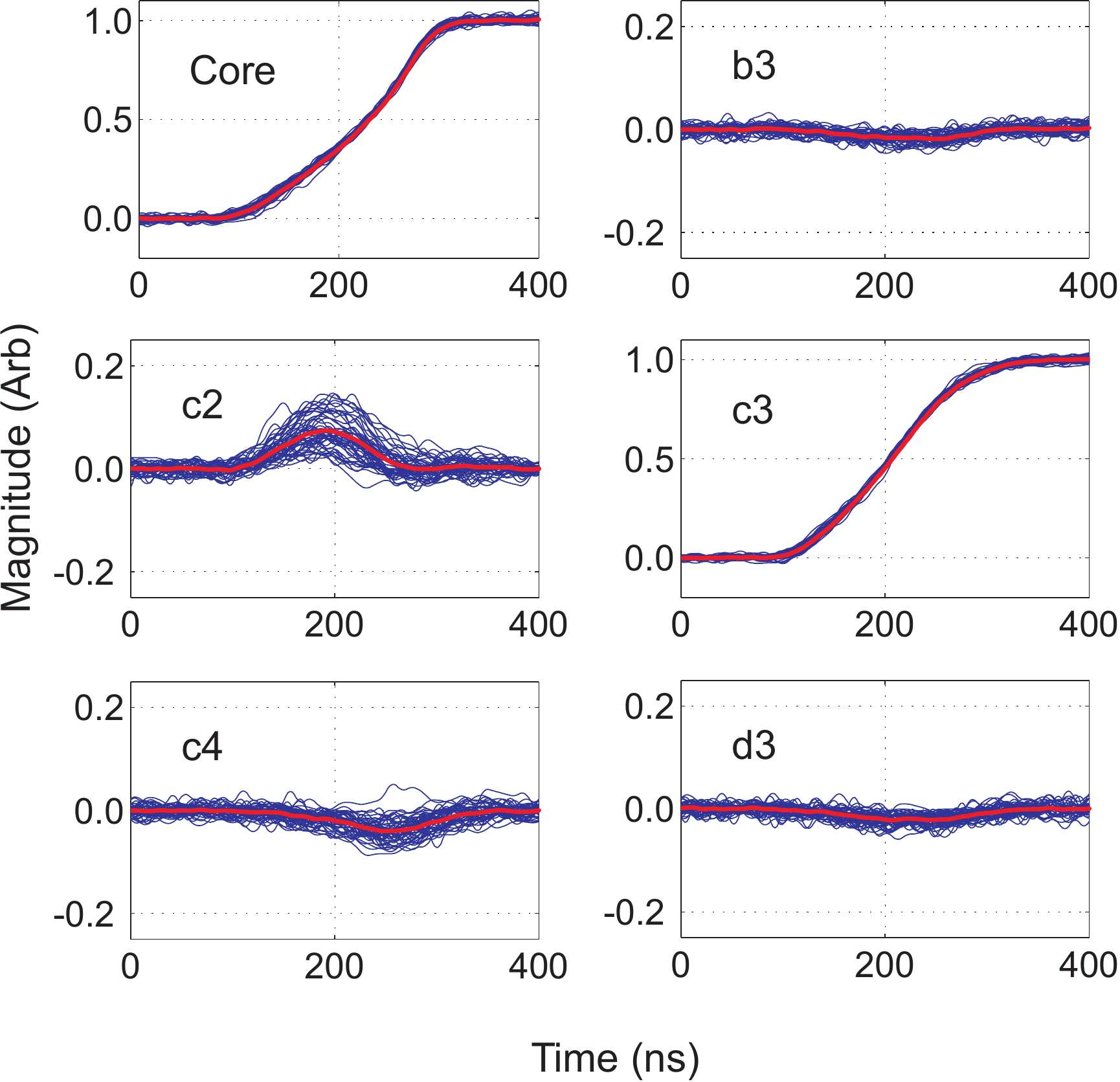}
    \caption{(Colour online) Average (thick red line) and all
      constituent (thin blue lines) pulse shapes for a typical
      interaction in segment c3 following the $\chi^2$ rejection (see
      text). The signal induced in the core and in the neighbouring
      segments b3, c2, c4 and d3 are also shown.}
    \label{fig:meanform}
  \end{center}
\end{figure}

Waveforms are then interpolated to allow for more accurate time
alignment. The pulse amplitudes are subsequently normalized in order
to have the same maximum amplitude for all of them.  Finally, the best
fit average waveforms corresponding to each location are obtained
through a $\chi^2$ minimisation procedure.  Only the central contact,
the segment with net charge deposition and its neighbours are
considered in the fit. ``Noisy'' events which give a large $\chi^2$
contribution are excluded from the fit procedure.  The final result is
exemplified in Fig.~\ref{fig:meanform}, where the average (thick red
line) and the constituent (thin blue lines) pulse shapes are shown for
a net charge deposition in segment c3. The effect of the cancellation
of the random noise across the pulses is clearly visible.  The
standard deviation of the baseline noise for the average pulses is
\SI{0.9}{\keV}, as opposed to \SI{4.7}{\keV} for single pulses.  The
pulse shapes illustrated in Fig.~\ref{fig:meanform} also demonstrate
the signal induced on the neighbouring segments b3, c2,c4 and
d3. These transient signals are those induced on adjacent electrodes
to the primary interaction due to the drift of the charge carriers
inside the germanium crystal.

\subsection{Orsay scanning system} \label{ss:oss}

The Orsay scanning system is based on the same concept as the
Liverpool system, i.e. a well collimated strong radioactive source, an
accurate moving system and an array of scintillator detectors to
define the $z$ position of the scattering through coincidence
measurements. The main difference is that the $z$ coordinate can be
continuously scanned due to the absence of the scatter collimators,
which are replaced by special tungsten collimators described below.

\begin{figure}[ht]
  \begin{center}
    \includegraphics[width=\columnwidth]{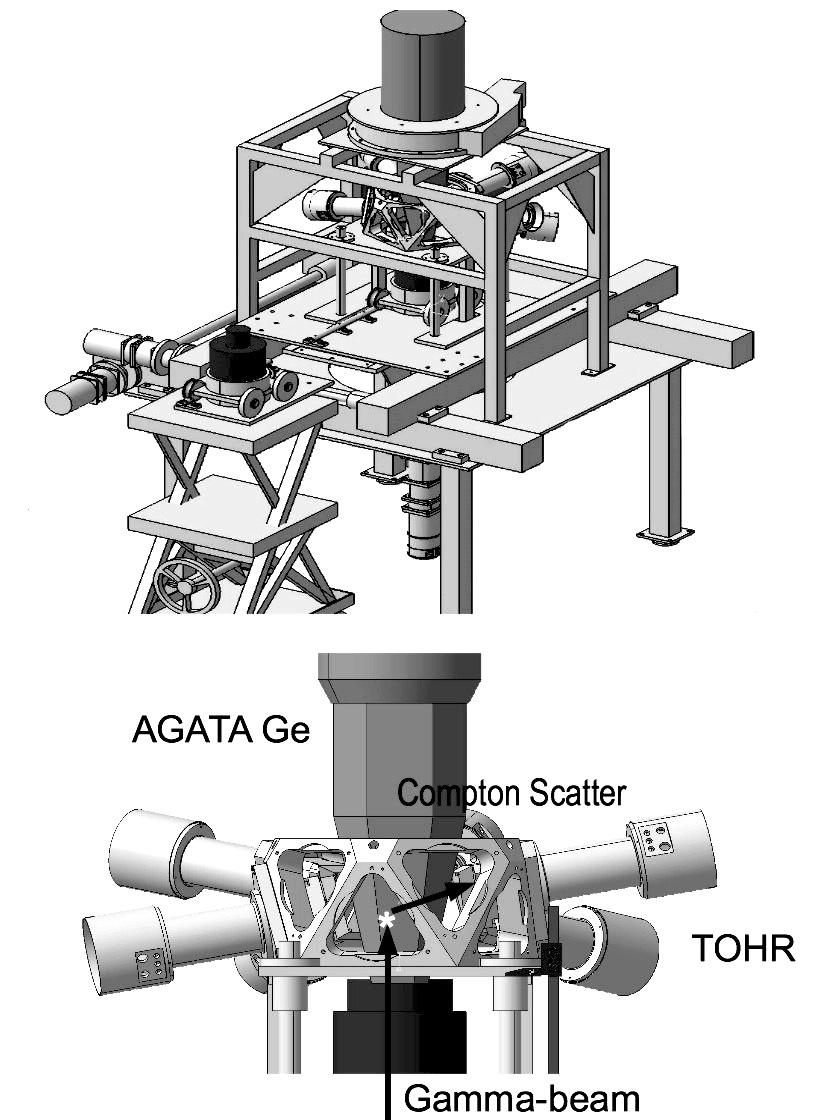}
    \caption{A schematic diagram showing the CSNSM Orsay scanning
      table assembly (top) and a closer view (bottom) showing the
      \num{6} NaI(Tl) detectors and the mechanical support for the
      TOHR, the AGATA detector and the collimated intense $^{137}$Cs
      source.  }
    \label{fig:orsay_scanning_system}
  \end{center}
\end{figure}

A schematic view of the scanning setup based at CSNSM Orsay is shown
in Fig.~\ref{fig:orsay_scanning_system} \cite{Ha2011}. A
\SI{477}{\mega\becquerel} $^{137}$Cs source is encapsulated in a
stainless steel cylindrical container with a diameter of \SI{4}{\mm},
a height of \SI{6}{\mm} and window thickness of \SI{0.4}{\mm}. This
container is inserted into a collimator made of densimet (W-Ni-Fe
alloy, density \SI[per=slash]{18.5}{\gram\per\cubic\cm}). The
$\gamma$ rays emitted by the source are collimated by a hole with a
diameter of \SI{1.6}{\mm} and a length of \SI{155}{\mm}.

Six modules of the TOHR (TOmographe Haute R\'esolution) detector are
used to perform the coincidence measurements. Each module is made of a
stack of \num{80} tungsten plates with a triangular shape and a
thickness of \SI{200}{\micro\metre}. These plates have
\SI{400}{\micro\metre} diameter holes positioned on a hexagonal
lattice and the geometry of each stack acts as a many slit collimator
(about 8000 slits) with a focal distance of $\sim$\SI{7}{\cm}.  At
the back of each stack of plates, there is a NaI(Tl) crystal for the
detection of $\gamma$ rays. The six modules are positioned in a
compact semi-circle at \SI{{\pm}10.2}{\degree} from the horizontal
plane around the AGATA detector, all having the same focal point in
the germanium crystal. The position of the common focal point can be
changed by translating the TOHR array or by turning the AGATA detector
about its central axis; thus allowing for a full 3D scan of the AGATA
detector. A more detailed description and measurements can be found in 
Ref. \cite{Ha2009}.

\subsection{Other scanning systems} \label{ss:gss}

Three additional scanning systems are under construction and
validation within the AGATA collaboration. At the University of
Salamanca a system, which is similar to the Liverpool and Orsay
systems, is being developed, while different concepts of scanning are
under validation in Strasbourg and at GSI. The technique used in
Strasbourg is based on the pulse shape comparison scan (PSCS)
principle \cite{Crespi2008}. In this setup the detector is rotated
instead of the source in order to scan the front and lateral sides of
the entire germanium crystal with a collimated source of $^{137}$Cs.

A new characterisation method has been developed at GSI
\cite{DomingoPardo201179,Goel2011591}. This approach combines the PSCS
principle with an advanced $\gamma$-ray imaging technique based on
positron annihilation Compton scattering \cite{Tickner2004,Gerl2004}.
The imaging is accomplished by means of a position-sensitive
scintillation detector (PSD) \cite{Domingo2009} coupled to a $^{22}$Na
source with an activity of \SI{300}{\kilo\becquerel}. In order to
reduce the background due to scattered $\gamma$ rays, the $^{22}$Na
source is placed in the centre of a tungsten collimator with two
conical openings, one of which is oriented towards the germanium
crystal, the other towards the PSD. The setup with the PSD, the source
and the collimator can be rotated around a vertical axis at which the
germanium crystal is located.  Two measurements are typically carried
out, in order to shine and image the entire crystal from two different
sides. The PSD is operated in coincidence with the AGATA crystal. For
each AGATA-PSD coincidence event, the traces from the core and all 36
segments are digitised and stored. The position of the $\gamma$-ray
interaction in the PSD is also recorded. The latter allows for the 3D
reconstruction of the trajectories of the two \SI{511}{\keV}
annihilation quanta, which are assumed to be collinear. By applying
the PSCS principle, the detector response for a particular
$\gamma$-ray interaction point $x$-$y$-$z$ in the AGATA crystal can be
determined by comparison of the two data sets of pulse shapes,
corresponding to the two scanned sides. The duration of a full scan of
an AGATA detector on a \SI{2}{\mm} pitch grid will be strongly reduced
compared to existing systems, down to about \num{10} days at
Strasbourg and \num{3} days at GSI.

\section{The AGATA infrastructure} \label{s:infra}
The AGATA infrastructure includes the mechanical structure and all
services to the detectors, mechanics, electronics and to the data
acquisition system to ensure that spectrometer operates reliably. It
includes the ``life-support system'' for the detectors providing the
cryogenic cooling, the low- and high-voltage power supplies, constant
monitoring, a user-friendly interface and reports on critical
situations (detector warm up, power losses, etc.). This system is
called the detector-support system (DSS).

\subsection{The AGATA mechanics}

AGATA requires a mechanical structure to accurately support the
detector elements and enable their safe insertion and removal. The
structure needs to be able to locate the detectors accurately with
minimal space between each ATC detector in order to maximise the solid
angle coverage.

At LNL, AGATA is located at the target position of the PRISMA
\cite{Stefanini2002217} magnetic spectrometer. The main design
constraint imposed by PRISMA is that both AGATA and PRISMA must be
free to rotate around the beam direction such that the optical axis of
PRISMA ranges from \ang{0} to \ang{117}. Both AGATA and PRISMA are
therefore mounted on a rotating platform that also supports the
front-end electronics (digitisers), power supplies and autofill
system. In addition, the arrangement at LNL allows the coupling to a
range of ancillary detectors for specific measurements. This setup,
including all the detector systems, is described in detail in
Ref. \cite{agata_installation_at_lnl}.

\begin{figure}[ht]
  \centering
  \includegraphics[width=0.90\columnwidth]{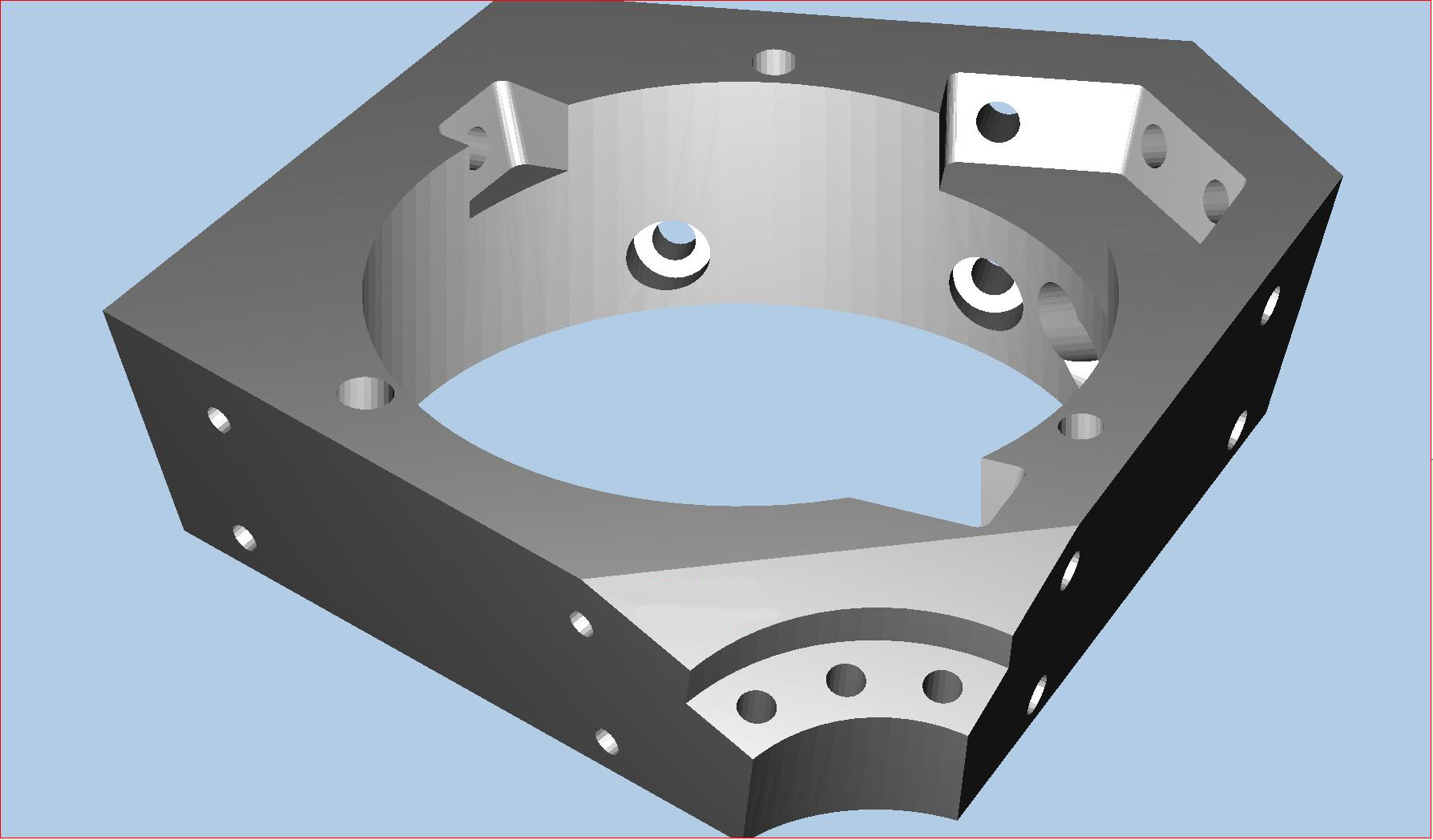}
  \caption{(Colour online) A computer aided design drawing of one of
    the main AGATA support flanges.}
  \label{fig:agata_flange}
\end{figure}

\begin{figure}[ht]
  \centering
    \includegraphics[width=0.90\columnwidth]{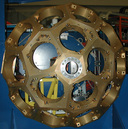}
  \caption{ (Colour online) A photograph of 15 flanges mounted at
    LNL.}
  \label{fig:agata_support_structure}
\end{figure}

The generic support structure for AGATA consists of a number of
identical flanges (Fig.~\ref{fig:agata_flange}), one for each detector
module.  These are assembled together to produce a solid structure as
shown in Fig.~\ref{fig:agata_support_structure}.  This generic support
structure is modular in concept and can be expanded up to the full
$4\pi$ system with 60 such flanges and will be used at all the
host laboratories.  The frame to support this structure will be
different at each site because of the details of the location and the
coupling to different spectrometers.

\begin{figure}[ht]
  \centering
  \includegraphics[width=0.99\columnwidth]{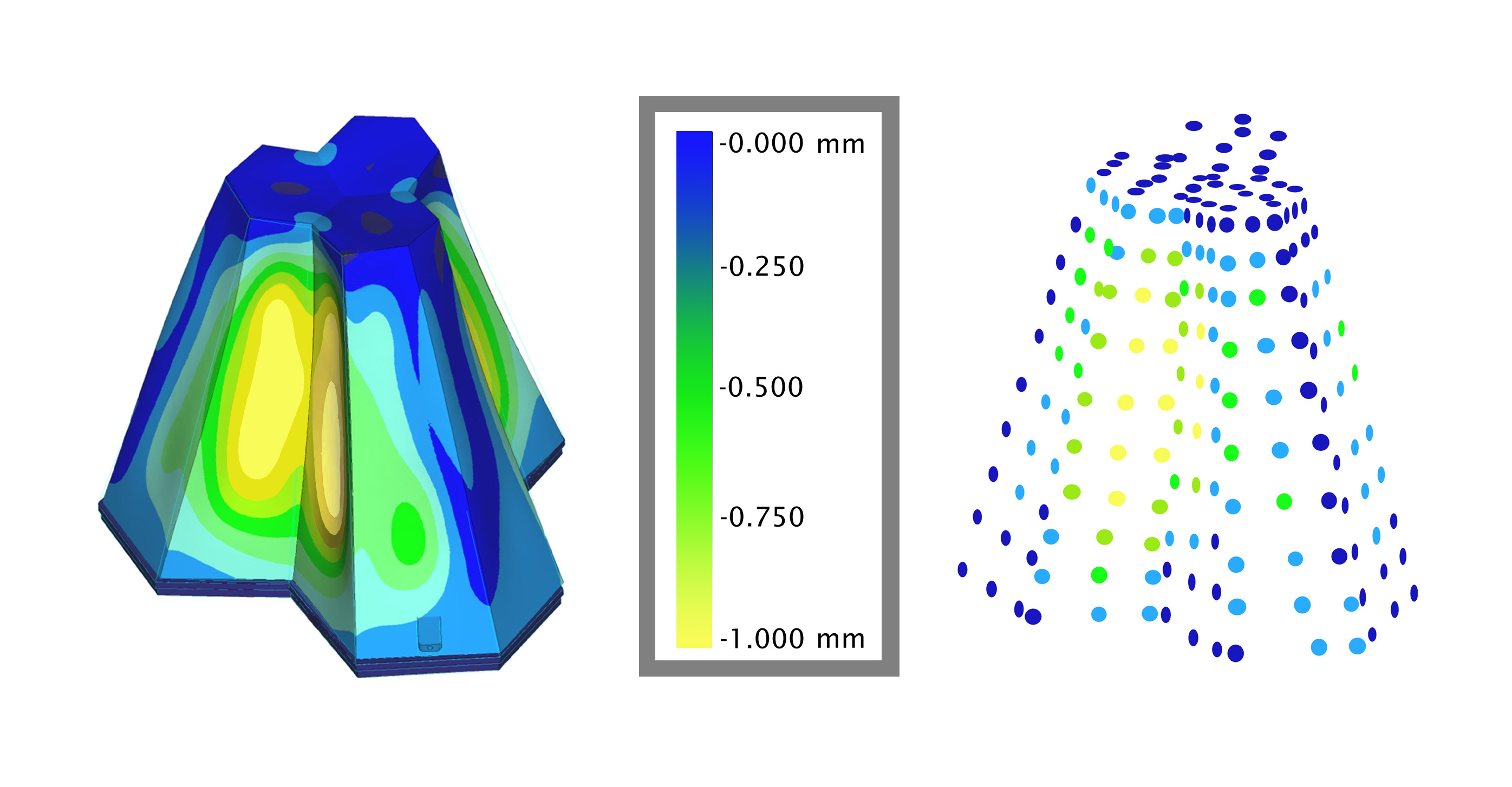}
  \caption{(Colour online) Right: measured endcap deflection of the
    detector module. Left: Finite element analysis deflection
    calculations of the endcap.}
  \label{fig:mechanics2}
\end{figure}

Accurate positioning of the detectors on the support structure is
achieved by using a system of \num{3} precision sliding rods, one of
which is threaded to enabled controlled insertion and removal.  The
accuracy of this system was checked using a high-precision coordinate
measuring machine.  Under operational conditions the endcap underwent
some deflection, which was modelled by finite element analysis
(Fig.~\ref{fig:mechanics2}). An adjustment mechanism for the detector
modules for each flange was therefore incorporated into the system.
This adjustment mechanism comprises \num{3} rings to provide for the
full \num{6} degrees of freedom adjustment.  In addition, a
stand-alone gauge was manufactured such that the detector module could
be adjusted to its final orientation before mounting into the main
support structure.  The detector module needs to be positioned within
\SI{0.1}{\mm} of its theoretical position, and so it was critical that
this gauge was made to a high accuracy.

\begin{figure}[ht]
  \centering
  \includegraphics[width=0.90\columnwidth]
                  {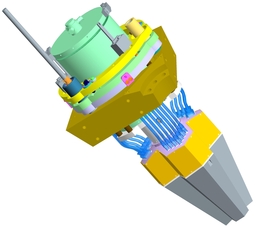}
  \caption{(Colour online) Computer aided design drawing of an ATC
    detector and the detector mounting arrangement comprising the main
    flange, 3 precision rods and the detector adjustment rings.}
  \label{fig:mechanics3}
\end{figure}

\subsection{Detector-support system}

The AGATA DSS consists of the low- and high-voltage power supplies,
the autofill system and an uninterruptable power supply (UPS) system.
The slow control of the DSS is managed by a system based on a
programmable logic controller (PLC), which is accessed and controlled
from a custom made graphical user interface.

\subsubsection{Low-voltage power supply unit}

This unit supplies low voltage power to the preamplifiers,
high-voltage modules, liquid nitrogen level measurement system,
digitisers and to the PROFIBUS-DP (Decentralized Peripherals) field
bus, which is used for controlling the power supplies.

The power consumption of the digitisers is much larger than of the
other units. In order to reduce the noise pick-up, the digitisers are
located as close to their power supplies as was practically feasible,
at a distance of about \SI{8}{\metre}.  The preamplifiers,
high-voltage modules and the liquid nitrogen level measurement systems
are about \SI{15}{\metre} from their related power supplies.

\begin{figure}[ht]
  \centering
  \includegraphics[width=0.99\columnwidth]{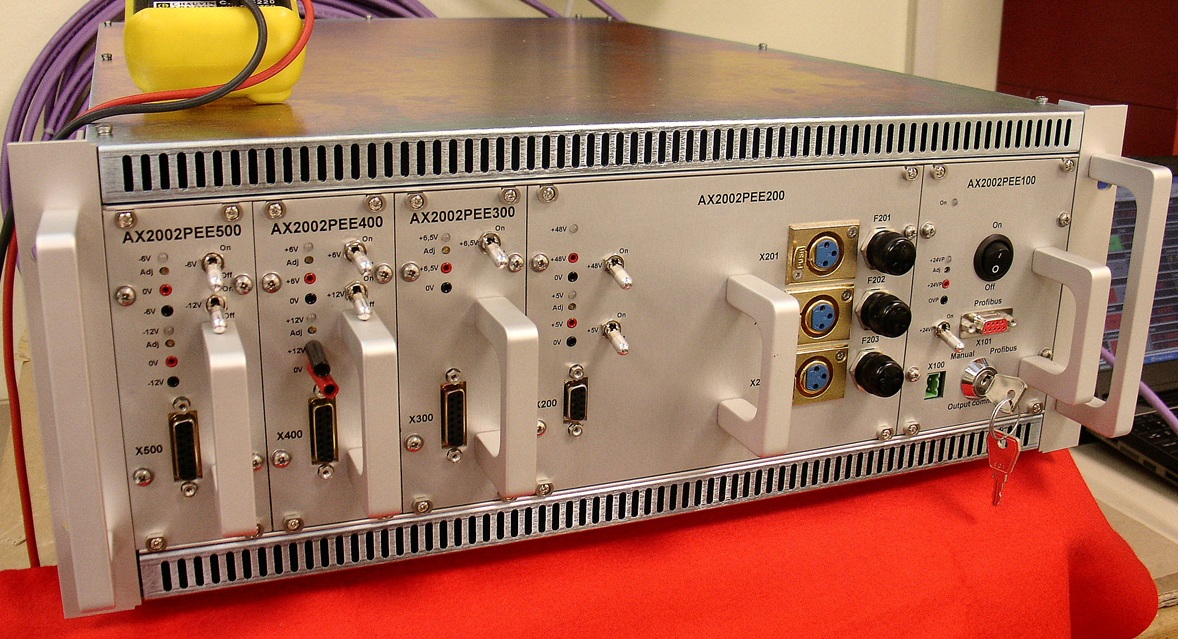}
  \caption{(Colour online) The AXIS low-voltage power-supply unit
    generating \SI{\pm 6}{\volt} and \SI{\pm 12}{\volt} to power the
    \num{111} preamplifiers of the ATC detector,
    \SI[retainplus]{+48}{\volt} and
    \SI[retainplus]{+5}{\volt} for three digitisers,
    \SI[retainplus]{+6.5}{\volt} for three HV Modules and
    \SI[retainplus]{+24}{\volt} for the PROFIBUS-DP
    network.}
\label{fig:LV_Crate}
\end{figure}

A low-voltage power supply from the company AXIS was chosen. It
consists of a 4U crate (see Fig. \ref{fig:LV_Crate}), which contains
all the needed power supplies: \SI{\pm 6}{\volt} and \SI{\pm 12}{V}
for the preamplifiers, \SI[retainplus]{+6.5}{\volt} for the
high-voltage modules, \SI{\pm 12}{\volt} for the liquid nitrogen level
measurement system, \SI[retainplus]{+48}{\volt} and
\SI[retainplus]{+5}{\volt} for the digitisers and
\SI[retainplus]{+24}{\volt} for the PROFIBUS-DP field bus.
All power supplies are linear, including the
\SI[retainplus]{+48}{\volt}, which represents the most
powerful module with a \SI{1440}{\watt} load.

The design of the low voltage supply for the \num{111} preamplifiers
uses a floating supply system, ensuring that the \SI{0}{\volt}
reference is at the detector itself and all return currents pass
through the supply to minimise the overall detector noise.  The
preamplifier load is the highest on the
\SI[retainplus]{+6}{\volt} line and, to a lesser extent on
the \SI{-6}{\volt} line.  The voltage drop across \SI{15}{\metre}
cable means that the \SI[retainplus]{+6}{\volt} and
\SI{-6}{\volt} voltages have to be regulated at the load. This has
been achieved to within a strict tolerance of \SI{\pm 0.05}{\volt}.

\subsubsection{High-voltage module}

Presently, the high voltages for the ATC detectors are provided by a
standard HV power supply produced by the company CAEN.  A development
is ongoing to equip each ATC detector with three HV modules, which
will generate locally the $\sim$\SI{5}{\kilo\volt} needed to bias
  the crystals; thus avoiding long HV cables and therefore reducing
  the pick-up noise.  In this new design, the high voltages are
  filtered, controlled and monitored directly at the ATC detector.
  The high voltages are produced from a low voltage
  (\SI[retainplus]{+6.5}{\volt}), which is provided by the AXIS unit.

Prototypes of the HV modules have been produced and they rely on a
compact HV unit made by the company ISEG.  The HV unit is able to
communicate with the PLC via the PROFIBUS-DP field bus by using
customised control electronics.  Additionally, a microcontroller
provides an autonomous bias shut down operation: each HV unit is able
to initiate a bias shut down if the temperature of the detector or the
current measured is too high.  The bias shut down due to too high
temperature is implemented by assigning one of the three HV modules of
the ATC detector to act as a master and to read the PT100 temperature
gauges of the detector.  For security and maintenance reasons, several
parameters are embedded in the microcontroller: the ID of the HV
module, the serial number, the self-calibration values of the ISEG
DC/DC converter, the maximum voltage, the maximum ramp-up and
ramp-down voltage rates and the set threshold value for the current
limit.

\begin{figure}[ht]
  \centering
  \includegraphics[width=0.99\columnwidth]{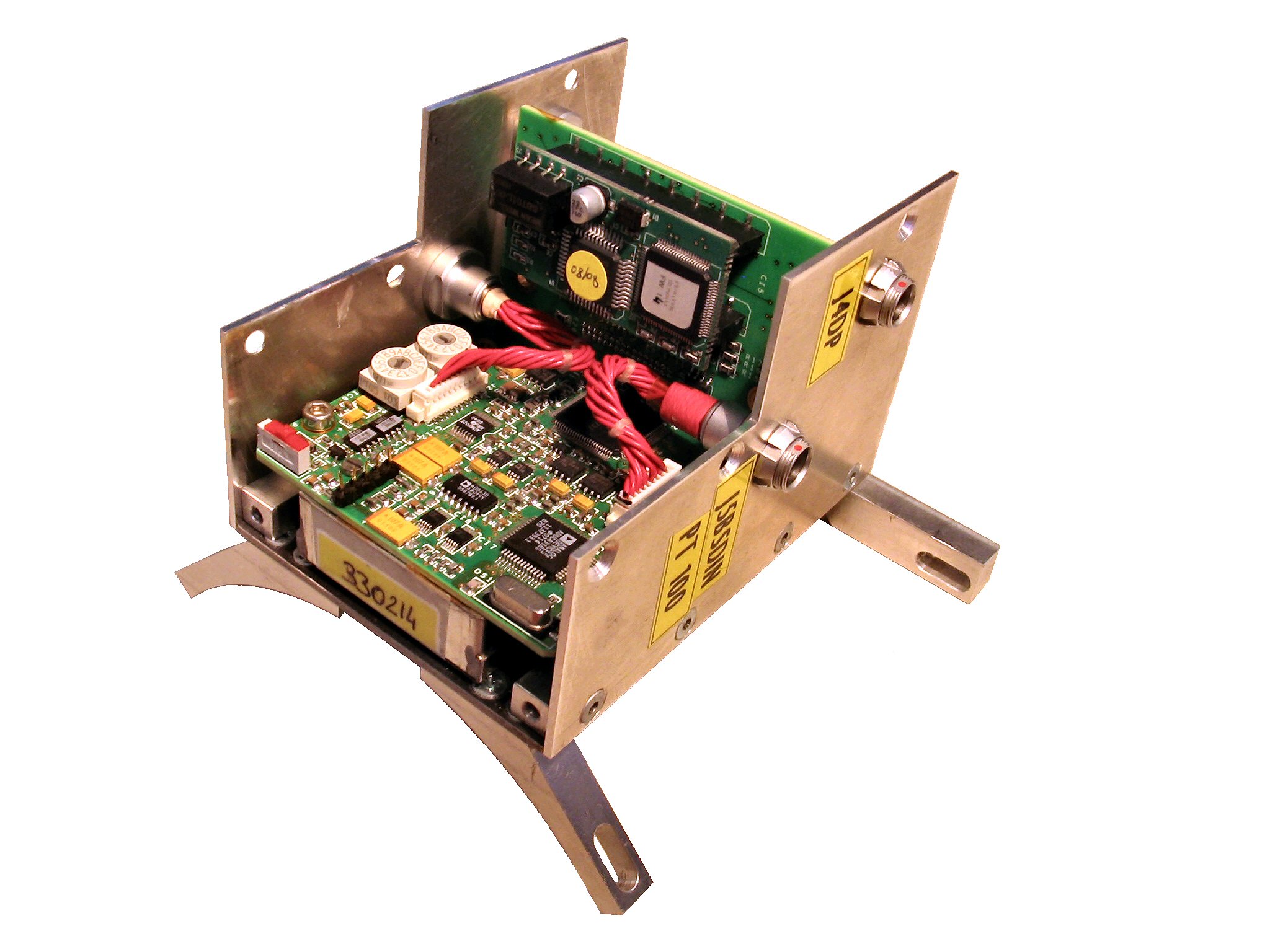}
  \caption{(Colour online) The AGATA high-voltage module composed of a
    compact ISEG DC/DC module and control-electronics boards including
    a microcontroller.  }
  \label{fig:HV_Box}
\end{figure}

A photo of the prototype HV module is shown in Fig. \ref{fig:HV_Box}.

\subsubsection{Autofill system}

The liquid nitrogen cooling of the ATC detectors is managed and
monitored by the AGATA autofill system. This system provides
information regarding the detector temperature and the amount of
liquid nitrogen in the detector Dewar \cite{Lersch2011133}. It
controls the filling cycle and operates all associated valves.

The autofill system is based on a PLC driven process running under the
PROFIBUS data acquisition and command standard. It performs the
regular filling of the detectors, forces the filling of any detector
if its temperature exceeds given threshold, issues warnings if any
parameter declines from its regular values and issues alerts if any of
the parameters is beyond the range of noncritical values. The system
is able to manage a direct filling from a pipeline as well as via
buffer tanks with the associated parameters, warnings and alarms. It
can run in fully automatic mode, in semi-automatic mode (manual start
of the fill but automatic follow up of the complete fill procedure;
remote action using the graphical user interface of the GUI), in
manual mode (full manual fill processed remotely using the graphical
user interface of the DSS) and in local mode (manual fill of the
detectors using keys on the autofill hardware in the experimental
hall).

Technically the autofill hardware is made of the following components:
\begin{itemize}
\item A PLC which controls the autofill routine.
\item A PROFIBUS crate, which contains the PROFIBUS terminals with
  various functions: PT100 readout, analogue readout
  (\SIrange{4}{20}{\mA}), digital input and output terminals, PROFIBUS
  watchdog, and the relevant power supplies. The signals from and the
  commands to the executors (valves, dialers) are sent via the valve
  control crate.
\item A Valve control crate, which contains \num{4} valve control
  cards, one master card and one dry contacts board. The valve control
  crate is presently under development. Parts of its functionality are
  mimicked using a system based on relays in the present setup.

\end{itemize}

\subsubsection{Uninterruptable power supply}

Many of the AGATA devices are protected by a UPS system. The power
consumption of the digitisers is much larger than of the
preamplifiers, which in turn is larger than the power consumption of
the autofill system. In case of a power failure, the DSS determines
when the power supply of each component is to be shut down and it
initiates the power-off procedure. The goal of the DSS is to save
enough power to enable a complete liquid nitrogen filling cycle.

\subsubsection{DSS slow control architecture}

\begin{figure}[ht]
  \centering
  \includegraphics[width=0.99\columnwidth]{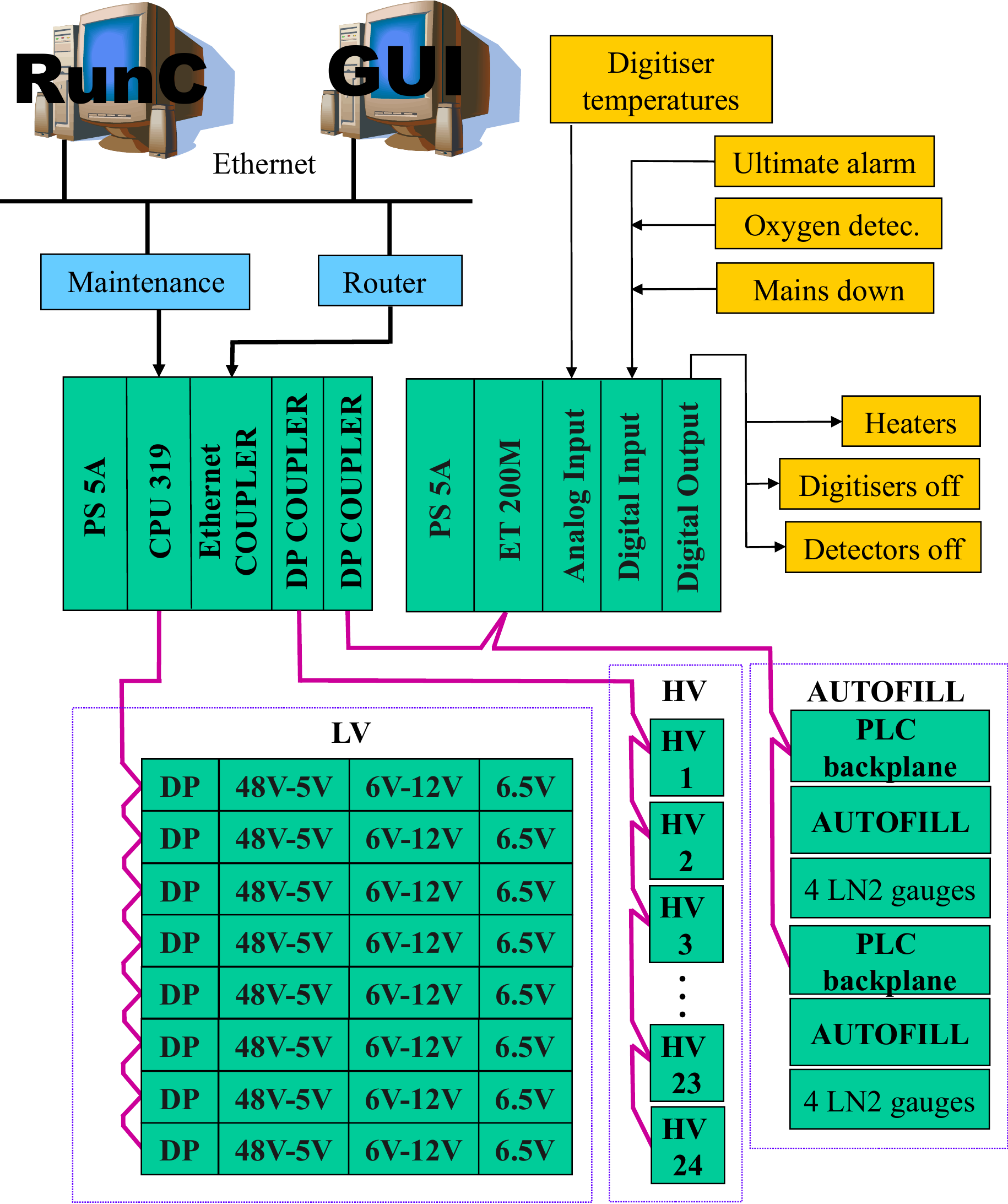}
  \caption{(Colour online) PLC architecture of the DSS. The PROFIBUS-DP
    field bus is indicated with red lines.
  }
  \label{fig:PLC_Architecture}
\end{figure}

As reliability and safety of the DSS are of paramount concern, a PLC
has been chosen to run the processes and to be the interface to the
graphical user interface of the DSS.  The PLC communicates with the
different elements of the DSS (autofill, low-voltage power supply and
high-voltage modules) via the PROFIBUS-DP field bus
(Fig. \ref{fig:PLC_Architecture}).

The communication between the PLC and the computer running the
graphical user interface is performed through the use of an OPC
(object linking and embedding for process control) server.

The maintenance server, which is based on an industrial PC, can be
accessed remotely. It gathers all data concerning the particular
process in order to be able to diagnose problems coming from the
process itself, from the PLC or from the field bus.

\subsubsection{Graphical user interface}

\begin{figure}[ht]  \centering
  \includegraphics[width=0.99\columnwidth]{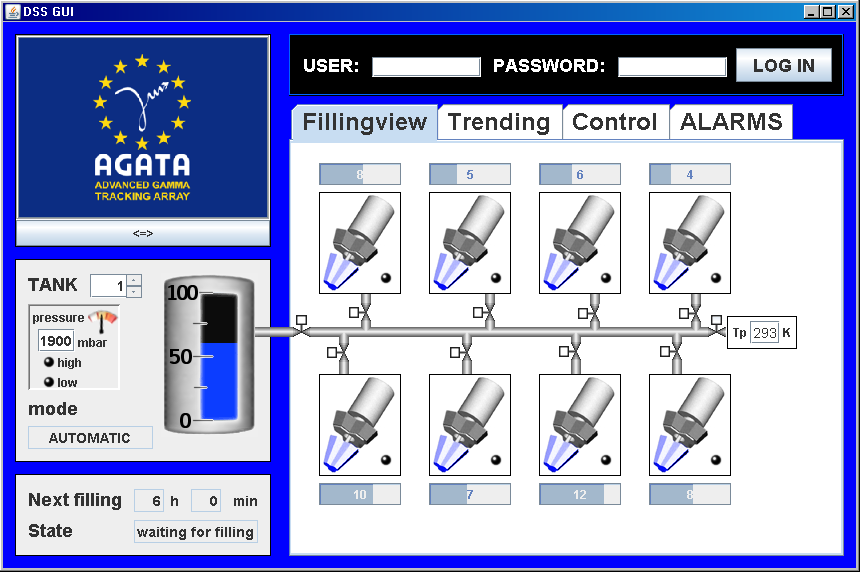}
  \caption{(Colour online) The graphical user interface of the AGATA
    detector-support system showing the autofill page. The autofill
    manages detectors in groups of 8 units. On the left hand side of
    the figure, the level of liquid in the main supply tank is shown
    as well as its internal pressure and the time remaining to the
    next programmed fill. On the right hand side, the colour of the
    detector endcap gives an indication of the detector temperature
    (blue for cold, orange for warming up, red for warm). Trending
    (real-time plot of any variable parameter), Control (parameter
    setting) and Alarms (alarm status) pages are also accessible from
    this GUI.}
  \label{fig:DSS_GUI}
\end{figure}

The graphical user interface (GUI) of the DSS is designed to control
and monitor the low voltages, high voltages and the autofill parts of
the DSS. To ensure a user-friendly and easy-to-use interface,
different requirements have been taken into account. Firstly, there
are different groups of users that demand different kinds of
information to be displayed and functions to be offered. Secondly, the
amount of data to be displayed is quite large. Therefore, it is
distributed among different display panels, such that only the
relevant part is shown (Fig.~\ref{fig:DSS_GUI}). The DSS GUI offers
various ways to access the whole set of data. For example, there are
three tables showing all parameters for one detector, user selected
parameters for all detectors or the raw data blocks from the PLC. A
graphical representation of historical data can be displayed. The user
can open multiple tabs and for each of them select the items to be
displayed and updated periodically.

The object-oriented approach allowed the separation of the two main
threads, the GUI and the PLC communication part.  Both threads
communicate with the DSS hardware interface, which holds all the data
and manages the reading and writing from both sides.

\subsection{Cable management}

The management of the many cables required to service, control and
extract signal information from the detectors is very important and
has to be taken into account in the overall design. The type, length,
weight and routing of cables has to be considered to enable detectors
to be installed and extracted easily. The cabling management system
will also be different at each host site.

\subsection{Grounding and electromagnetic compatibility}

To provide the best signal-to-noise ratio for the detector signals,
especially for their use in pulse-shape analysis, grounding is of
utmost importance. The design of the grounding includes the mechanics,
the detectors and their preamplifiers, the digitisers and all hardware
of the DSS. The rest of the AGATA system (pre-processing electronics
and data acquisition system) is optically isolated from these
front-end components.

It is mandatory that the AGATA grounding system shields the various
components from both low- and high-frequency perturbations.
Low-frequency disturbances ($\sim$\SI{50}{\Hz}) cause energy
resolution degradation in the detectors, but they are generally easy
to filter out. High-frequency perturbations (from $\sim$\SI{1}{\MHz}
up to $\sim$\SI{100}{\MHz}) are generally more difficult to be
filtered out, and they affect the performance by distorting the pulse
shape of the signals, which might have a significant impact on the
quality of the pulse-shape analysis process. A mesh grounding system
is used for AGATA with the whole system grounded to a common voltage,
which is provided by a large common conductive plate. The grounding of
the mechanical structure is formed by interconnected conductive
components using as short and thick grounding shunts as possible. In
addition, the power distribution to the front-end components of the
array is ensured via a single UPS. The measured electromagnetic
compatibility performance of the AGATA front-end electronics is such
that the \SI{50}{\Hz} noise is less than \SI{100}{\micro\volt} RMS and
the high-frequency noise in the range from \SIrange{0.1}{100}{\MHz} is
less than \SI{5}{\mV} RMS.

\section{Front-end electronics} \label{s:fee}
The objective of the AGATA front-end electronics (FEE) is to digitise
the signals from each crystal, process them in real time to establish
when the crystals detected a $\gamma$ ray (marking the time with a
timestamp), determine the amount of energy deposited by the
interaction of the $\gamma$ ray in each segment and extract the total
$\gamma$-ray energy from the core contact of the crystal. The
positions of the $\gamma$-ray interactions are calculated in the
pulse-shape analysis (PSA) farm (see section \ref{s:psa}). Therefore,
the AGATA FEE supplies a short trace of the digitised leading edge of
each pulse along with the energy and the timestamp. The design of the
AGATA FEE has been a challenge due to the high acquisition rates
specified for the system. The rate specifications and the measured
values for the different subsystems are summarised in Table
\ref{tab:rates}.

\begin{table}          
  \centering
  \caption{Count-rate specifications and current limits in kHz after the 
    front-end electronics, global trigger and pulse-shape
    analysis farm. The rates are given per crystal and for the setups
    with 15 and 180 crystals (AGATA $4\pi$) with the nominal 
    source-detector distance (\SI{23.5}{\cm}).
  }
  \label{tab:rates}
  \begin{threeparttable}
  \begin{tabular}{|c|c|c|c|}
    \hline
    System      & Per          & 15          & 180 \\
                & crystal      & crystals    & crystals \\
    \hline
    FEE         & 50\tnote{a}  & 750         & 9000    \\
    (detectors) &              &             & \\   
    \hline
    Global      & 50           & 750         & 3000\tnote{b} \\
    trigger     &              &             & \\   
    \hline
    Pulse-      & 5\tnote{c}   & 75\tnote{c} & 3000 (M$_\gamma=1$)\tnote{d} \\     
    shape       &              &             &  300 (M$_\gamma=30$)\tnote{d} \\
    analysis    &              &             & \\   
    \hline
  \end{tabular}
  \small{
  \begin{tablenotes}
  \item[a] With the FWHM value degraded by about \SI{50}{\percent}
    compared to the nominal FWHM value obtained with a single crystal
    counting rate in the range \SIrange{10}{20}{\kHz}.
  \item[b] Limit defined by the check idle cycles.
  \item[c] Current limits of PSA processing. Writing the signal traces to disk
    results in a reduction of the rate to \SIrange{1}{3.5}{\kHz} per crystal.
  \item[d] Full AGATA specifications rates with on-line PSA and tracking.
  \end{tablenotes}
  }
  \end{threeparttable}
\end{table}

The PSA requires each digitizing sampling ADC to be aligned with all
the others and that time-stamped data based on a common clock are
generated. AGATA clock distribution and time stamping is linked to a
global-trigger mechanism, which allows either free running
time-stamped operation or hardware-triggered operation depending on
whether data-rate reduction is needed in order not to saturate the PSA
processor farm. The data sent to the PSA contain the same parameters,
including timestamps, in both cases. The only difference is whether
part of the data are rejected prior to the PSA, for example because of
a multiplicity condition or a requirement for coincidences with one or
more ancillary detectors.

\begin{figure}[ht]
  \centering
  \includegraphics[width=1.0\columnwidth]{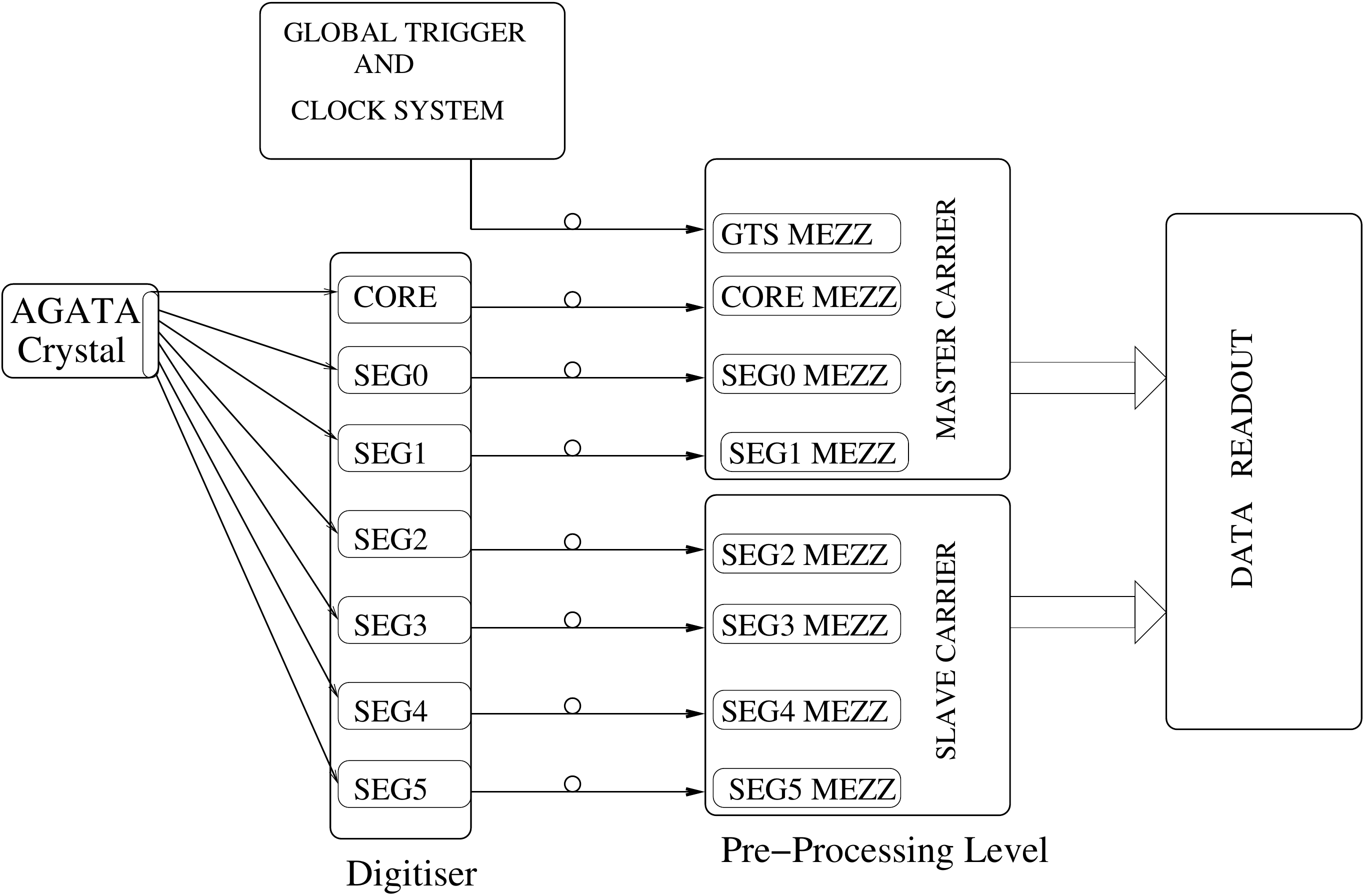}
  \caption{Schematic view of the AGATA front-end electronics and data
    readout system.}
  \label{fig:feescheme}
\end{figure}

Fig.~\ref{fig:feescheme} shows the schematic design of the AGATA FEE
and data-readout system, which treats each crystal (one core plus 36
segment outputs) as a separate entity. Within that entity the core is
treated differently to the segments. The core signal is formed by the
sum of the charge released in all the interactions in the
crystal. Therefore, it can be used as a trigger for the whole
crystal. The AGATA electronics for each crystal consists of the
following components:
\begin{itemize}
\item One digitiser comprising \num{6} segment cards, \num{1} core
  card, \num{2} power-supply cards and \num{2} control cards.
\item Two pre-processing carrier cards in the Advanced
  Telecommunications Computing Architecture (ATCA) card format, each
  containing \num{4} Common Mezzanine Cards (CMC) with PCI Express
  readout to the PSA farm. Seven CMC mezzanines correspond to the
  \num{6} segment and \num{1} core card in the digitiser and one
  contains the interface to the global trigger and clock system.
\end{itemize}
The various elements of the FEE are connected through optical fibres
in order to achieve appropriate data-transmission rates and to
maintain good electrical isolation.
 
\subsection{The digitiser} \label{ss:digitizer}

\begin{figure}[ht]
  \centering
  \includegraphics[width=\columnwidth]{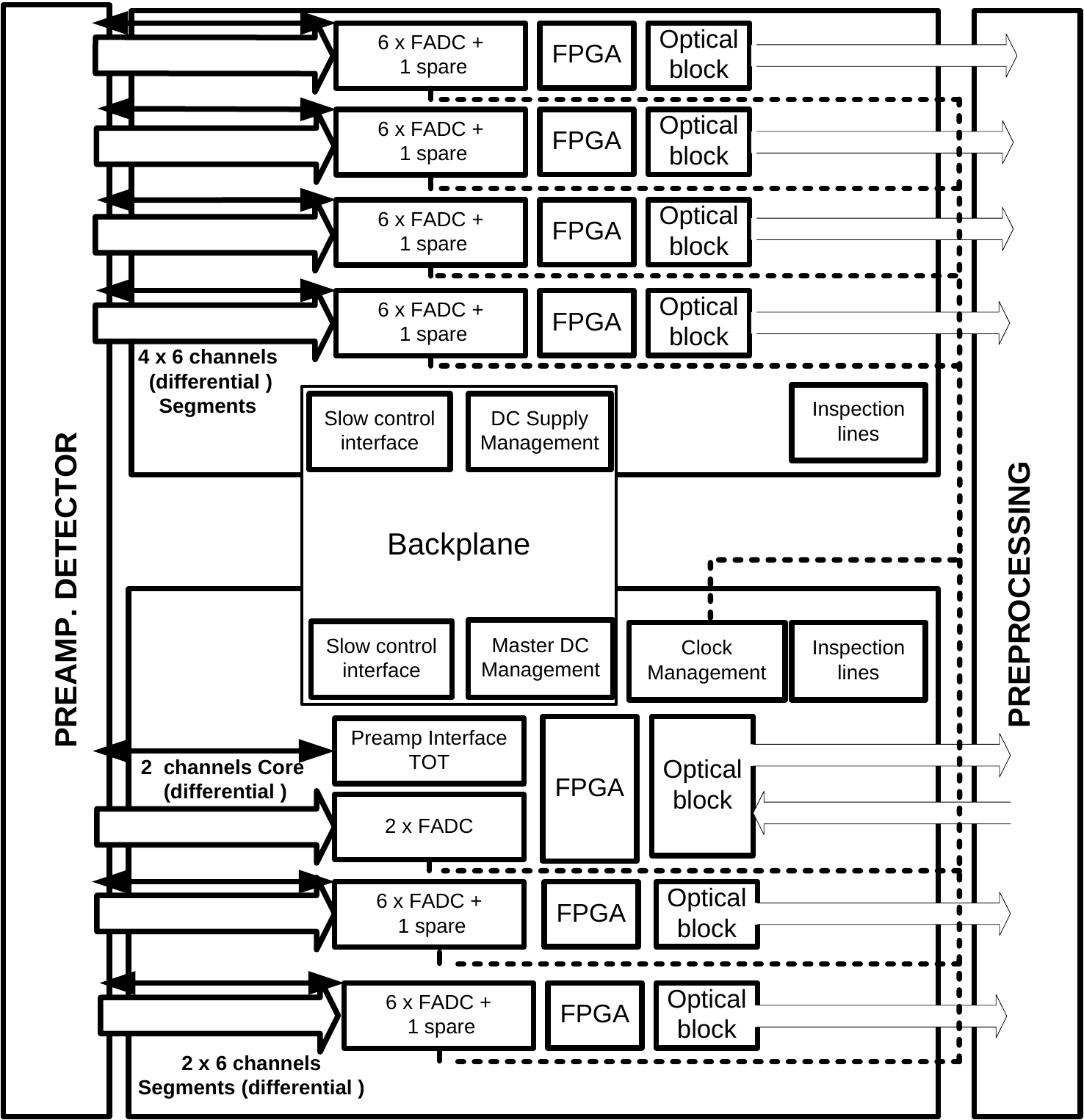}
  \caption{Block diagram of the AGATA digitiser.}
  \label{fig:digitizer_block_diagram}. 
\end{figure}

\begin{figure}[ht]
  \centering
  \includegraphics[width=\columnwidth]{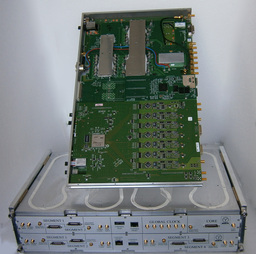}
  \caption{(Colour online) Photograph of the AGATA digitiser.}
  \label{fig:digitizer_photo}. 
\end{figure}

The principal goal of the digitiser is to interface the detectors with
the AGATA signal-processing system. In order to do that, the digitiser
module performs the following tasks:
\begin{itemize}
\item Receives all \num{37} preamplifier outputs from one crystal.
\item Digitises the input signals at a rate of \SI{100}{\MHz}, using
  \num{14}-bit ADCs.
\item Serialises the ADC data and transmits it over optical fibres to
  the pre-processing electronics.
\item Implements a constant-fraction discriminator (CFD) algorithm in
  an FPGA to generate an isolated fast-logic signal to be used in
  hardware triggers of ancillary detectors.
\item Implements a time-over-threshold (TOT) algorithm using
  preamplifier inhibit signals (see section \ref{ss:preamp}).
\item Provides spare digitiser channels and inspection lines for
  maintainability and diagnostics.
\item Provides interfaces for re-programming, control in an
  electronics and detector workshop, as well as an interface for slow
  control.
\end{itemize}
The AGATA specifications require that the digitiser module is mounted
less than \SI{10}{\metre} from the detector's preamplifiers to minimise signal degradation. The
digitiser is housed in a water cooled box of the size \SI{30}{\cm}
$\times$ \SI{14}{\cm} $\times$ \SI{55}{\cm} and it contains two
modules with the electronics for two crystals.  A block diagram of the
digitiser is shown in Fig.~\ref{fig:digitizer_block_diagram} and a
photograph of a printed circuit board for one module is shown in
Fig.~\ref{fig:digitizer_photo}.  The main building blocks of the
digitiser are: the differential analogue input, the flash analogue to
digital converter (FADC) sampling blocks, the field programmable gate
arrays (FPGAs), which receive the FADC data and serialise it, the
optical transceiver blocks, the global clock receiver block and the
monitoring block. The monitoring block provides control of the spare
channels and inspection lines via a slow control link to the GUI based
system control software. The temperature of the various parts of the
digitiser electronics is also accessible by the slow control and a
local over-temperature shutdown is implemented in the control/power
supply card.

The analogue input buffer adds the optional offset signal and also
includes an anti-aliasing filter before the FADC. The FADCs convert
the signals into \num{14} bits at \SI{100}{\MHz}; \num{15} data bits
are sent to the FPGA for serialisation and transmission as
\num{16}-bit data over the optical fibres. Bits \num{0} to \num{13}
are FADC data, bit \num{14} is the FADC overrange and bit
\num{15} is used as the synchronisation pulse from the GTS (for a
description of the GTS, see subsection \ref{ss:gts}).

Although the main purpose of the FPGAs in the segment and core cards
is to serialise data, the FPGA of the core card also implements two
algorithms for real-time data processing. The first is a digital CFD,
which is used as a trigger for ancillary detectors (output both on an
isolated connector and over a fibre). The CFD algorithm runs in the
digitiser rather than in the pre-processing because of the latency of
the fibre links (several \SI{100}{\ns}) and the need for a prompt
output. The second algorithm is a TOT algorithm that measures the time
during which the preamplifier inhibit signal is asserted when it
receives an over-range energy input. Since the preamplifier recovers
from overload by discharging a capacitor using a constant current
source the period of overload (TOT) is a measure of the energy in the
crystal when the preamplifier is over-range (see section
\ref{ss:preamp}).  A special serial fibre protocol is used to indicate
when FADC data is replaced by TOT data.

One \num{8} channel fibre-optic transceiver link (\num{4} RX, \num{4}
TX) is used to interconnect the core FADCs serialised data and the
timing and control of the pre-processing subsystem. The ``transmit''
channels are allocated to the core \SI{100}{\MHz} FADC data (two
channels), the core CFD logic signal (one channel), and the global
clock calibration feedback path (one channel). The ``receive''
channels are used for the global clock signal, the synchronisation
signal, and the analogue offset control for the core electronics. A
\num{12} channel multi-fibre transmits the \num{6} serialised segment
FADC data streams from each of the segment cards.  Each analogue
channel from the detector can be inspected before and after the digitisation.

One analogue line is provided before and two after coding per group of
\num{6} segments channels. The global clock function reconstructs a
high quality clock with a very small jitter ($<$\SI{7}{\pico\second})
from the clock received through one optical receiver within the core
transceiver.

The digitiser is able to work in stand-alone mode when a specific
firmware and GUI interface (called SAMWIZE) are used. In this mode it
is possible to visualise the recorded traces from the detector input
signal (oscilloscope mode). It is also possible to record up to
\num{6} traces per segment and one per core. The Multi-Channel
Analyzer mode is capable of histogramming up to \num{6} segment
channels and one core in parallel. This mode was particularly useful
for developing and testing the core energy and constant-fraction
discriminator (trigger) block of the pre-processing firmware.

\subsection{The pre-processing electronics}
\label{ss:preprocessing}

The pre-processing system reduces the data volume from the digitisers
by a factor of about \num{100} by extracting and processing data from
the digitiser's data stream only for the segments which have
registered a detection of a $\gamma$-ray interaction. Further
filtering (triggering) can optionally be performed in conjunction with
the GTS from which a clock for the digitiser and the timestamp
information is also derived. The pre-processing sends the filtered
data to the PSA farm.  The processing rate for traces in the segments
is the same as in the core.  The triggering is always made by the
core, so the core contact electronics is the master and the segments
are controlled by it.

The pre-processing hardware de-serialises and processes the incoming
data streams and stores the traces. The sampling speed is
\SI{100}{\MHz} in the reconstructed data streams, so the
pre-processing hardware also uses \SI{100}{\MHz} clock rates for
incoming data (some internal clocks are running at \SI{200}{MHz}). The
GTS interface provides the system clock and a trigger system.

\begin{figure}[ht]
  \centering
  \includegraphics[width=1.0\columnwidth]
                  {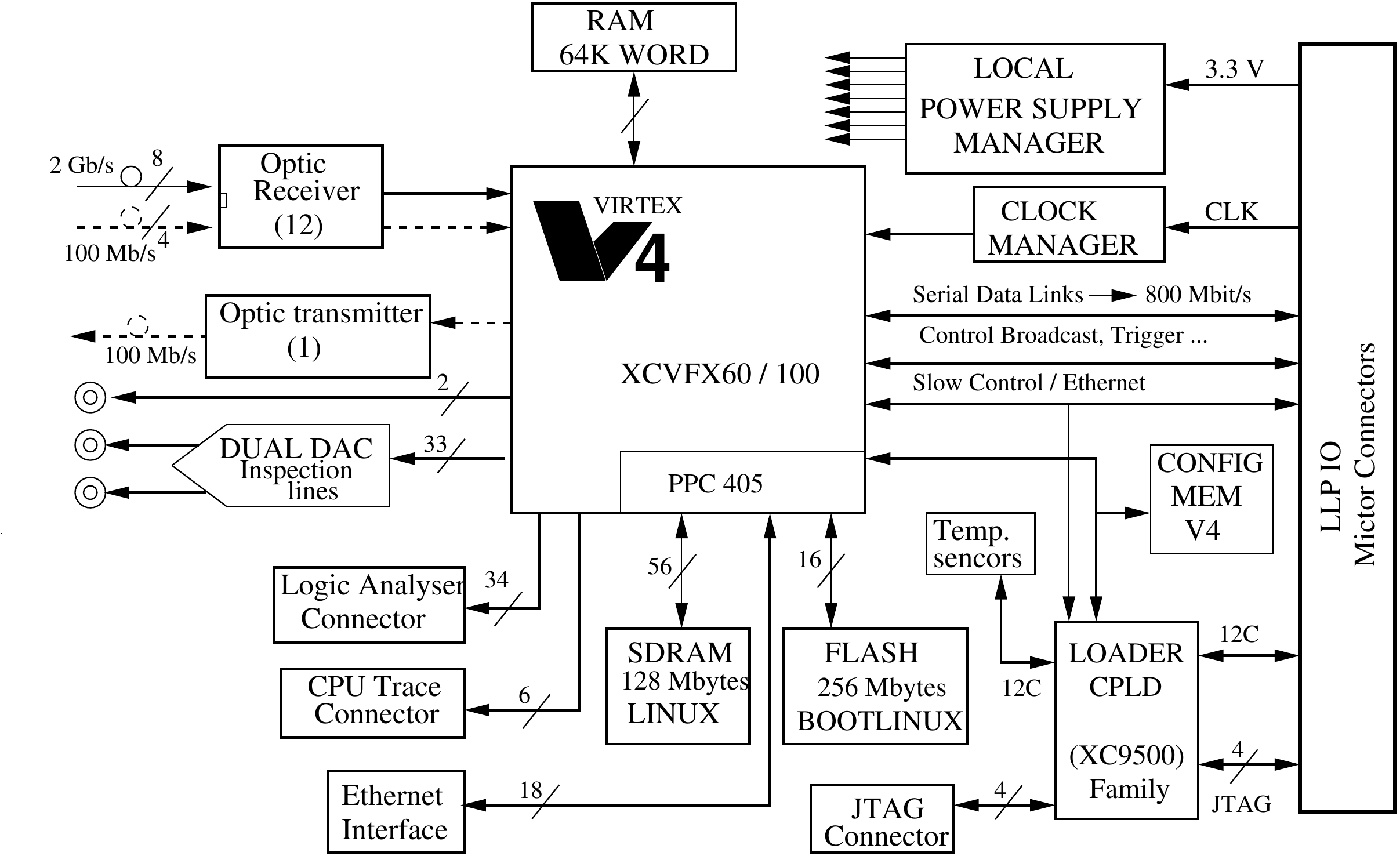}
  \caption{Block diagram of the AGATA pre-processing mezzanines.
  }
  \label{fig:preprocessing_mezzanine_block_diagram}
\end{figure}

\begin{figure}[ht]
  \centering
  \includegraphics[width=1.0\columnwidth]{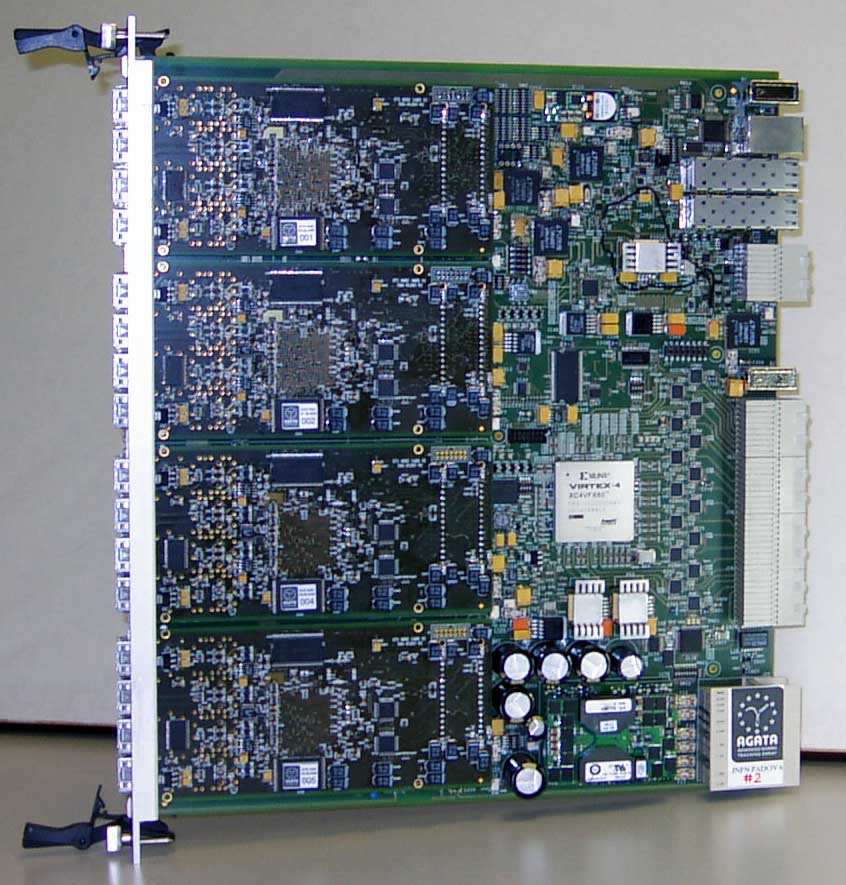}
  \caption{(Colour online) Photograph of the AGATA pre-processing ATCA
    carrier card.}
  \label{fig:preprocessing_carrier_photo}
\end{figure}

\begin{figure}[ht]
  \centering
  \includegraphics[width=1.0\columnwidth,angle=180]
                  {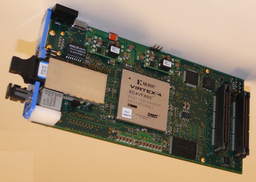}
  \caption{(Colour online) Photograph of the AGATA pre-processing
    mezzanine card.
  }
  \label{fig:preprocessing_mezzanine_photo}
\end{figure}

The AGATA trigger system can be used to reduce the counting
rate. Where rate reduction (see Table~\ref{tab:rates}) is not
required, the pre-processing runs in triggerless mode, which means
that all the processed data are sent to the PSA farm. In this case a
software trigger is performed after PSA and tracking. The maximum
delay (latency) which can be accommodated in the pre-processing
hardware while waiting for a trigger decision is limited by the
maximum of the trace length, which is \SI{20}{\micro\second}. Lower
values can be configured in the trigger system, but coincidences
separated by more than \SI{20}{\micro\second} are detected by using
software triggers.

The local trigger of the core signal is a digital-trigger algorithm
operating on the data stream from the core contact of the
crystal. When this trigger finds a pulse in the data it generates a
local trigger output which indicates to all the segment electronics
that they should also extract a trace from the data stream. Traces are
stored locally, within each pre-processing channel. The traces are
held in each channel's local memory for up to \SI{20}{\micro\second}
while the GTS makes a decision. An event can be either accepted or
rejected. For events which are accepted, the pre-processing stores a
trace of the digitised leading edge of the pulses from the core and
from all 36 segments in a buffer waiting to be sent to the PSA farm.

In addition to selecting useful portions of the incoming data stream
using a trigger algorithm, the pre-processing also applies the moving
window deconvolution (MWD) algorithm \cite{mwd,tnt} on the incoming
data streams to determine the $\gamma$-ray energy deposited in each
segment and in the whole crystal. The MWD filter
output is sampled a programmable time after the trigger algorithm
detects the start of a digitised pulse in the incoming core data
stream. The energy parameter sampled from the MWD filter is stored
along with a data trace showing the samples taken during the leading
edge of the pulse and a timestamp indicating when the trigger
happened.

Core and segment mezzanines have PowerPC processors with
an embedded Linux operating system housing all the drivers and
applications for the card control and online temperature monitoring.
A slow control GUI, based on a web server architecture, has been
developed for the control and online monitoring of the status of the
mezzanine cards. The block diagram of the pre-processing mezzanine is shown
in Fig. \ref{fig:preprocessing_mezzanine_block_diagram}.  Photographs of
the printed circuit boards  of the carrier and mezzanine cards are shown in
Fig.~\ref{fig:preprocessing_carrier_photo} and
\ref{fig:preprocessing_mezzanine_photo}, respectively. Carrier cards, implemented
in the ATCA standard \cite{atca}, house the CMC format mezzanines,
which read data from the core and segment cards in the digitiser and a
GTS mezzanine, which communicates with the GTS system.

\begin{figure}[ht]
  \centering
  \includegraphics[width=1.0\columnwidth]{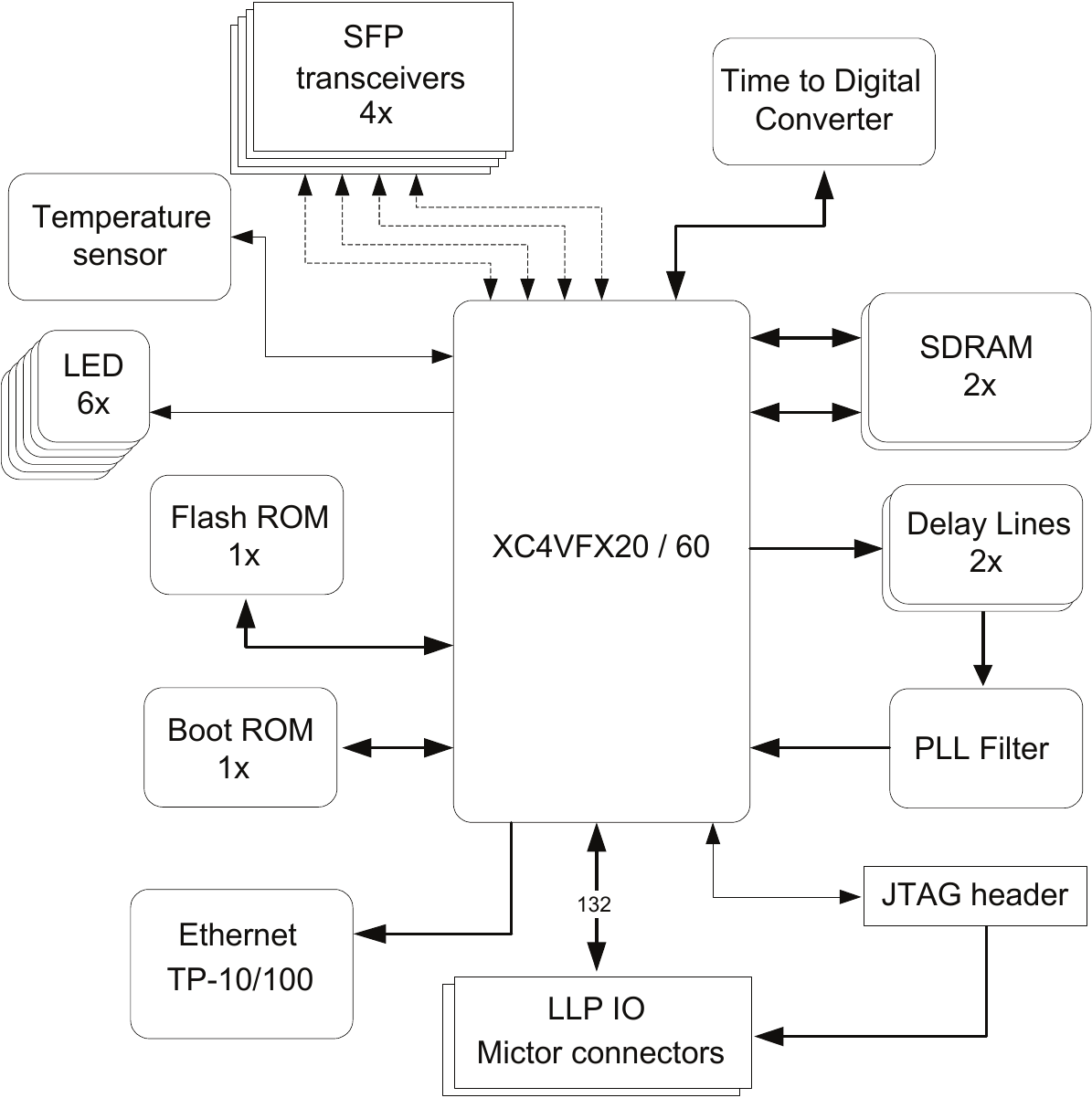}
\caption{Block diagram of the GTS mezzanine.}
  \label{fig:gts_mezzanine}
\end{figure}

\begin{figure}[ht]
  \centering
  \includegraphics[width=0.45\columnwidth]{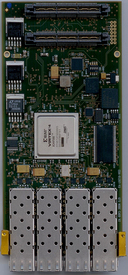}
  \includegraphics[width=0.45\columnwidth]{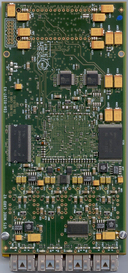}
  \caption{(Colour online) Photographs of the two sides of the GTS mezzanine.}
  \label{fig:gts_photo}
\end{figure}

A block diagram of the GTS mezzanine is shown in
Fig.~\ref{fig:gts_mezzanine} and a photograph of the printed circuit
board in Fig.~\ref{fig:gts_photo}.  The GTS mezzanine receives the
global clock, aligns it locally within the pre-processing and then
uses a dedicated data path in the core mezzanine to align the clock in
the digitiser. It is also used to accept or reject local triggers. The
core and segment mezzanines receive data from the digitiser and
process the data using digital filter algorithms. The mezzanines are
read out via the carrier, all data being concentrated into a single
FPGA per carrier before transmission via PCI express to the PSA farm
on demand by the PSA. The carrier reads each mezzanine at
\SI[per=slash]{100}{\mega\byte\per\second}. A different FPGA
handles the trigger interconnections within the carrier. Trigger
connections between the two carriers handling one crystal use a
dedicated backplane link (TCLK). The ATCA carrier cards accept \num{4}
CMC format mezzanines. The connection between the CMC and the carrier
card is achieved by two Mictor connectors, each with \num{114} pins.

\subsection{The Global trigger and synchronisation (GTS) system}
\label{ss:gts}
	
Data synchronisation is an important aspect in the operation of the
trigger and readout systems of AGATA. Tracking and PSA require the
concurrent digitisation of preamplifier signals of the 36
segmented Ge crystals composing the array. Therefore, the design of
the front-end readout and level-1 (L1) trigger in AGATA follows a
synchronous pipeline model: the detector data are stored in pipeline
buffers at the global AGATA frequency, awaiting the global L1
decision. The L1 latency must be constant and match the pipeline
buffer length. The whole system behaves synchronously and
synchronisation at different levels and in different contexts has to
be achieved and monitored for proper operation of the system.
 
In AGATA each crystal is considered as a separate entity and, from the
point of view of the data acquisition system, the whole detector may be
considered as the aggregation of synchronised data supplied by
individual crystals, possibly disciplined by a global trigger
primitive.

The data from the core contact are processed for event
detection and either an L1 trigger request or a local trigger is
generated. The choice between the two behaviours is done upon
configuration, the former corresponding to an effective way to reduce
front-end data rates in cases where any one of the stages of the
readout chain is unable to perform at the actual data throughput.
From the logical description of the front-end operation given above it
turns out that a certain number of global time-referenced signals are
needed. The most important of these are: the common clock, the global
clock counter, the global event counter, the trigger controls, the
trigger requests and the error reports. In AGATA, the transport medium
of all these signals is shared by use of serial optical bi-directional
links connecting the FEE of each crystal with a central global trigger
and synchronisation control unit in a tree-like structure; thus
actually merging together the three basic functionalities of
synchronisation distribution, global control and trigger processing.

The common clock is a \SI{100}{\MHz} digital clock supplied by a
central timing unit and used to clock the high-speed optical
transceivers reaching the FEE of every crystal. At the crystal
receiving side the clock is reconstructed and filtered for jitter. The
clock signals of each crystal may be equalised for delay and phase;
thus accounting for different fibre lengths and different crystal
locations in the array.

The global clock counter is a \num{48}-bit digital pattern used to tag
event fragments before front-end buffer formatting.  The pattern is
the actual count of the global clock. It will be used by the PSA
processing and by the global event builders to merge the event
fragments into one single event. The global event counter is a
\num{24}-bit digital pattern used to tag event fragments before
front-end buffer formatting. The pattern is the actual count of the L1
validations.  The trigger control must guarantee that subsystems are
ready to receive every L1 accept delivered. This is essential in order
to prevent buffer overflows and/or missed trigger signals when the
crystals are not ready to receive them. In either case, the
consequence would be a loss of synchronisation between event
fragments.

The trigger requests are generated by the core contact signal from the
AGATA detectors by asserting a signal that is transmitted via the
high-speed serial links of the GTS network upwards to the central
trigger unit. All the trigger requests collected from the crystals at
each global clock cycle form a pattern that can be processed centrally
for multiplicity or coincidence with ancillary detectors. The result
of this processing stage constitutes the L1 validation. Error reports
indicate abnormal conditions such as buffer overflows, local faults,
built-in self tests, etc., and can be reported centrally for proper
corrective actions.

\begin{figure}[ht]
  \centering
  \includegraphics[width=1.0\columnwidth]{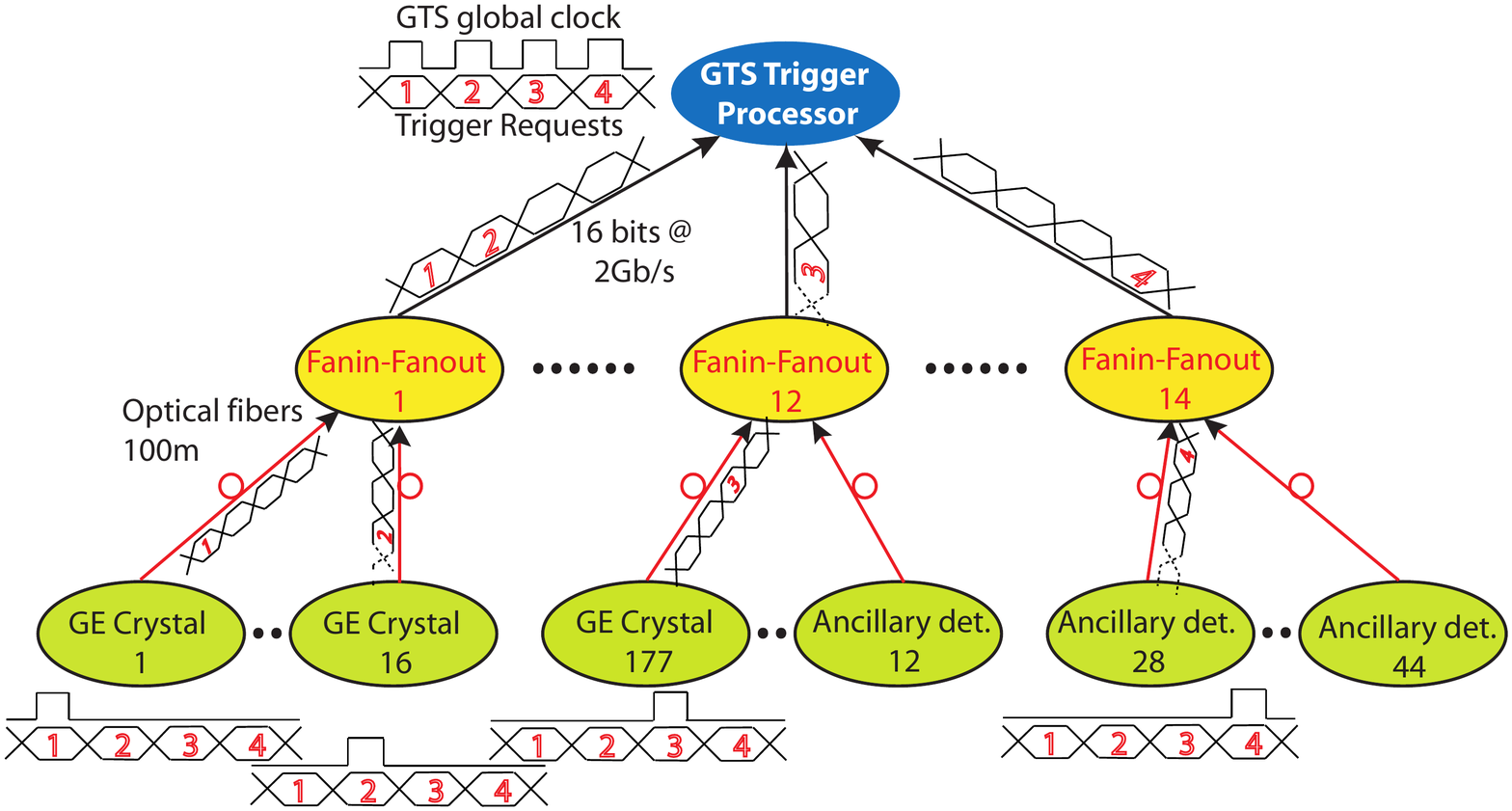}
  \caption{(Colour online) Illustration of the topology of the GTS
    tree.  Any detector node can produce at any time a trigger
    request, which is labeled with the clock and transmitted upwards
    in the tree, through the fan-in-fan-out architecture to the global
    trigger processor.}
  \label{fig:feegts}
\end{figure}

The AGATA GTS has a tree topology as shown in Fig.~\ref{fig:feegts}.
The tree originates from the root node, which at the same time acts as
the source of all global information (clock, timestamps, commands, L1
validations) and all the trigger requests.  It also
performs fast monitoring of signals and services requests coming
from the crystals.
To solve the problems of building a bi-directional, high-capacity and
high-speed network tree that drives hundreds of nodes which are
displaced by several tens of meters, a certain number of technological
issues have been addressed. Among these issues are the fan-out of a
source synchronous transmission, noise immunity, low error rate and
throughput. The GTS tree is composed of five different elements: the
root node, the backplane, the fan-in-fan-out nodes, the fibre
connections and the mezzanine interface.

At the root node, a fast optical link forwards all the collected
trigger requests to the trigger processor, which is a high-performance
and fully-pipelined custom-built processor that sorts incoming
requests, computes multiplicities and prompt or delayed coincidences
among seven user defined trigger partitions.  The trigger processor is
fully programmable in the C programming language by the user through
an application programming interface and it is controllable via
a standard PCI express interface.

\section{Coupling of complementary instrumentation} \label{s:ancillary}
The coupling of large $\gamma$-ray detector arrays, such as EUROBALL
\cite{Simpson1997} and GAMMASPHERE \cite{Lee1990}, to complementary
detectors has played a major role in spectroscopic investigations in
the last decade.  The AGATA array will be used with radioactive or
high-intensity stable beams and in most experimental conditions, in
order to fully exploit its capabilities, it will be essential to
couple it to devices providing complementary
information. Additionally, the foreseen use of AGATA at different
facilities makes some of the complementary devices (e.g. beam-tracking
devices in the case of fragmentation facilities) absolutely necessary
for the normal operation of the array.

Within the AGATA project, dedicated new complementary detection
systems will be developed in addition to adapting existing ones. The
use of fully digital sampling electronics is desirable in the future.
However, in order to utilise the many existing systems based on
conventional analogue readout electronics an interface to the GTS
system has been developed. This interface is called AGAVA (AGATA VME
Adapter). Since nearly all of the presently available detectors within
the AGATA community have front-end/readout systems based on the VME or
VXI standard, the AGAVA interface has been developed in the VME
standard with full compatibility with the VXI readout modes.

\subsection{General description of AGAVA}
\label{ss:agava}

The AGAVA Interface is a one-unit wide A32/D32 type VME/VXI slave
module. It is the carrier board for the GTS mezzanine card used in
AGATA for the global trigger and timestamp distribution.

\begin{figure}[ht]
  \begin{center}
    \includegraphics[width=\columnwidth]{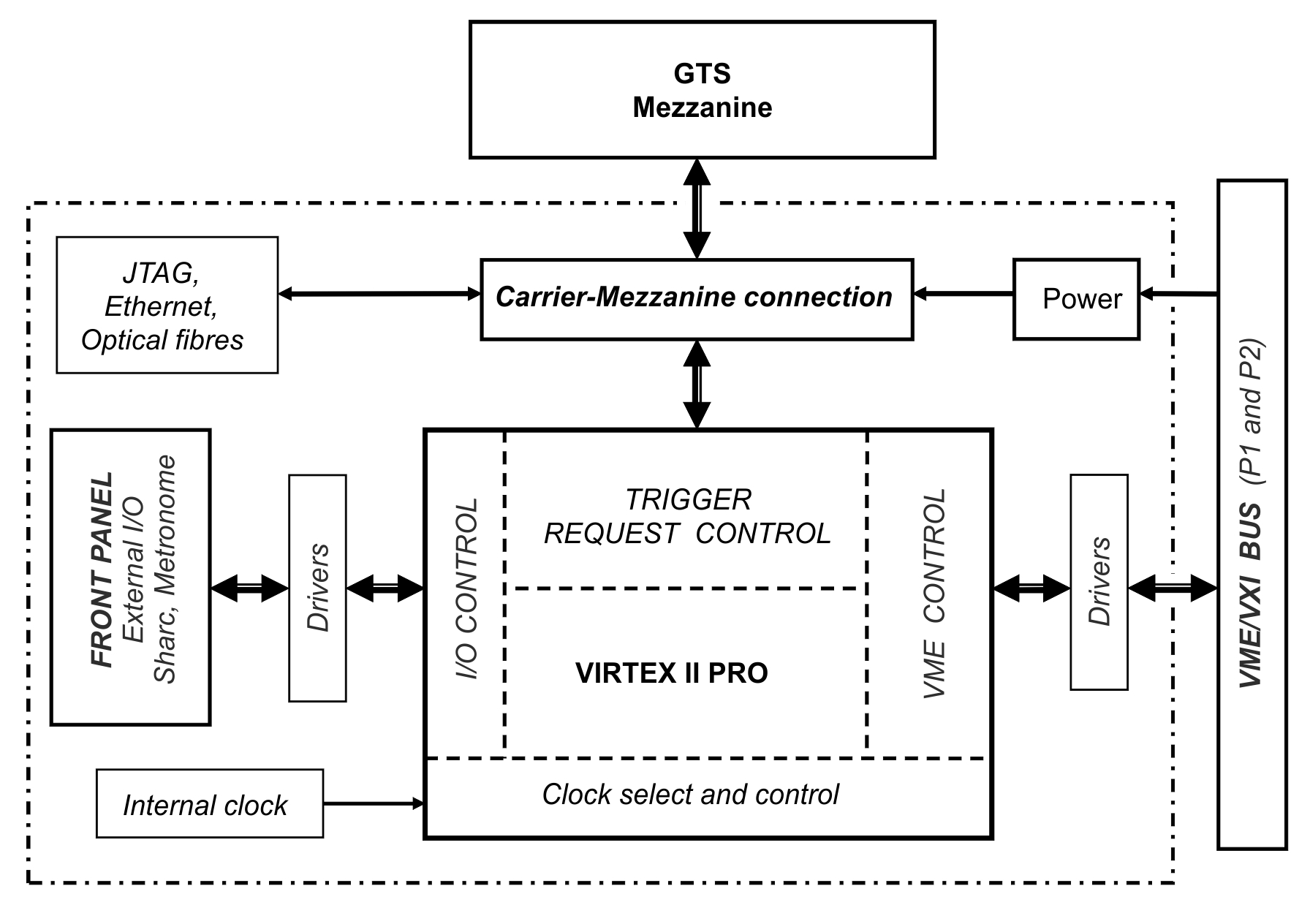}
   \caption{Block diagram of the AGAVA interface.}
    \label{fig:agava1}
  \end{center}
\end{figure}

\begin{figure}[ht]
  \centering
  \includegraphics[width=\columnwidth]{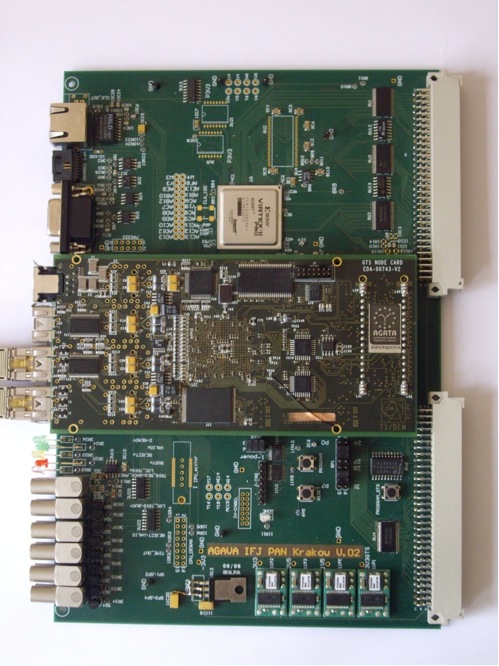}
  \caption{(Colour online) Photograph of the AGAVA module with the GTS
    mezzanine card.}
  \label{fig:agava_photo}
\end{figure}

The main purpose of the AGAVA interface is to merge the AGATA
timestamp-based system with conventional readout 
based on VME or VXI, which is used for example by EXOGAM \cite{Simpson2000}.
AGAVA also has the necessary connectors to interface
with the VME Metronome and SHARC \cite{metronome1} link
systems. The logic process is controlled by an FPGA of the type
Virtex II Pro. The block diagram of the AGAVA interface
is shown in Fig.~\ref{fig:agava1}.
The AGAVA module (see Fig.~\ref{fig:agava_photo}) includes all
necessary connections for the
trigger cycle and for a total data readout system
\cite{metronome1}.
It contains also a passive Ethernet interface, which
provides a direct connection to the GTS mezzanine card.

\subsection{The operation of AGAVA}

The AGAVA module supports VME access, provides interfacing with the
GTS mezzanine and provides input/output signals at the front-panel
Lemo connectors. Two VME access modes have been implemented in AGAVA:
the VME single-access read or write operations, according to the
standard VME handshake rules, and the chained block transfer (CBLT)
readout mode on the VME bus. Standard access can be used both for
board configuration and readout, while the CBLT mode (available in
many commercial VME ADC and TDC front-end modules) has been introduced
to speed up data readout.  The CBLT mode makes use of a common address
specified during the initialisation phase in a dedicated register for
all participating modules in the chain. The VME CPU reads their data
in blocks of variable lengths starting from the module defined as the
first in the chain, followed by one or more intermediate modules and
finished by the one configured as the last one. The CBLT mode is used
for faster readout and is helpful for event building. The CBLT readout
mode of AGAVA has been extensively tested. The AGAVA interface also
implements a VXI EXOGAM-like readout.

The interface to the GTS mezzanine provides the full
trigger-request/validation/rejection cycle and acquires also the clock
tag and the event number, the latter only in case of a validated
cycle.

The AGAVA module receives, through the front-panel input, the trigger
requests from the complementary detector (named here \textit{external}
trigger) that is passed as a single \SI{10}{\ns} wide pulse to the GTS
system (an \textit{internal} trigger request can be generated for
testing purposes). After the trigger request is passed to the GTS
system, the AGAVA module receives the local trigger signal and tag and
waits for the validation or rejection signal and tag coming from the
GTS System.  The latency time to receive the validation or rejection
depends on the GTS tree architecture and is of the order of
\SI{12}{\micro\second}.

Information is stored in registers or RAM memory (depending on
the firmware functionality mode). Once the information is stored, the
AGAVA module sets the data ready flag to inform the VME or VXI system
that the data can be readout by the CPU module and transferred to the
event builder. The busy flag is set after accepting every trigger
request. The release of the busy (i.e. ending the dead time for new
triggers) strongly depends on the VME/VXI readout mode and speed.
In case of high trigger rates, the backpressure front-panel
input can be used to reduce the trigger request to the GTS
system.

In order to support the wide range of complementary instruments,
two different AGAVA functionalities have been introduced: common dead-time
mode and parallel-like mode.

In the common dead-time mode only one trigger at a time is
accepted. The trigger request, acknowledged only if the AGAVA module
is not in busy status from the previous cycle, sets the busy state
until all data are readout by the CPU module.

In the parallel-like mode, when a trigger request arrives on the
front-panel input, the AGAVA module sets and holds its busy status
only until the local trigger and local trigger tag are received from
the GTS mezzanine card and are stored in the AGAVA multi-event RAM. A
new trigger request can be accepted afterwards by the AGAVA module
avoiding the GTS latency time and with a sizable reduction of the
total system dead time. The flow of validation or rejection triggers
and tags received by the AGAVA module are compared with the local ones
present in the multi-event RAM and if they match correctly the VME/VXI
CPU can read the relative data.

\section{Data acquisition} \label{s:daq}
\begin{figure}[ht]
  \begin{center}
    \includegraphics[width=\columnwidth]{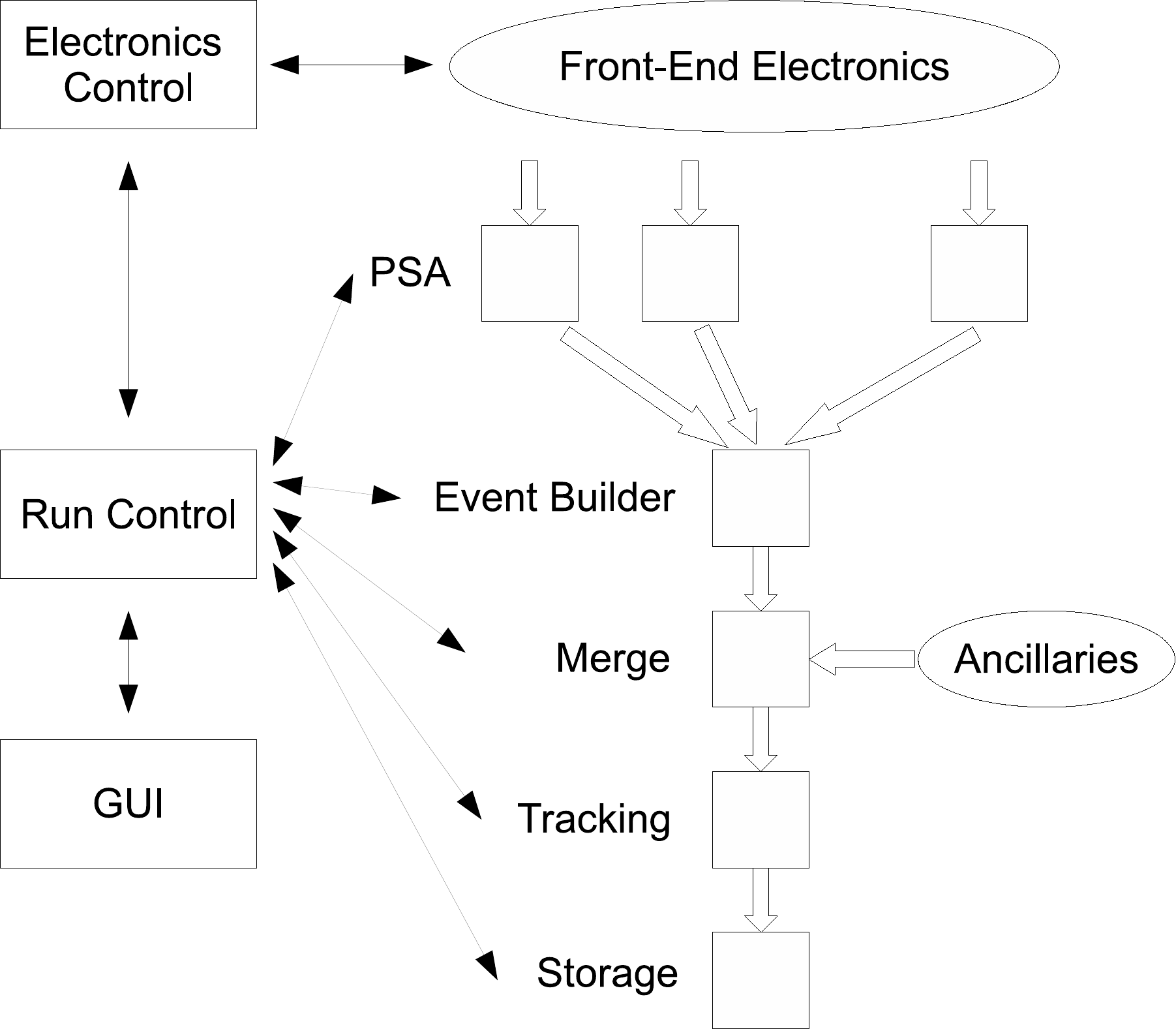}
    \caption{Layout of the AGATA data acquisition system.}
    \label{fig:daqlayout}
  \end{center}
\end{figure}

As shown in Fig.~\ref{fig:daqlayout}, the data acquisition (DAQ)
system receives the preamplifier traces from the FEE and processes
them into several stages up to the storage of reconstructed events;
see section \ref{ss:dataflow} for a description of the data
flow. Services described in section \ref{ss:services} are needed to
control and monitor the whole system including the electronics. For a
complex instrument such as AGATA, large computing, network and storage
capabilities are needed, as will be discussed in section
\ref{ss:daqhw}.

\subsection{Data flow}
\label{ss:dataflow}

The data flow integrates the algorithms needed to process the
information from the interaction points. At the end of the chain, data
have to be provided to the users in the AGATA Data Flow (ADF) library
\cite{adf}. After the PSA, data from all crystals are merged together
taking into account the physics correlations provided by the
pre-processing electronics. These correlations can be based on
timestamp or event number depending on which AGATA trigger mode is
used (see section \ref{s:fee}) If additional detectors are coupled to
the AGATA array, the corresponding data flow is assembled in the same
way. After this procedure tracking is performed to reconstruct the
$\gamma$-ray trajectories.  This procedure is then followed by the
data storage on a large local disk array before being sent to Grid
Tier1 computing centres based at INFN-CNAF in Bologna (Italy) and at
CC-IN2P3 in Lyon (France). It is also possible to write data to disk
at each step of the data flow.

Because of the high rate specifications (\SI{50}{\kHz} per crystal,
see Table \ref{tab:rates}) and the required processing power and large
data transfer bandwidths, the algorithms have to be distributed over
several computers (CPUs and cores) At this high rate the storage of
\SI{200}{\byte} per segment and core gives a data rate of about
\SI[per=slash]{400}{\mega\byte\per\second} per crystal at the
output of the FEE.  The data flow and embedded algorithms are managed
by the Narval \cite{Grave2005,narval} framework. Narval is a
highly-distributed data acquisition system running across a network
and acting like a single program that transports the data to the
storage of the reconstructed events.  Narval is based on actors
corresponding to separate processes that receive and send out data at
any stage of the data-flow chain.  It is developed in Ada for a high
reliability and safety level and it is very flexible since the actors
can load C and \CC\ shared libraries.

The event-builder and merge algorithms are written in Ada while the
PSA and tracking algorithms are written in \CC. The originality of the
DAQ system relies on the development of the ADF framework \cite{adf},
which has been designed in order to facilitate the separation between
data transport and algorithm development. ADF provides the interface
between Narval and the algorithms (PSA and tracking). It also allows
to encode/decode the data at various stages of the data flow. ADF can
be either used with Narval or as a stand-alone library for data replay
and it is part of the GammaWare package \cite{gammaware} (see section
\ref{s:data_analysis}).

In order to survey the data acquisition system, Narval provides a
``spy'' mechanism by which it is possible to sample pieces of the data
flow from each actor for control, monitoring and online analysis
purposes.  The spy is available to any client connected to the DAQ
services network.

\subsection{Services}
\label{ss:services}

A set of services are provided to control and monitor the whole system
including the detector electronics. The run control centralises
information from different subsystems of AGATA. It also collects
messages, errors, and provides an ELOG logbook \cite{elog} for the
users. A GUI is provided for the electronics and run controls.

\subsubsection{Run control}

The main purpose of the run control (RC) is to control and monitor the
DAQ components. It coordinates the large number of activities that are
necessary to achieve an operational state for the AGATA detector
including the DAQ system.  Actions like initialisation, setup of
several components, start and stop of the data acquisition, are
performed by the operator through the RC system.  The RC interacts
with the electronics control system, which is in charge of the control
of the hardware devices. The RC also provides the monitoring of the
data acquisition (input rate, buffer occupancy, error rate, etc.),
error report, error handling functionalities and logging capabilities.
The RC is based on the middleware produced by GridCC \cite{gridcc}.
To communicate with all the controlled components a standard WSDL (Web
Services Description Language) interface has been defined.

\subsubsection{The Cracow graphical user interface} \label{sss:cracow}

The GUI known as ``Cracow''~\cite{Grebosz2007251} allows the user to
perform commands such as loading the configuration, starting and
stopping the DAQ system, etc.  All the possible states of the DAQ
state machine are reflected in the GUI by a set of buttons that allow
particular actions to be performed. Cracow is an independent tool,
which communicates with the RC through the web services. It can be run
on any computer connected to the AGATA network.  As mentioned before,
Narval is a system whose actors (algorithms) are highly distributed,
which means that the user can create actors on many different
processors.  A configuration of these actors is called a topology.
The Cracow GUI can show a working diagram of the currently loaded
topology. The graph is constructed based on information delivered
online by the actors. They send information about themselves and about
their direct collaborators.

\subsubsection{Spectrum server and viewer}

The different algorithms implemented in the DAQ provide spectra or
matrices. In order to collect histograms from the various sources, the
GRU spectrum server (GANIL Root Utilities) \cite{gru} has been
implemented. As soon as a root histogram is added in the GRU database,
it is immediately available on the network and can be displayed by any
client like ViGRU \cite{gru} or the ``Cracow'' GUI described above.

\subsubsection{Electronics control}

The electronics control (EC) system is an ensemble of software tools
that allows to setup of all the electronics subsystems (digitisers,
pre-processing, GTS and possibly ancillary electronics) and to monitor
some key parameters.  The main functionalities to be provided by the
EC system are the following: the description of the system insuring
its coherency, the localisation of the servers controlling each
subsystem, the initialisation of the different devices with the
correct values, the saving of all the setup parameters for the whole
electronics or a part if necessary, the restoration of a previously
saved setup, the monitoring of key parameters in the different boards,
the handling of error/alarm events passing them to the RC. The EC
system has to insure the integrity of the system and the
synchronisation of the state machine of the different subsystems. It
accepts a set of simple commands from the RC (setup, go, stop, get
state, etc.).

The EC system is designed with a client/server approach and with
partitions accordingly to each type of hardware.  There is one EC
subsystem with its engineer-oriented GUI for each type of hardware,
i.e. one EC subsystem for digitisers, one for carrier boards, one for
core/segment mezzanines, one for GTS and one for ancillary
electronics. In order to centralise the different electronics
subsystems, a Global Electronics Control (GEC) based on the ENX
\cite{enx} framework has been designed.  The GEC acts as a coordinator
between the different subsystems and the RC.

\subsection{Hardware implementation}
\label{ss:daqhw}

The system first implemented for AGATA at LNL has been designed to be
scalable to cope with the full array specifications, easily movable
between different host sites and easily maintained by the
collaboration.  The AGATA DAQ hardware (servers, disk, local network,
etc.) and software should be seen from the host laboratory as a black
box.

On the hardware side, there is a big requirement for computing power
to run algorithms hosted by the data acquisition (PSA, event-builder,
merge, tracking). Each crystal is attached to one PSA server.  Since
the PSA processing is the most demanding part of the system, Narval
distributes events on the different cores of each PSA server. There is
also one server for event building, one for merging with ancillary
devices, one for tracking, three disk servers, two for slow control,
two for DAQ box control, and two for data analysis. All servers are
identical and connected to a KVM (Keyboard Video Mouse) switch to ease
the hardware maintenance.  The model chosen is the
IBM\textregistered\ x3550 1U server (from initial to M3 versions) with
2 quad cores Intel\textregistered\ Xeon\textregistered\ CPUs each,
which allows the PSA algorithms to be run at a crystal rate of about 4
kHz. The Debian distribution \cite{debian} of the GNU/Linux operating
system is installed on all machines.

AGATA generates a large data flow, which must be stored at very high
rate, especially when traces are stored. The local disk array of AGATA
consists of 4 disk servers. Three of them are
attached to a SUN\textregistered\ Storage Tek\texttrademark\ 6540
fibre channel system with 112 TB of disk (90 TB using RAID 6) and
shares data using the GPFS (General Parallel File System) clustered
file system. A fourth server attached to the storage system provides
the interface to the Grid for further data storage or analysis (see
section \ref{s:data_analysis}). Many services are needed to handle
the DAQ box: monitoring, documentation (Wiki documentation mirrored on
the main DAQ web site \cite{daqweb}), network control, installation
server, hardware monitoring using Zabbix \cite{zab}, etc.  These
services are hosted on a virtual Xen \cite{xen} machine. A backup
system is installed on a SUN\textregistered\ Fire\texttrademark\ X4560
system with 8 TB of disks to save periodically the vital elements of
the DAQ box.

In order to reduce the dependency particular to the host laboratories,
basic network services are provided in the DAQ box: DHCP (Dynamic Host
Configuration Protocol), DNS (Domain Name System), software
distribution, etc. External users (developers, system administrators)
can access the DAQ box through a VPN (Virtual Private Network).

Networking is an important issue having in mind the high data rates
foreseen in forthcoming AGATA experimental campaigns. The
communication between nodes is organised around two different
networks, one for data acquisition called DAQ network and the other
for network services called DAQ services.

The data acquisition infrastructure can be hosted in one or several
buildings. For example at LNL, the Tandem accelerator building hosts
the data flow and DAQ services machines while the Tier 2 computer room
hosts all the disk servers. Both rooms are connected with
\SI{3}{\giga\byte} optical links.

\subsubsection{Algorithm for compression of traces} \label{sss:traces}

\begin{figure}
  \centering
  \includegraphics[width=\columnwidth]{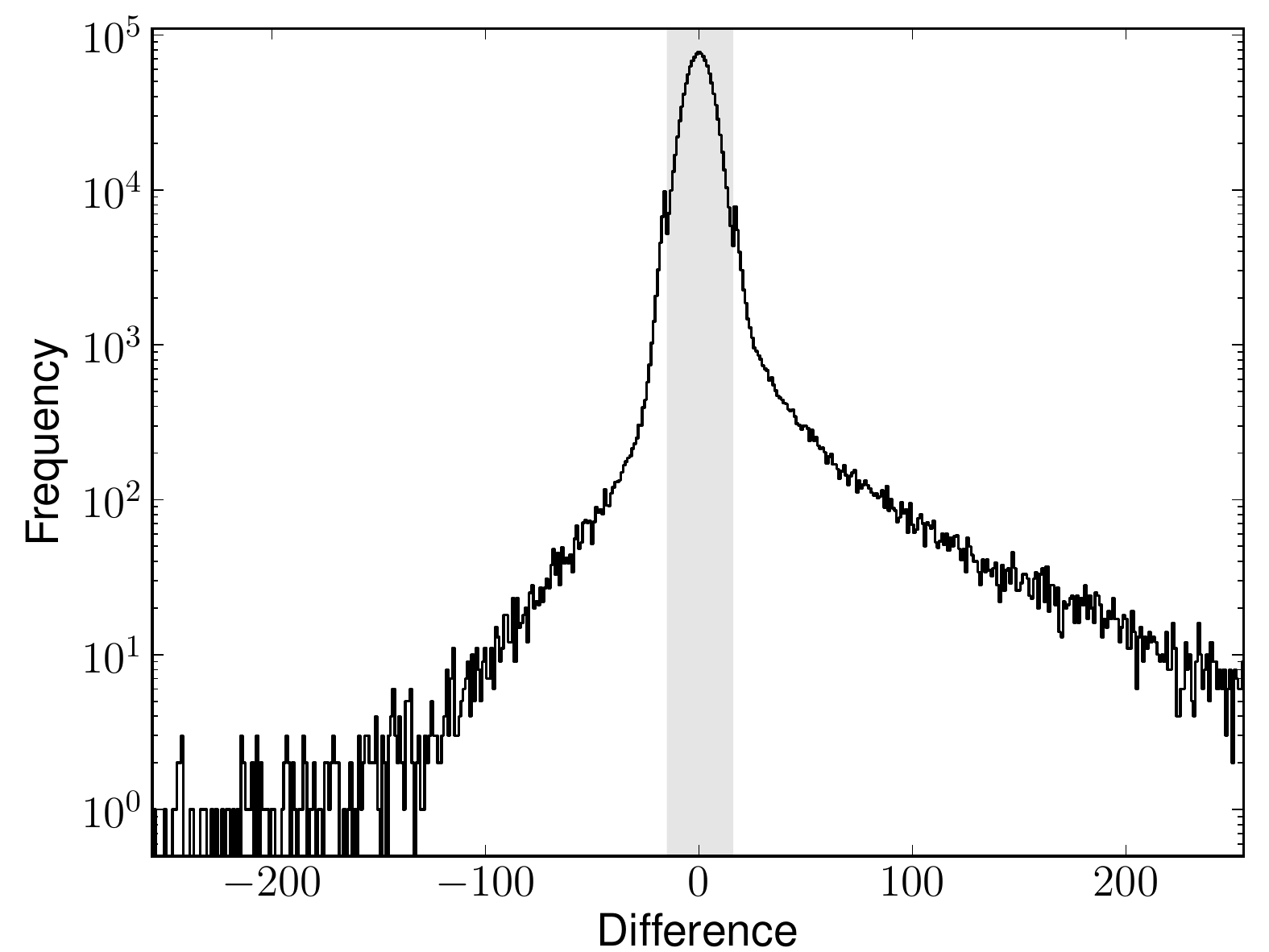}
  \caption{(Colour online) Typical difference spectrum of recorded
    traces.  The parameter on the horizontal axis is the difference in
    the value of two consecutive sampling points. The shaded region,
    which marks difference values from \num{-31} to
    \num[retainplus]{+32}, contains \SI{93}{\percent} of the
    data. The histogram contains data from about 13 million sampling
    points.}
  \label{fig:diff_spec}
\end{figure}

In order to reduce the size of the stored traces a simple algorithm
has been devised to compress the traces to about one half of their
original size. The compression algorithm is based on the observation
that the difference between the measured voltages of two consecutive
sampling points is small. Fig.~\ref{fig:diff_spec} shows a typical
spectrum of differences between neighboring sampling points. The
spectrum is broad with a very sharp peak around zero difference. The
mean value of the distribution is positive because the traces are
recorded at the beginning of the pulse to cover the leading edge of
the signal.

The sampling points have a \num{14}-bit resolution and are stored by
the pre-processing electronics in \num{16}-bit unsigned integer
values. The trace length is typically 100 points.  In
Fig.~\ref{fig:diff_spec} one may see that about \SI{93}{\percent} of
the difference spectrum is contained in the region between \num{-31}
and \num[retainplus]{+32}.  This fact will be used for the
compression as several of these small differences can be packed into
\num{16} bits. The chosen storage format is the following:
\begin{itemize}
\item The first sampling point of the trace is always stored with its
  original value. It becomes the first reference point.
\item If three consecutive sampling points after the reference point
  differ from their respective previous sampling point by values
  between \num{-15} and \num[retainplus]{+16},
  the three differences are encoded in
  \num{3 x 5} bits in a \num{16}-bit unsigned integer with the most
  significant bit set to \num{0}.
\item If one sampling point after the reference point differs from the
  reference by a value between \num{-31} and \num[retainplus]{+32},
 the difference is
  encoded in an \num{8}-bit unsigned integer with the two most significant
  bits set to the binary value \num{10}.
\item Otherwise, the \num{14}-bit value is stored in a \num{16}-bit unsigned
  integer with the two most significant bits set to the binary value
  \num{11}.
\item Finally the reference point is advanced to the last point stored
  so far.
\end{itemize}
For most of the sampling points of a trace, three consecutive points
from originally three \num{16}-bit words can be packed into a single
\num{16}-bit word.  For traces with \num{100} points excluding the
headers of the data, a compression of about \SI{38}{\percent} is
achieved. The compression of data containing both sampling points and
data headers is typically about \SI{44}{\percent}.

\section{Pulse-shape simulations} \label{s:pss}
In order to realise the real-time localisation of the scattering
sequence following a $\gamma$-ray interaction inside an AGATA
detector, the experimentally digitised pulse shapes will be compared
with a basis data set. For online pulse shape analysis (PSA) to be
implemented successfully, it has been estimated that more than
\num{30000} basis sites per crystal for a \SI{2}{\mm} Cartesian grid
\cite{Venturelli2005} are required. The basis can be obtained from
either experimental or simulated data. However, presently the time
prohibitive nature of the experimental methodology (see section
\ref{s:detchar}), means that the only practical way to generate a full
basis is by calculation.

The AGATA collaboration has therefore developed electric field
simulation codes such as Multi Geometry Simulation (MGS) package
\cite{Medina2004}, the Java AGATA Signal Simulation (JASS) toolkit
\cite{Schlarb2011_jass} and the IKP Detector Simulation and
Optimisation method now called the AGATA Data Library (ADL)
\cite{Bruyneel2006764, Bruyneel2006774, Bruyneel2006, adl}, in order
to facilitate the realisation of the on-line PSA.  These codes have
been used to generate variants of a single crystal basis and an
earlier code originally developed for the MARS project
\mbox{\cite{Kroll2001227}} was used to optimise the physical
segmentation scheme in the depth of the AGATA crystal.

\begin{figure}[ht]
  \begin{center}
    \includegraphics[width=\columnwidth]{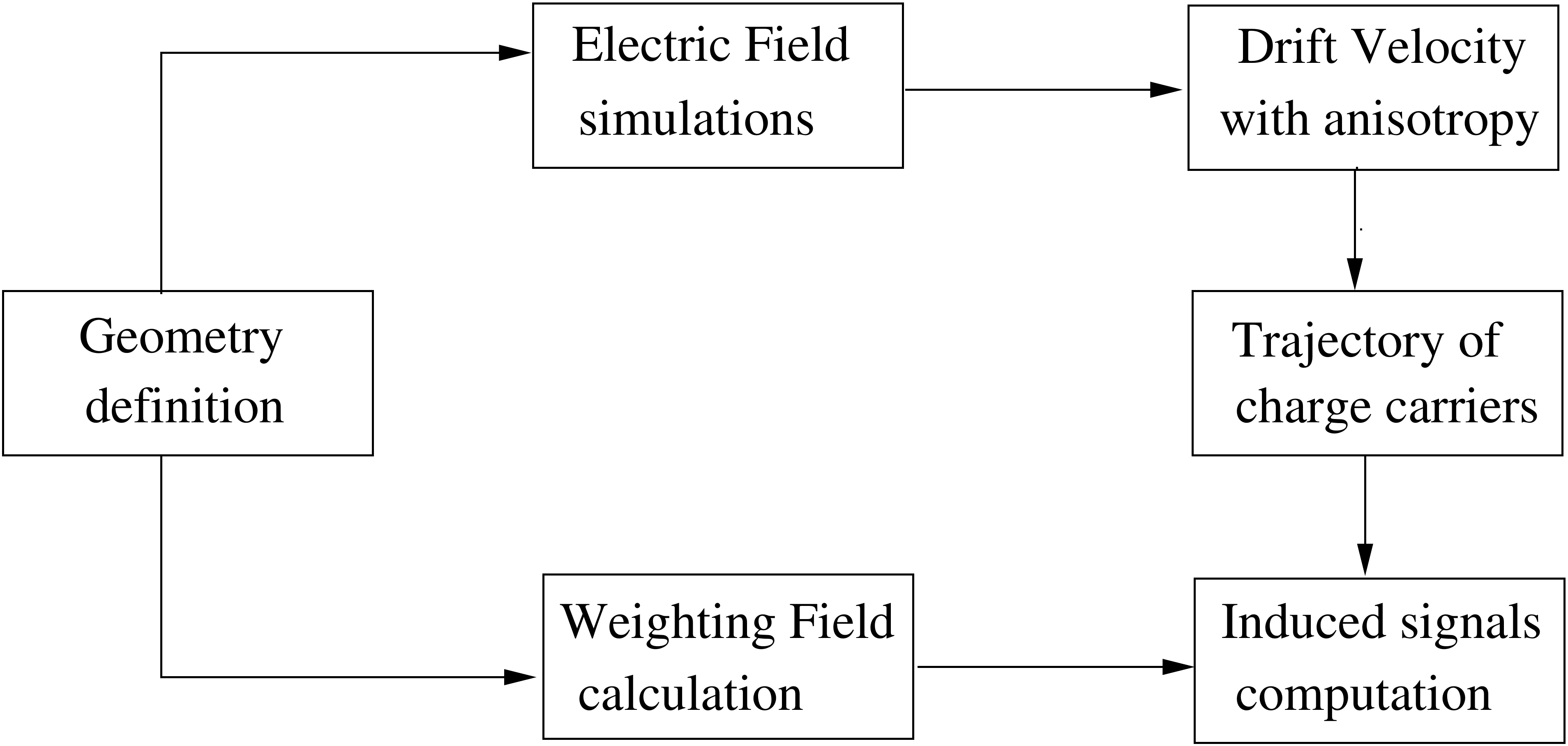}
    \caption{ Data flow diagram for the simulation of the expected
      pulse shapes at the contacts of any arbitrary HPGe detector
      geometry. For a given detector, the crystal volume can be
      divided into a cubic matrix of lattice sites. Values for the
      electric potential, electric field and weighting field are
      calculated at each position. The drift velocity matrices are
      calculated from the electric field matrix. The detector response
      for a given interaction site is calculated by tracking the
      trajectory of the charge carriers through the weighting field
      \cite{Radeka1988}.  }
    \label{fig:psc02}
  \end{center}
\end{figure}

In general, the electric field simulation codes use finite-element
methods for solving partial differential equations, such as the
solution to Poisson's equation.  Environments such as FEMLAB and
DIFFPACK can also be adapted for providing an analytical solution to
the complex electric field distribution inside the AGATA crystals. The
codes utilise the approach shown in block format in
Fig.~\ref{fig:psc02}. The calculations are performed on a user
specified 3D grid that maps a given detector volume. Results from each
stage of the calculations are stored in matrices. The values at each
point in the matrices are then recalled to generate the pulse shape
response, as determined by the trajectories of the charge carriers
through the weighting field \cite{Radeka1988}.

The MGS package \cite{Medina2004} utilises MatLab's matrix environment
to derive the expected pulse-shape response at the contacts of any
geometry of HPGe detector. The stand-alone package has been compiled
for use with both Linux and Microsoft Windows operating systems. MGS
was developed as an alternative specialised solution to the commercial
packages.

\begin{figure}[ht]
  \begin{center}
    \includegraphics[width=0.9\columnwidth]{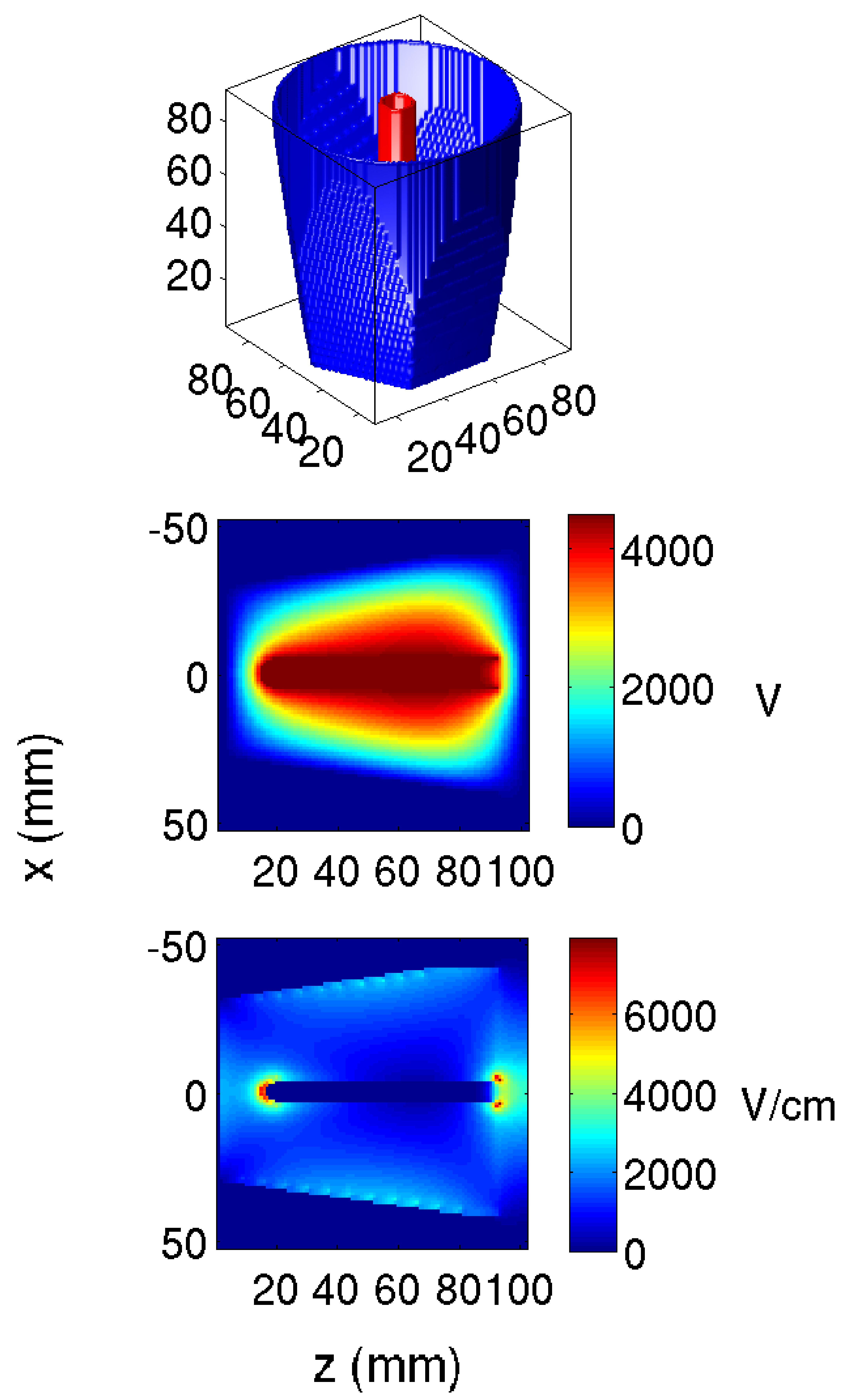}
    \caption{(Colour online) An MGS simulation of the AGATA symmetric
      prototype crystal geometry (top). Electric potential (middle)
      and electric field values (bottom) in the $z$-$x$ plane at $y =
      \SI{51}{\mm}$ (centre of the crystal). The results show the
      decrease in potential and field strength as a function of
      increasing radius from the central anode.  }
    \label{fig:psc03}
  \end{center}
\end{figure}

The top panel of Fig.~\ref{fig:psc03} shows the geometry of the AGATA
symmetric prototype crystal as generated by MGS.  The middle panel
shows the electric potential and the bottom panel the electric field
values in the $z$-$x$ plane at $y = \SI{51}{\mm}$ (centre of the
crystal). The results show the decrease in potential and field
strength as a function of increasing radius from the central
anode. The figure clearly illustrates the complex nature of the
electric field within the closed-end coaxial geometry of the AGATA
crystal.

At high electric fields and low temperatures the charge carrier drift
velocities in germanium become anisotropic \cite{Mihailescu2000350}.
They depend on the electric field vector with respect to the
crystallographic lattice orientation. The electron drift velocities,
$v_{\textnormal{e}}$, are saturated at field strengths
$>$\SI{3000}{\volt\per\cm} for the <100> and <110> directions, and at
$\sim$\SI{4000}{\volt\per\cm} for the <111> direction. The hole
drift velocities $v_{\textnormal{h}}$ are saturated for fields
$\sim$\SI{4000}{\volt\per\cm} along all three major crystallographic
axes. The models for anisotropic drift of the electrons and holes in
$n$-type HPGe detectors implemented in MGS have been derived from work
published in \cite{Bruyneel2006764} and \cite{He2001250},
respectively.

Before the induced current at each electrode can be calculated, the
weighting potentials and weighting fields must be generated. The
calculation is performed with a null space charge density and with
\SI[retainplus]{+1}{\volt} on the sensing electrode with all
other electrodes grounded. For a given interaction position, the
charge-pulse response observed at any electrode depends on the
trajectory of the charge carriers through the weighting potential of
that electrode.

\begin{figure*}[ht]
  \begin{center}
    \includegraphics[width=0.90\textwidth]{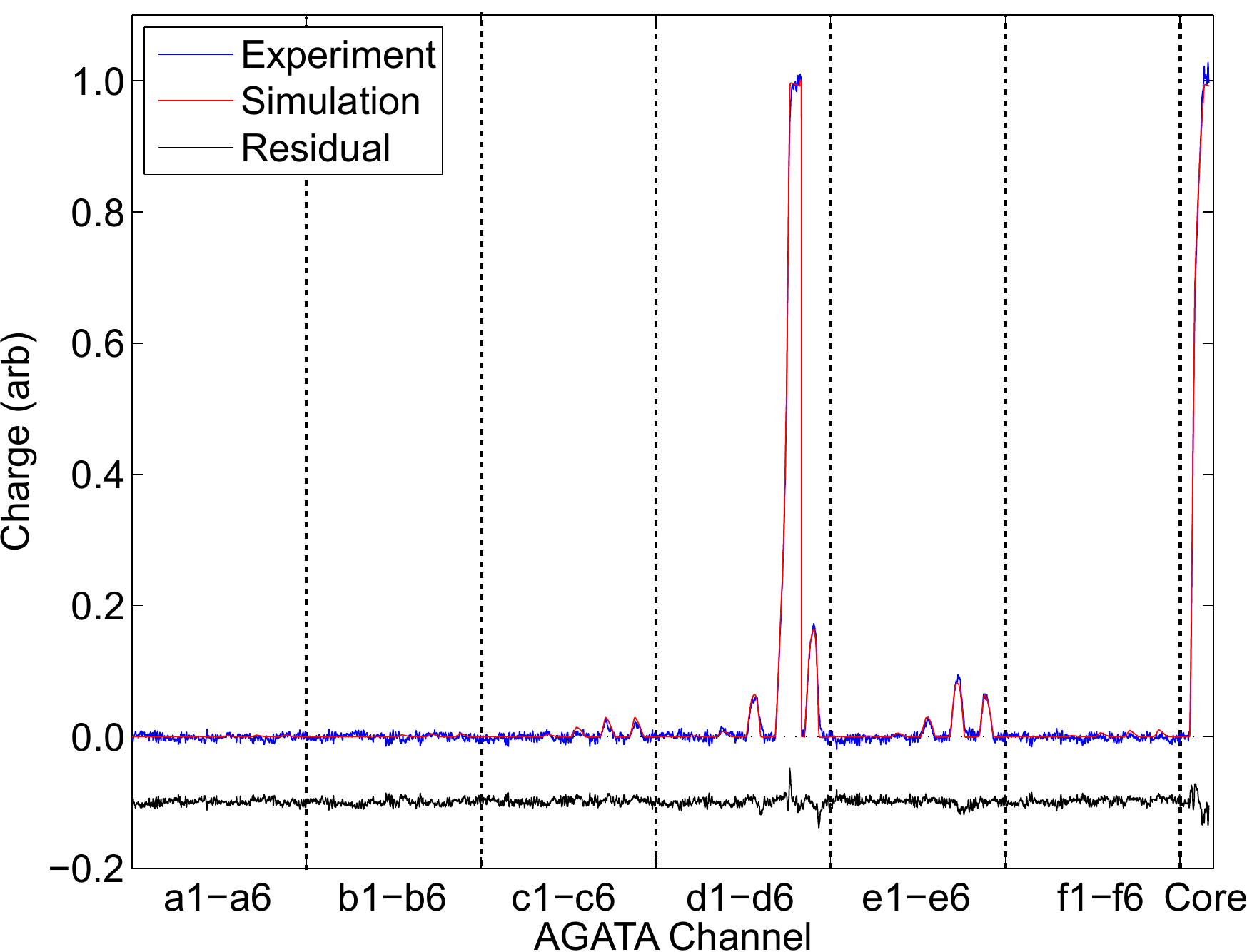}
    \caption{(Colour online) A figure showing the \num{37} charge
      pulses from the C001 detector for an interaction in a "true
      coaxial" region of the crystal in segment d5. The average
      experimental pulses are shown in red, the MGS calculated
      responses are shown in blue and the difference in magnitude
      between the two pulse-shape data sets is shown in black. The
      difference is shifted down by \num{0.1} for clarity.  These data
      were obtained from a coincidence scan using a $^{137}$Cs source
      of \SI{662}{\keV} $\gamma$ rays. The AGATA detector records the
      \SI{374}{\keV} $\gamma$ rays that have been Compton scattered by
      \ang{90}.}
    \label{fig:mgs_waveforms}
  \end{center}
\end{figure*}

The experimental characterisation data sets have been used to adjust
and optimise the parameters of the code.  Fig.~\ref{fig:mgs_waveforms}
shows an example comparison between the average experimental pulses
collected from AGATA crystal C001 and the ones calculated with MGS.

The JASS toolkit~\cite{Schlarb2011_jass} provides a Java-based
pulse-shape simulation using a very similar logic structure as MGS
shown in Fig.~\ref{fig:psc02}.  In addition, it allows for a much
higher granularity in the grid as well as a more precise description
of the complex geometry of the AGATA detectors and the processes at
the segment boundaries.  In connection with an extremely accurate
interpolation method the description is very precise, especially in
the critical regions of the crystals. With its automated generation of
the signal basis for the AGATA detectors an excellent agreement
between simulated signals and the experimentally characterised
detector signals has been found.

The ADL package \cite{Bruyneel2006764, Bruyneel2006774, Bruyneel2006,
  adl} allows for the simulation of the position-dependent detector
response to $\gamma$-ray interactions. This library was written in C
and it is used to create a pre-calculated set of position-sensitive
detector responses.  Such sets are used as a lookup table to translate
online acquired signals into position information. ADL makes use of
realistic anisotropic electron- and hole-mobility models specially
developed for germanium. The code has a unique capability to simulate
partially depleted detectors and their capacitances and this feature
allows for the space-charge reconstruction in highly-segmented
detectors from capacitance-voltage
measurements~\cite{Birkenbach2011176,Bruyneel201192}.  The code was
recently also successfully used to simulate the position-dependent
collection efficiency within the AGATA detectors. Such efficiencies
can be used to correct for trapping effects by exploiting the high
position sensitivity of the detectors \cite{Bruyneel2010}. The
approach utilises the radial dependence of the trapping magnitude
which allows a parametric correction to be applied.

A code based on the chain of free open-source software
OpenCASCADE~\cite{OCC}, gmesh~\cite{Geuzaine2009},
libmesh~\cite{Kirk2006}, and GSL~\cite{GSL}, called AGATAGeFEM
~\cite{Ljungvall2005PhD} has also been developed.  The open-source
software are used to model the geometry, to create a mesh adapted as a
function of the field shape for the solving of the partial
differential equations, to solve these equations, and finally to solve
the problem of transporting the charges inside the crystal and the
formation of the signals. The response of the preamplifier and
crosstalk are included. A pulse-shape database calculated using this
code has yet to be tested with real data.

\section{Pulse-shape analysis} \label{s:psa}
The task of pulse-shape analysis (PSA) is to identify with high
precision the location of the individual interaction points and the
corresponding energy deposits of a $\gamma$ ray.  A $\gamma$ ray will
normally have a chain of interactions in the shell of germanium
detectors (e.g. \num{3} to \num{4} at \SI{1.3}{\MeV}). There can be
more than one interaction in one detector segment and/or the $\gamma$
ray can be scattered to another segment of the same crystal or to an
adjacent detector, or even across the shell.  The accuracy of the
location of these interaction points has to be better than \SI{5}{\mm}
(FWHM).  Such a high precision is required in order to perform the
tracking process with high efficiency.

The location and energy determination must be performed by algorithms
that are fast enough for real time application, with the computing
power available.  Indeed, the high experimental event rate multiplied
by the data rate of the digitiser produces an information flux far
beyond the realistic disk storage capacities in terms of volume and
recording speed rate. Data from a set of \num{37 x 100} ADC samples
delivered by the digitiser are reduced through PSA to few parameters
of interest. These are determined, by comparing the detected pulse
shapes to a calculated reference basis, in real time. This provides
the three dimensional interaction position, energy, time and a
confidence in the quality of the determined fit.

A set of algorithms, such as grid search
\cite{Venturelli2005,Schlarb2008}, genetic algorithms
\cite{Kroll2006691}, wavelet decomposition, 
and a matrix method \cite{Olariu2007,Olariu2006}, optimised for
different types of events, have been developed.  A first operational
version of the PSA code, a stepwise refined grid-search
\cite{Venturelli2005}, has already been implemented in the final
Narval environment. Other faster algorithms, which have been
extensively tested with simulated data and are suited to more complex
event structures, will be implemented in the near future. It is
expected that the performance of PSA will continuously improve during
the project due to refined algorithms and increased computing power
but especially due to improvements in the quality of the reference
signals.

The approach based on a genetic algorithm aiming for a full
decomposition was found to be too slow for online processing
\cite{Kroll2006691}. Therefore, the collaboration focused initially on
the grid-search algorithm \cite{Venturelli2005}, which due to its
simplicity is the most robust among the signal decomposition
codes. 

\begin{figure}[ht]
  \begin{center}
    \includegraphics[width=\columnwidth]{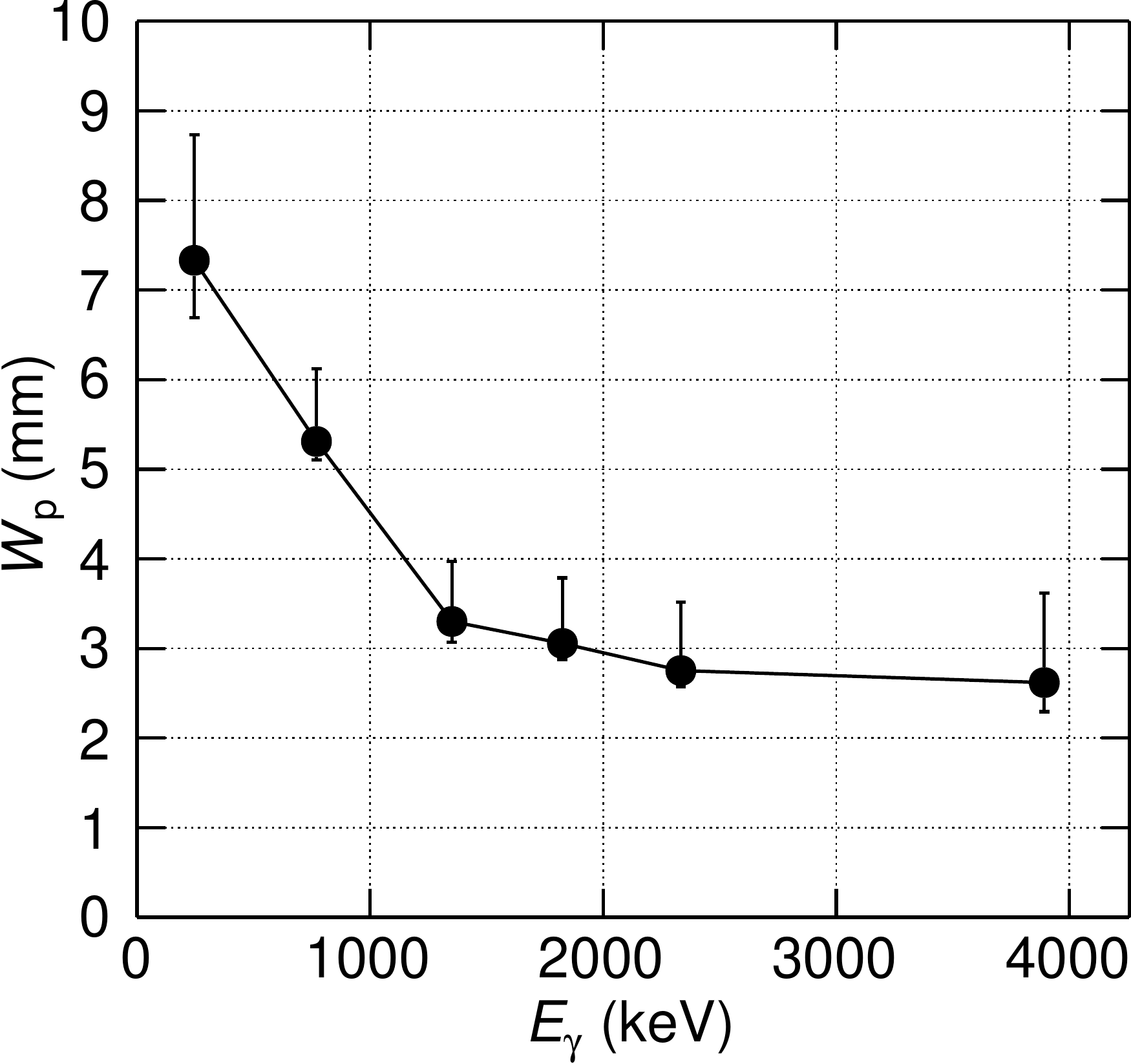}
    \caption{Interaction position resolution $W_{\textnormal{p}}$
      (FWHM) versus $\gamma$-ray energy $E_{\gamma}$ extracted from
      the data obtained in the first AGATA in-beam commissioning
      experiment based on an inverse kinematics fusion-evaporation
      reaction with a \SI{64}{\MeV} $^{30}$Si beam and a thin $^{12}$C
      target. The error bars include both statistical and systematic
      errors.  See \cite{Soderstrom2011} for further details.  }
    \label{fig:wp_vs_egamma.pdf}
  \end{center}
\end{figure}

The code based on the grid-search algorithm has been validated with
experimental data obtained in two in-beam experiments using techniques
based on Doppler effects of $\gamma$ rays emitted by nuclei in flight
\cite{Recchia2009555,Soderstrom2011}. The first experiment was
performed at IKP Cologne with the prototype ATC detector, which
consisted of three symmetric HPGe crystals \cite{Recchia2009555}. The
second experiment was the first in-beam AGATA commissioning experiment
performed at LNL with the first production series ATC detector, which
consisted of three asymmetric HPGe crystals \cite{Soderstrom2011}. The
measured interaction position resolution versus $\gamma$-ray energy
obtained in the latter experiment is shown in
Fig.~\ref{fig:wp_vs_egamma.pdf}.  The grid-search algorithm has also
been validated in a measurement with the setup at LNL using a
$^{22}$Na source \cite{Klupp2011}.

It was shown that in the case of single-hit events the grid-search
algorithm already gives a sufficiently good Doppler
correction. However, it is not well suited to more complex events, the
reason being that in the case of multiple interactions in a crystal,
or especially in a segment, the analysis effort strongly increases
with the number of degrees of freedom. However, the grid-search
algoritm has been used successfully with in-beam data to handle
multiple interactions in the same crystal as along as the segments
with transient signal do not overlap.

Several fitting procedures as mentioned above have been developed in
order to deal with these types of events.  The currently favoured two
methods are, the fully informed particle swarm \cite{Schlarb2008} or
the SVD (singular-value decomposition) matrix method
\cite{Olariu2007,Olariu2006,Desesquelles2005,Desesquelles2009a}. The
former method is faster but the latter gives a better precision on the
location of the hits. All these new algorithms still have to be
validated with AGATA data. Given the absence of a one-for-all
algorithm a dispatcher code, choosing the most adequate PSA algorithm
for a given event, will also be implemented soon.

Table \ref{tab:psa01} gives an overview on the performance of
different algorithms derived from tests using simulated data. The CPU
performance is quoted relative to the grid-search algorithm
(\SI{3}{\ms} per event) for single interactions and the particle-swarm
algorithm (\SIrange{2}{5}{\ms} per event) for multiple interactions.

\begin{table*}[ht]
  \label{tab:psa01}
  \begin{center}
    \caption{Overview of the performance of different PSA algorithms
      derived from tests using simulated data. The performance is
      quoted relative to the grid-search algorithm. The position
      resolution is quoted in as FWHM values.  
    }
    \begin{threeparttable}
     \begin{tabular}{|l|c|c|c|c|}
      \hline
      \multirow{3}{*}{Algorithm} & \multicolumn{2}{c}{Single interaction} & 
        \multicolumn{2}{|c|}{Multiple interactions} \\
      \cline{2-5}
      & CPU time & Resolution & CPU time & Resolution \\
      & [norm/GS] & [mm] &  [norm/PS] & [mm]  \\
      \hline
      Grid search (GS) \cite{Venturelli2005} & $1$ & $2$ & $-$ & $4$\tnote{a} \\
      \hline
      Extensive GS \cite{Schlarb2008} & $2.7$ & $1$ & $6\cdot10^4$ & $4$ \\
      \hline
      Particle swarm (PS) \cite{Schlarb2008} & $0.1$ & $2$ & $1$ & $4$ \\
      \hline
      Matrix method \cite{Olariu2007,Olariu2006} & $6.7$ & $2.4$ & $10$ & $10$ \\
      \hline
      Genetic algorithm \cite{Kroll2006691} & $330$ & $1.9$ & \num{2e2} & $8.1$ \\
      \hline
      Binary search & $0.06$ & $1$ & Not adapted & Not adapted \\
      \hline
      Recursive subtraction \cite{Crespi2008PhD,Crespi2007459} & 
      Not evaluated & $3$\tnote{b} & Not evaluated & $5$\tnote{b} \\
      \hline
      Neural network \cite{Schlarb2008} & $2$ & $1.5$ & Not adapted & Not adapted \\
      \hline
      Wavelets & Not evaluated & $2.3$ & Not adapted & Not adapted \\
      \hline
    \end{tabular}
    \begin{tablenotes}
    \item[a] Only neighbouring segment hits.
    \item[b] Only radial coordinate.
    \end{tablenotes}
    \end{threeparttable}
  \end{center}
\end{table*}

After the energy calibration, which is done by the front-end
electronics, the influence of noise, pedestal, time jitter and cross
talk needs to be understood in order for the experimental signals to
be successfully compared with the basis signals

Concerning the time jitter, different time-shift determination methods
have been considered: supplementation of the basis with time-shifted
signals, Neural Network determination \cite{Schlarb2008},
Kolmogorov-Smirnov determination \cite{Desesquelles2009b}, Taylor
expansion and the substitution of residue minimisation by chi-square
minimisation \cite{Desesquelles2009542,Desesquelles2009c}. The first
method has the drawback of requiring larger signal bases, the latter
method reduces the influence of the time jitter but does not measure
it.  Using simulated data, the Neural Network algorithm has shown to
be very fast and provides a resolution better than \SI{2}{\ns} at
computing times in the order of \SI{10}{\micro\second}. Taking into
account only the core signal, this algorithm seems to be very robust
and hardly affected by noise and cross talk discussed below. The
Kolmogorov-Smirnov and Taylor expansion methods are purely
algebraic. They also give excellent results in terms of computing
speed, robustness and precision.

The cross-talk effect is present in any segmented detector. It induces
energy shifts and decreases the hit-location precision as it mixes the
transient signals. Transient signals are those induced on adjacent
electrodes to the primary interaction due to the drift of the charge
carriers inside the germanium crystal. Cross-talk contributions can
also appear between segments of different detectors. Only negligible
cross-talk components were observed between a firing detector and the
core signal of a neighbouring detector
\cite{Bruyneel2006,Bruyneel2009196}. The effects of cross talk are
described in details in section \ref{s:det}. Two quite similar
correction methods have been used. The first one includes the
cross-talk effect into the basis signal using a cross-talk matrix. The
second corrects the measured signal by the cross-talk effect. These
corrections permit the lowering of the FWHM error on the position by
about \SI{1}{\mm}.

Another difficulty in the determination of the precise interaction
position arises from the fact that the germanium crystal may be
slightly displaced or tilted inside its capsule. It has been shown
that this problem may be addressed prior to the experiment using the
scanning tables or within the final setup using a radioactive source
at fixed points and the $\gamma$-imaging method \cite{Recchia2008}.

In order to facilitate the maintenance and development of algorithms
used for AGATA two levels of abstraction are used for the PSA. The
first is the ADF library \cite{adf} that is responsible for coding and
decoding data in the data flow. The second layer of abstraction is a
set of \CC\ classes, which provide, via inheritance, simple means of
``attaching'' a PSA algorithm to the Narval DAQ. This system has been
successfully tested with the grid-search algorithm. Due to the fact
that no cross correlations between the signals of different AGATA
crystals has been found so far, all the PSA relevant algorithms
operate on the signals of each individual crystal in parallel on
different machines. This allows the use of several computers in
parallel to analyse different events. Currently, it is foreseen to use
\num{2} processors with \num{4} cores each to process the data of one
crystal which will improve the PSA performance by a factor of \num{2}
to \num{4}.

Finally the events have to be reassembled according to their
timestamps and a tracking algorithm is applied in order to disentangle
the coincident interaction points and to determine the total energy
and the emission direction of those $\gamma$ rays that have been fully
absorbed in the germanium shell. Absolute positions of the individual
crystals, tilting angles and target position corrections enter at this
stage.

\section{Gamma-ray tracking} \label{s:grt}
The aim of tracking algorithms is to reconstruct the trajectories of
the incident photons in order to determine their energy and
direction. To do this, the algorithms must disentangle the interaction
points identified in the detectors and establish the proper sequences
of interaction points. Tracking algorithms can be divided into two
classes: algorithms based on back tracking \cite{vanderMarel1999538}
and algorithms based on clustering and forward tracking
\cite{Schmid199969}. Both are related to the particular properties of
the interaction of photons with matter.  In forward-tracking
algorithms, the first step of the procedure is to group interaction
points into clusters in ($\theta$, $\phi$) space. Back tracking, on
the other hand, is based on the observation that the final
photoelectric interaction after a $\gamma$-ray scattering sequence
usually falls into a narrow energy band. Starting from interaction
points in that energy range the algorithm tracks back towards the
original emission point using the physical characteristics of the
interaction process and selects the most probable interaction scheme.

For photon energies of interest (from about \SI{10}{\keV} to about
\SI{20}{\MeV}), the main physical processes that occur when a photon
interacts in germanium are Compton scattering, Rayleigh scattering,
pair creation and the photoelectric interaction. Since Compton
scattering is the dominant process between \SI{150}{\keV} and
\SI{10}{\MeV}, all tracking algorithms are based on the properties of
this interaction.

Within the AGATA project, the development and optimisation of tracking
algorithms have been performed. Tracking has also been used to
evaluate the performance of PSA. Another part of the work related to
tracking has been the use of tracking techniques to investigate the
background-suppression capabilities from sources of radiation that do
not originate from the target.

\subsection{Development of tracking algorithms}

The development of tracking algorithms and the improvement of existing
ones is an ongoing process. Realistic simulated data sets, produced
with the \geant\ AGATA code \cite{Farnea2010331}, have been and are
systematically used to test, compare and improve the performance of
tracking algorithms. This code treats the physics of the energy
deposition correctly by generating and propagating all the possible
secondary particles and by taking into account the momentum profile of
electrons in germanium \cite{LopezMartens2004454}. In order to produce
more realistic data, the interaction points separated by less than
$d_{\textnormal{res}}= \SI{5}{\mm}$ are packed together, the positions
of the interaction points are randomly shifted in all direction ($x$,
$y$ and $z$) by sampling a Gaussian energy-dependent uncertainty
distribution and an energy threshold of \SI{5}{\keV} is applied to all
the interaction points within each segment.

It has been shown that forward-tracking algorithms are more efficient
than back-tracking ones \cite{LopezMartens2004454}. This is because
back-tracking algorithms rely on the identification of the last
(photoelectric) interaction point in a scattering sequence, which
generally loses its original characteristics after PSA: it is poorly
localised and/or packed with other interaction points.  However,
ultimately the optimal $\gamma$-ray tracking algorithms may require
the use of a combined approach where for example back tracking is used
to recover events missed by the clustering techniques.

\subsection{Clustering techniques}

In the Mars Gamma-Ray Tracking (MGT) \cite{Bazzacco2004248} and the
Orsay Forward Tracking (OFT) \cite{LopezMartens2004454} codes, both
based on the forward-tracking technique, points are grouped into
clusters according to their relative angular separation. In order to
increase the tracking performance for high photon multiplicities
without decreasing the performance at lower multiplicities, the
maximum allowed separation angle is set to depend on the number of
interaction points in the event.

A clustering technique based on the Fuzzy C-Means algorithm has been
developed \cite{fuzzycmeans}. After running this algorithm, the
optimal positions of the centres of the clusters and the degree of
membership of each point to each of the clusters are obtained. The
next step in the code is the ``defuzzification'', making possible the
use of standard validation procedures. Another clustering method,
called Deterministic Annealing Filter (DAF), has also been developed
\cite{Didierjean2010188}. In this method, the clustering problem is to
perform the minimisation of the distortion criterion, which
corresponds to the free energy related to the annealing process in
statistical physics.

The detection efficiency and peak-to-total obtained by the
above-mentioned algorithms for \SI{1}{\MeV} photons emitted at the
centre of AGATA are summarised in Table \ref{tab1_track}.

\begin{table}[htdp]
  \caption{Simulated full-energy efficiency and peak-to-total (P/T) of
    AGATA for cascades of \num{30} (and \num{1}) \SI{1}{\MeV}
    photons. In all the cases, the data is packed and smeared in the
    standard way and a \SI{5}{\keV} energy threshold has been
    applied.}
  \begin{center}
    \begin{tabular}{|c|c|c|}
      \hline
      \multirow{2}{*}{Algorithm} & Efficiency & P/T \\
                & [\%] & [\%] \\
      \hline
      MGT & 28(43) & 49(58) \\
      OFT & 24(37) & 54(68) \\
      Fuzzy C-Means & 27 & 46 \\
      DAF & 26(36) & 47(69) \\
      \hline
    \end{tabular}
  \end{center}
  \label{tab1_track}
\end{table}

\subsection{Effective distances}

In all tracking philosophies, the photon trajectories are tracked with
the help of the Compton scattering formula and with the ranges and
cross sections of the photoelectric, Compton and pair-production
interactions. To compute the appropriate ranges of photons in
germanium, the effective distances in germanium between interaction
points as well as between the interaction points and the source need
to be computed. The problem can be solved analytically if the detector
geometry is approximated to a $4\pi$ germanium shell. This
approximation was validated with the \geant\ simulation code
\cite{Farnea2010331}, in which the exact geometry of AGATA is defined,
i.e. the shape of the crystals, encapsulation, cryostats, the empty
spaces and the distances between all these elements.

\subsection{PSA and front-end electronics}

It has recently become clear that PSA algorithms have difficulties in
dealing with events in which there is more than one interaction point
per segment. This is because the parameter space to search becomes
very large and because the properties of the response function of the
detector can lead to more than one solution. Although algorithms to
determine the number of interaction points in each segment have been
developed and have shown encouraging results in the coaxial part of
the crystal \cite{Crespi2007459}, the effect of systematically
assuming only one interaction point per segment has been implemented.
No drastic drop in performance is observed, not even when the recoil
velocity is \SI{50}{\percent} of the speed of light, which yields a
higher concentration of interaction points because of the Lorentz
boost.  This result is understood since the average of the
distribution of the number of interaction points per segment is close
to \num{1} and varies only slightly with incident photon energy.

Due to electronic noise, signals corresponding to a deposited energy
below a certain value will not be detected in the segments of
AGATA. The effect of this energy threshold on the tracking efficiency
has been investigated. The loss in full-energy efficiency with respect
to the case of a \SI{1}{\keV} threshold is found to be
\SI{1}{\percent} (\SI{5}{\keV} threshold), \SI{8}{\percent}
(\SI{10}{\keV} threshold), \SI{13}{\percent} (\SI{20}{\keV} threshold)
and \SI{29}{\percent} (\SI{50}{\keV} threshold). The data from AGATA
has shown that most of the full-energy efficiency can be recovered by
using the core-energy information.

\subsection{Neutrons}

The influence of the elastic and inelastic interactions of neutrons on
tracking performance has been found to be significant for low to
medium $\gamma$-ray multiplicities especially on the P/T ratio
\cite{Ljungvall2005379}. As far as pulse shapes are concerned, it has
been shown that neutrons and $\gamma$ rays yield similar signals in
nonsegmented closed-end coaxial and planar $n$-type HPGe crystals,
even though the interaction mechanisms are different
\cite{Ljungvall2005553}. Distinguishing neutrons and $\gamma$ rays by
time of flight methods requires a timing resolution better than about
\SI{5}{\ns}, which will be hard to achieve in the case of
low-amplitude signals. Fingerprints of the neutron interaction points
have been found, which may be used to reduce the background due to
neutrons without too much loss in full-energy efficiency
\cite{Atac2009554}.

\subsection{Scattering materials}

The effect of passive materials in the inner space of the array,
especially in connection with the used of complementary
instrumentation, has been thoroughly discussed in
Ref.~\cite{Farnea2010331}. The conclusion presented in this work is
that at medium or high $\gamma-$ray energies, the presence of
scattering materials does not affect the performance of a tracking
array more than in the case of a conventional array. However, at low
energy Rayleigh scattering changes the direction of the incident
photon, which directly affects the tracking process.

\subsection{Imaging}

The performance of PSA is generally tested by checking how well
experimental spectra can be Doppler-corrected since this procedure
directly depends on the attainable precision in locating each
interaction point. Another method has recently been devised.  It
relies on Compton imaging, which does not need any beam nor complex
experimental setup. Instead it is assumed that the quality of the
image reconstruction obtained through the knowledge of the Compton
scattering sequence in the detector can provide information on the
interaction position resolution. The position and energy of the
interaction points were extracted using the grid-search algorithm
\cite{Venturelli2005} assuming only \num{1} interaction point per
segment. Since the incident energy is known, the first scattering
angle can be extracted by selecting events according to their total
energy and number of interaction points. This angle defines a
scattering cone in 3D space. If more than one event is analysed, all
the cones should overlap in a single point corresponding to the source
of the $\gamma$ rays. Placing the source far away from the detector
reduces the problem to the forming of a 2D image on the surface of a
sphere, which can be represented in a ($\theta$, $\phi$)
plane. Comparison of the experimental and simulated $\theta$ and
$\phi$ image profiles of a $^{60}$Co source has yielded an interaction
position resolution of about FWHM $= \SI{5}{\mm}$
\cite{Recchia200960}, which is comparable with the result obtained
from in-beam experiments
\cite{Recchia2009555,Soderstrom2011,Soderstrom_phd_thesis_2011}.
Attempts, using simple algorithms, to evaluate the capability of the
AGATA detectors to reject $\gamma$ rays originating from locations
that are different from the target position have shown partial
success~\cite{Doncel2010614}.

\subsection{Integration into the DAQ}

Tracking is performed on \num{1} or \num{2} nodes of the PC farm of
the AGATA DAQ, which runs the Narval system. The integration of the
tracking and other algorithms into Narval were done by using the ADF
library \cite{adf}. The goal is to make the data format transparent to
the algorithms, i.e. ADF allows each algorithm to access the data it
needs without knowing the structure of the data flow. It also provides
a virtual Narval environment in which to test and debug algorithm
codes in stand-alone mode. OFT and MGT are currently the only tracking
algorithms, which have been integrated into the Narval
architecture. The online performance of OFT has been optimised and
meets the specifications.

\subsection{Tracked $^{60}$Co spectrum}

\begin{figure}[ht]
  \centering
  \includegraphics[angle=90,width=1.0\columnwidth]
                  {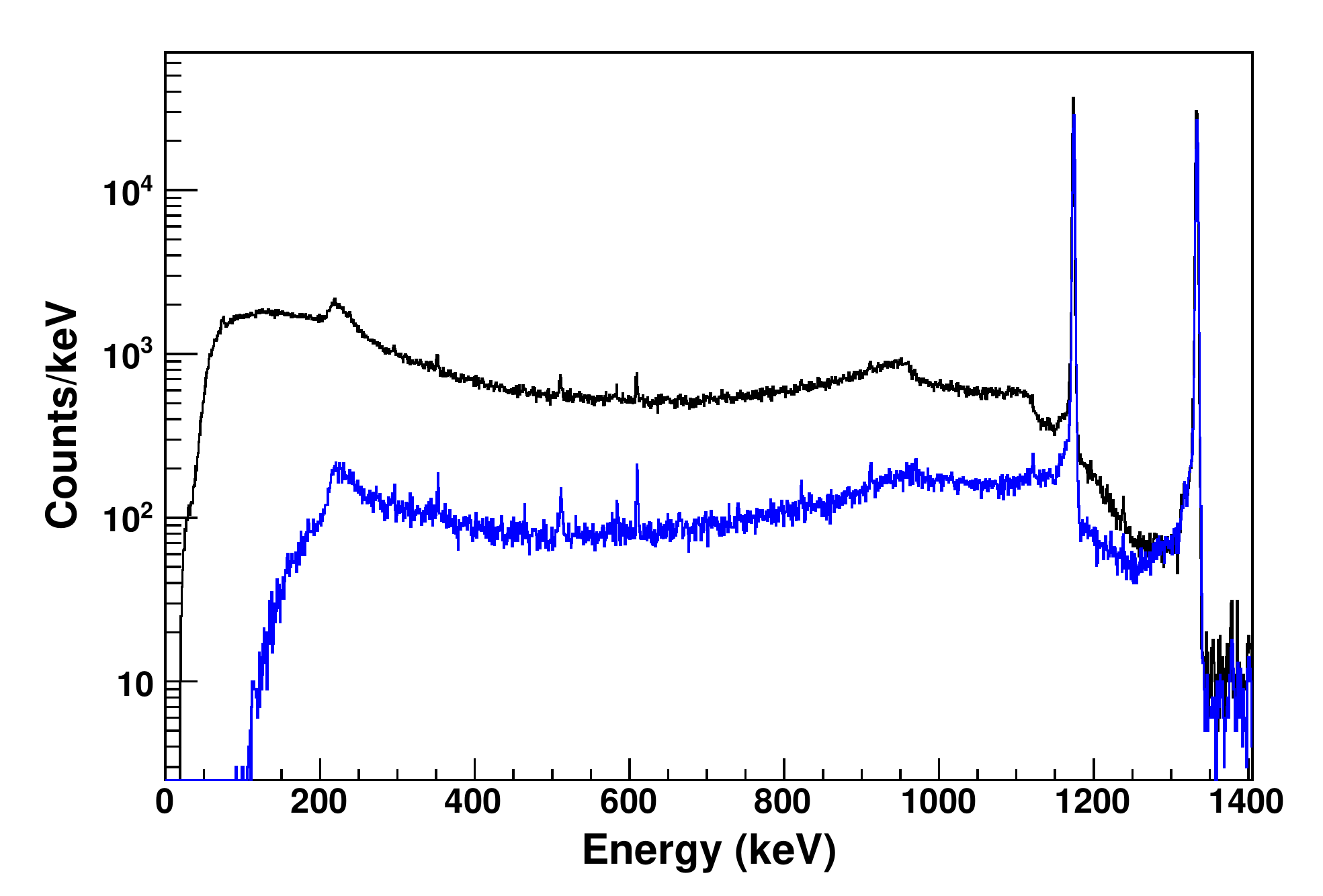}
  \caption{(Colour online) $^{60}$Co spectrum measured with three
    ATC detectors. Black histogram (greater number of counts):
    obtained by using the gain matched core signal of the
    crystals. Blue histogram (fewer number of counts): tracked
    spectrum excluding single interaction points. See text for further
    details.  }
  \label{fig:60co_spectrum.jpg}
\end{figure}

The results of a measurement with a $^{60}$Co source with three
ATC detectors are shown in Fig.~\ref{fig:60co_spectrum.jpg}. The
source was placed in the focal point of AGATA at a distance of
\SI{235}{\mm} from the front face of the crystals. The black histogram
shows a ``raw'' crystal spectrum obtained by summing all the
individual gain-matched energy depositions in the crystal, corrected
for segment threshold effects with the core-energy signal.  The blue
histogram is a tracked spectrum obtained by OFT and by excluding all
single interaction points. The PSA was performed using the grid-search
algorithm (see section~\ref{s:psa}) with the ADL pulse-shape database
(see section~\ref{s:pss}).
The P/T of the raw spectrum is \SI{16.8}{\percent} using a
low-energy cut-off of \SI{0}{\keV}.  The P/T of the
tracked spectrum is \SI{53.9}{\percent} and
\SI{54.6}{\percent} using \SI{0}{\keV} and
\SI{200}{\keV} cut-off values, respectively.  The tracking
efficiency, defined as the ratio of the number of counts in the
full-energy peak in the tracked spectrum to the number of counts in
the raw crystal spectrum, is \SI{84.5}{\percent}.

\section{Data analysis} \label{s:data_analysis}
As described in section \ref{s:daq}, at the end of the AGATA DAQ
chain, coincidences between tracked $\gamma$ rays and ancillaries are
provided. Data analysis needs to be performed at several stages,
including replaying the full experiment from the raw traces. To help
in these tasks a package called GammaWare (GW) \cite{gammaware} has
been developed.

The GW package is divided into several sub-projects: \textit{Core},
\textit {Physics}, \textit{GEM} (Gamma-ray Event Monte Carlo),
\textit{Tools}, \textit{ADF}, \textit{ADFE} (Extended ADF library). A
dynamic library, written in \CC, is associated to each specific
part. All libraries, except \textit{ADF}, are based on the
\ROOT\ framework \cite{Brun199781}.  As a collaborative software, GW
is available through the open-source version control system
\texttt{subversion} and uses a bugtracker system.  User's
documentation and web documentation, made by \texttt{doxygen}, can be
found at the AGATA data-analysis web site \cite{gammaware}.

The GW package adds facilities to \ROOT\ that are specific to analysis
of $\gamma$-ray spectroscopy data. Only the most significant add-ons
are described here.  The \textit{Core} library contains facilities
common to all sub-projects.  The \textit{Physics} library defines
\CC\ objects specific to $\gamma$-ray spectroscopy analysis such as
level schemes.  As any ROOT object, a level scheme can be displayed
and saved in ROOT files.  They can be built graphically or imported
from ENSDF \cite{ensdf} and RadWare \cite{Radford1995297} files.  As
an example, using the graphical level scheme display, gates are
selected and then applied on correlated spaces, which are (virtual
\CC) interfaces to any kind of system storing $\gamma$-ray
coincidences ($\gamma^{n}$). So far $\gamma$-$\gamma$ ($\gamma^{2}$)
like matrices are implemented but the system is designed to be
extendable to higher coincidence orders. From level schemes, cascades
of discrete $\gamma$ rays can be randomly generated and these cascades
can be fed into \geant\ simulations: the \textit{GEM} sub-project is
devoted to physics generators. In the \textit{Tools} library many
useful facilities can be found. It is the place where links to other
data analysis frameworks are implemented, allowing for instance to
ease exchanges of 1D and 2D spectra.

\begin{figure}[ht]
  \centering
  \includegraphics[width=1.0\columnwidth]{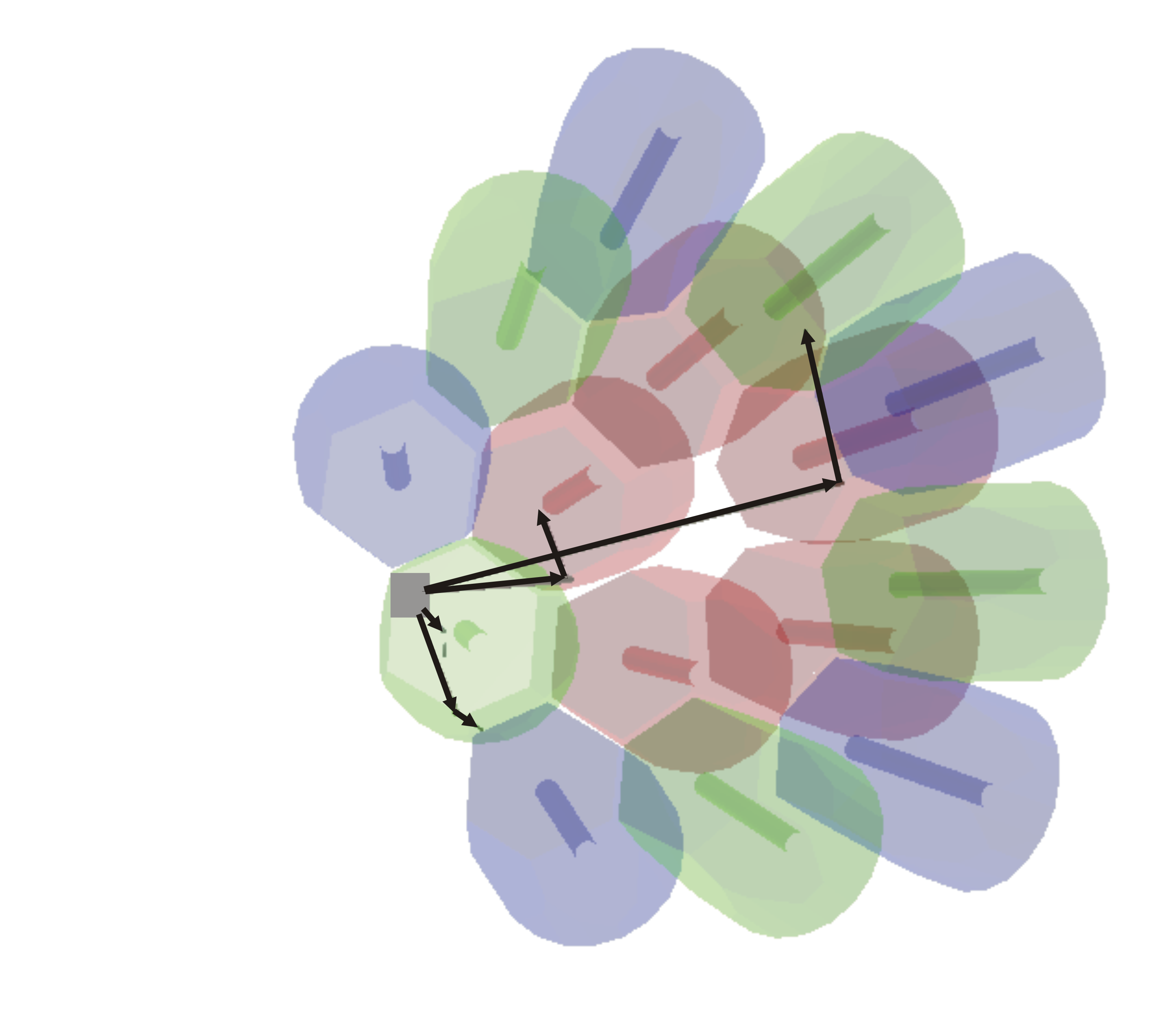}
  \caption{(Colour online) Online tracking Watcher display showing
    typical Compton scattering events in the \num{15} Ge crystals of
    AGATA, as provided by the tracking algorithm. Only the first and
    second interaction points are displayed.}
  \label{fig:DA_AGATA}
\end{figure}

The GW package includes the \textit{ADF} library (see section
\ref{s:daq}), which can be used also in stand-alone mode. Such a
library contains a complete virtual interface to any actor processing
the AGATA data flow. It defines the format of the data that are
exchanged between the AGATA algorithms: they are encapsulated in
frames with a trigger mechanism to deliver to a particular algorithm
only the specified frames.

All the \CC\ actors processing the data flow are grouped in a package
consisting of several dynamic libraries that are loaded into Narval.
The actors can also be used in any other framework, in particular
ROOT/GW, as well as linked to build a standard executable
program. Some actors can process the data flow and moreover perform
first stage analyses (counting rates, histograms, even traces saved in
\ROOT\ or ASCII files) to check online that the system is running
properly.

In the GW package, the \textit{ADF} library is extended through the
\textit{ADFE} sub-project bringing the AGATA like data to the ROOT/GW
environment. In particular Watchers are defined there. They are light
tasks dedicated to specific frames (output of an algorithm) running
independently and easy to plug whatever the data flow structure is. A
typical goal of the Watchers is to build histograms/graphs, to
calibrate events from ancillary detectors. It is also used to save
data in \ROOT\ trees and to display events in 3D geometries via the
ROOT/EVE facility. This is illustrated in Fig.~\ref{fig:DA_AGATA} in
which Compton events as provided on-line by the tracking algorithm can
be visualised. The spy mechanism proposed by Narval (section
\ref{s:daq}) allows the same Watchers to be used online/offline and
allows analysis to be performed at all stages of the data flow:
signals (core and segments), hits distributions, raw data from the
AGAVA card, counting rates, etc.

The virtualisation provided by the ADF library allows for the
development of emulators, i.e. actors connected together within a
common environment in various topologies as it is done in Narval
\cite{Grave2005}. Contrary to Narval, emulators only run so far on
single computers and are not efficient in case of complex situations
despite the fact that some of them run in multi-threading mode.
However, as they are written in C/\CC, it makes the development,
testing, and debugging phases of the AGATA data-processing algorithms
easier.  Most of the actors are shared libraries that can be plugged
in Narval or in emulators.  For efficiency reasons the event builder
and the merger running online are written in Ada so they cannot be
used as such for offline purposes. To overcome this problem, specific
\CC\ actors have been developed to reproduce event builders and
mergers as they act online. Emulators provide also elegant solutions
to build almost any kind of complex analysis chain. Simple ones are
used in the GW package to launch Watchers. More complete ones have
been developed as for example the emulator of the whole topology as
defined for the actual experiments.

\subsection{Data processing on the Grid} \label{ss:grid}

The AGATA collaboration has adopted the Grid technologies \cite{Grid}
and has created its own virtual organisation (VO), namely
vo.agata.org. The Grid computing resources (tape, disk and CPU) are
presently provided by Lyon, Strasbourg, Paris and Valencia and are
shared and accessed by the VO members under agreed common policies.

As mentioned in section~\ref{s:daq}, the raw data containing the
preamplifier traces are recorded on a Grid tape-storage system at
Tier1 sites for future off-line replay using, for instance, more
sophisticated PSA and tracking algorithms than those implemented to
run on-line. About \SI{150}{\tera\byte} of data have been produced
during the commissioning and the first experiments performed with up
to 15 crystals during the period from March 2009 to September 2011,
with an average value of \SI{8}{\tera\byte} of data collected per
experiment.

Besides the storage space question, partly discussed in
\cite{Mendez_CLOSER_2011}, the transfer of the raw data to local
institutes for analysis, as well as their processing, are also
critical issues as they are very time and resource consuming.  As the
raw data have been stored on the Grid, then it is advisable to use
also the Grid computing power (CPU) to replay the data, by emulating
all or part of the DAQ system processing stages.

It has been demonstrated that replaying raw data on the Grid using an
emulator is feasible without the need of changing the software to run
on the Grid \cite{Kaci_Braga_2010}. This is the case as long as copies
of the input data files are previously downloaded into the node where
the emulator is running, so the data are accessed locally. However, in
the framework of designing a Grid computing model for data management
and data processing for AGATA \cite{Kaci_AGATA_GCM}, more developments
have been recently tested in order to access directly the data on the
Grid storage \cite{Kaci_INGRID_2011}. For this, the emulator software
has been modified by including the necessary functions from the Grid
File Access Library (GFAL) in order to interact with the Grid storage
resource managers, through the appropriate protocols.

\begin{figure}[ht]
  \centering
 \includegraphics[width=0.88\columnwidth]
                 {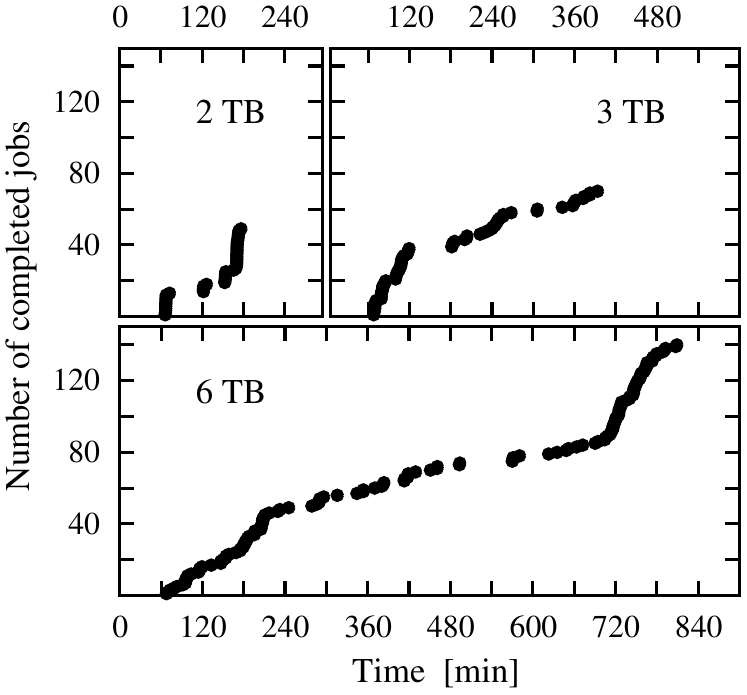}
  \caption{Completed jobs of data processing for \num{3} tasks run on
    the Grid as a function of time. Each job runs an instance of the
    emulator to replay part of the AGATA raw data set of interest.}
\label{fig:grid_completed_jobs}
\end{figure}

In the following, an example is presented in order to illustrate the
performances in execution times for replaying raw data on the Grid.
Three tasks of \num{49}, \num{70} and \num{140} jobs, respectively,
have been run on the IFIC/GRID-CSIC e-science infrastructure, each job
executing an instance of the emulator that processes PSA and tracking
on a fraction of the considered raw data set. It is worth noting that
this infrastructure provides a distributed file system, namely Lustre,
which allows direct access to the data located at the Grid storage of
the site, without using GFAL. A total of respectively \num{2}, \num{3}
and \num{6}~TB of commissioning data (\num{147}, \num{210} and
\num{420} data files of \num{14}~GB each, about \num{e8} events) have
been processed. Fig.~\ref{fig:grid_completed_jobs} shows the evolution
with time of the execution of the jobs for the above defined
tasks. The average values of time measured for the completeness of the
tasks, obtained from re-submitting them to Grid several times, are
\num{3}, \num{7} and \num{15} hours to process \num{2}, \num{3} and
\num{6}~TB data, respectively. These results indicate that \num{8}~TB
of data, produced in a typical experiment, can be processed in less
than one day.

\section{Summary and outlook} \label{s:summary}

The realisation of the AGATA spectrometer is a result of many
technological advances. These range from advances in Ge detector
technology, digital DAQ systems, signal decomposition and $\gamma$-ray
interaction reconstruction, and in many areas of the infrastructure
needed to support and operate such a complex device.

\begin{figure*}[ht]
  \begin{center}
    \includegraphics[width=0.9\textwidth]
                    {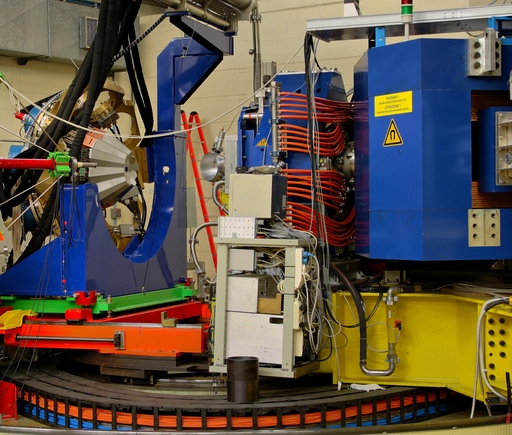}
    \caption{(Colour online) The AGATA and PRISMA spectrometers at
      LNL. The beam from the accelerators enters from the left. A part
      of the beam line, close to AGATA and half the scattering chamber
      has been removed. The quadrupole and dipole magnets of PRISMA
      are seen on the right-hand side of the photograph.  }
    \label{fig:agata_and_prisma_panoramic_view}
  \end{center}
\end{figure*}

The AGATA spectrometer is now fully operational in its first physics
campaign at INFN LNL in Legnaro, Italy, utilizing the wide range of
stable beams available.  A view of AGATA at the target position of the
PRISMA spectrometer is shown in
Fig.~\ref{fig:agata_and_prisma_panoramic_view}.  AGATA is designed to
be a peripatetic instrument and will move between major laboratories
in Europe to take advantage of the range of different beams and
equipment at each laboratory and of the resulting scientific
opportunities. AGATA will be operated in a series of campaigns, the
one first at LNL, and subsequently at the GSI facility in Germany and
the GANIL laboratory in France. At GSI, AGATA will be used at the exit
of the Fragment Separator (FRS) to study very exotic nuclei produced
following high-energy fragmentation and secondary Coulomb excitation.
At GANIL it will use the wide range of radioactive ions from the
coupled cyclotrons and SPIRAL. During these first three physics
campaigns the array will continually increase its efficiency as more
detector systems are added. The first stage is to build up the system
to 60 detector crystals, and then proceed towards the full
implementation of the $4\pi$ AGATA. Subsequent physics campaigns
will take advantage of the new radioactive beams available as
facilities such as FAIR, SPIRAL2, SPES and HIE-ISOLDE come online.

AGATA will have an enormous impact on nuclear physics research in
particular the exploration of nuclear structure at the extremes of
isospin, mass, angular momentum, excitation energy, and
temperature. This radically new device will constitute a dramatic
advance in $\gamma$-ray detection sensitivity that will enable the
discovery of new phenomena in nuclei, which are only populated in a
tiny fraction of the total reaction cross section or that are only
produced with rates of the order of a few per second or less. The
unprecedented angular resolution afforded by its position sensitivity
will facilitate high-resolution spectroscopy with fast and ultra-fast
fragmented beams giving access to the detailed structure of the most
exotic nuclei that can be reached. In addition, the capability to
operate at much higher event rates will allow the array to be operated
for reactions with intense $\gamma$-ray backgrounds.

The instrumentation and technical advances driven by this work, and
the knowledge gained by those involved, is also important in a wide
range of applications. These advances have potential impact in areas
such as medical imaging systems, homeland security, environmental
monitoring and the nuclear industry.

In addition to the technical advances, AGATA represents a tremendous
human achievement in the successful collaboration of several hundred
personnel in \num{12} countries and over \num{40} laboratories and
institutes across Europe. The collaboration is now excited with the
prospect of using this spectrometer and to capitalise on its discovery
potential for the understanding of the atomic nucleus.

\section{Acknowledgments} \label{s:ack}
AGATA and this work is supported by the European funding bodies and
the EU Contract RII3-CT-2004-506065, the German BMBF under Grants
06K-167 and 06KY205I, the Swedish Research Council and the Knut and
Alice Wallenberg Foundation, UK EPSRC Engineering and Physical
Sciences Research Council, UK STFC Science and Technology Facilities
Council, AWE plc, Scientific and Technological Research Council of
Turkey (Proj. nr. 106T055) and Ankara University (BAP
Proj. nr. 05B4240002), the Polish Ministry of Science and Higher
Education under Grant DPN/N190/AGATA/2009, the Spanish MICINN under
grants FPA2008-06419 and FPA2009-13377-C02-02, the Spanish
Consolider-Ingenio 2010 Programme CPAN (contract number
CSD2007-00042), and the Generalitat Valenciana under grant
PROMETEO/2010/101.  A.~Gadea and E.~Farnea acknowledge the support of
MICINN, Spain, and INFN, Italy, through the AIC10-D-000568 bilateral
action.






\bibliographystyle{./jn-nima}
\bibliography{./agata_references}

\begin{thebibliography}{100}
\providecommand{\url}[1]{\texttt{#1}}
\providecommand{\urlprefix}{}

\bibitem{SharpeySchafer1988293}
J.~Sharpey-Schafer and J.~Simpson, Prog. Part. Nucl. Phys. 21 (1988) 293 --
  400.

\bibitem{Beausang1996}
C.~Beausang and J.~Simpson, J. Phys. G {22} ({1996}) {527--558}.

\bibitem{Beck1992443}
F.~Beck, Prog. Part. Nucl. Phys. 28 (1992) 443 -- 461.

\bibitem{Simpson1997}
J.~Simpson, Z. Phys. A 358 (1997) 139--143.

\bibitem{Lee1990}
I.~Lee, Nucl. Phys. A {520} ({1990}) {C641--C655}.

\bibitem{Deleplanque1999292}
M.~A. Deleplanque et~al., Nucl. Instr. Meth. A 430 (1999) 292 -- 310.

\bibitem{Lee2003}
I.~Y. Lee, M.~A. Deleplanque, and K.~Vetter, Rep. Prog. Phys. 66 (2003) 1095.

\bibitem{Lee2004}
I.~Lee et~al., Nucl. Phys. A {746} ({2004}) {255C--259C}.

\bibitem{Eberth2001389}
J.~Eberth et~al., Prog. Part. Nucl. Phys. 46 (2001) 389 -- 398.

\bibitem{Habs1997111}
D.~Habs et~al., Prog. Part. Nucl. Phys. 38 (1997) 111 -- 126.

\bibitem{Kroll2001227}
T.~Kröll and D.~Bazzacco, Nucl. Instr. Meth. A 463 (2001) 227 -- 249.

\bibitem{Lieder2001279}
R.~Lieder et~al., Nucl. Phys. A 682 (2001) 279 -- 285.

\bibitem{Lieder2001399}
R.~Lieder et~al., Prog. Part. Nucl. Phys. 46 (2001) 399 -- 407.

\bibitem{Farnea2010331}
E.~Farnea et~al., Nucl. Instr. Meth. A 621 (2010) 331 -- 343.

\bibitem{Eberth1996135}
J.~Eberth et~al., Nucl. Instr. Meth. A 369 (1996) 135 -- 140.

\bibitem{Agostinelli2003250}
S.~Agostinelli et~al., Nucl. Instr. Meth. A 506 (2003) 250 -- 303.

\bibitem{Wiens2010223}
A.~Wiens et~al., Nucl. Instr. Meth. A 618 (2010) 223 -- 233.

\bibitem{Lersch2011133}
D.~Lersch et~al., Nucl. Instr. Meth. A 640 (2011) 133 -- 138.

\bibitem{Pullia2006}
A.~Pullia, F.~Zocca, and G.~Pascovici, IEEE Trans. Nucl. Sci.  (2006)
  2869--2875.

\bibitem{Zocca2009}
F.~Zocca et~al., IEEE Trans. Nucl. Sci.  (2009) 2384--2391.

\bibitem{Pascovici2008}
G.~Pascovici et~al., WSEAS Trans. Circ. Syst. 7 (2008) 470--481.

\bibitem{Bruyneel2009196}
B.~Bruyneel et~al., Nucl. Instr. Meth. A 599 (2009) 196 -- 208.

\bibitem{Bruyneel200999}
B.~Bruyneel et~al., Nucl. Instr. Meth. A 608 (2009) 99 -- 106.

\bibitem{Eberth2008283}
J.~Eberth and J.~Simpson, Prog. Part. Nucl. Phys. 60 (2008) 283 -- 337.

\bibitem{Descovich2005535}
M.~Descovich et~al., Nucl. Instr. Meth. A 553 (2005) 535 -- 542.

\bibitem{Lazarus2004}
I.~Lazarus et~al., IEEE Trans. Nucl. Sci. 51 (2004) 1353--1357.

\bibitem{Dimmock2009a}
M.~Dimmock et~al., IEEE Trans. Nucl. Sci. 56 (2009) 1593--1599.

\bibitem{Dimmock2009b}
M.~Dimmock et~al., IEEE Trans. Nucl. Sci. 56 (2009) 2415--2425.

\bibitem{Ha2011}
T.~Ha et~al., Eur. Phys. J. A  (2011), submitted.

\bibitem{Ha2009}
T.~Ha, {Charact\'erisation des d\'etecteurs d'AGATA et Etude de
  l'Hyperd\'eformation Nucl\'eaire dans la R\'egion de Masse 120}, Ph.D.
  thesis, Universit\'e Paris-Sud 11, Orsay, France (2009).

\bibitem{Crespi2008}
F.~C.~L. Crespi et~al., Nucl. Instr. Meth. A 593 (2008) 440--447.

\bibitem{DomingoPardo201179}
C.~Domingo-Pardo et~al., Nucl. Instr. Meth. A 643 (2011) 79 -- 88.

\bibitem{Goel2011591}
N.~Goel et~al., Nucl. Instr. Meth. A 652 (2011) 591 -- 594, symposium on
  Radiation Measurements and Applications (SORMA) XII 2010.

\bibitem{Tickner2004}
J.~Tickner, M.~Currie, and G.~Roach, Appl. Radiat. Isot. 61 (2004) 67--71.

\bibitem{Gerl2004}
J.~{Gerl} et~al., Nucl. Instr. Meth. A 525 (2004) 328--331.

\bibitem{Domingo2009}
C.~Domingo-Pardo et~al., IEEE Trans. Med. Imaging 28 (2009) 2007 --2014.

\bibitem{Stefanini2002217}
A.~Stefanini et~al., Nucl. Phys. A 701 (2002) 217 -- 221.

\bibitem{agata_installation_at_lnl}
A.~Gadea et~al., Nucl. Instr. Meth. A  (2011), in print.

\bibitem{mwd}
A.~Georgiev and W.~Gast, IEEE Trans. Nucl. Sci. {40} ({1993}) {770--779}.

\bibitem{tnt}
L.~Arnold et~al., IEEE Trans. Nucl. Sci. {53} ({2006}) {723--728}.

\bibitem{atca}
{PICMG 3.0 Revision 2.0 AdvancedTCA Base Specification},
  \urlprefix\url{http://www.picmg.org/}.

\bibitem{Simpson2000}
J.~Simpson et~al., Acta Phys. Hung. New Ser.-Heavy Ion Phys. {11} ({2000})
  {159--188}.

\bibitem{metronome1}
I.~Lazarus et~al., IEEE Trans. Nucl. Sci. {48} ({2001}) {567--569}.

\bibitem{adf}
O.~Stezowski, {AGATA Data Flow Library},
  \urlprefix\url{http://agata.in2p3.fr/doc/ADF_DesignProposal.pdf},
  unpublished.

\bibitem{Grave2005}
X.~Grave et~al., 14th IEEE-NPSS Real Time Conf.  (2005) 119--123.

\bibitem{narval}
{Narval web site}, \urlprefix\url{{http://narval.in2p3.fr/}}.

\bibitem{gammaware}
O.~Stezowski and the AGATA Data Analysis~Team, {GammaWare User's Guide},
  \urlprefix\url{http://agata.in2p3.fr/doc/GwUserGuide.pdf}, unpublished.

\bibitem{elog}
{ELOG web site}, \urlprefix\url{http://midas.psi.ch/elog/}.

\bibitem{gridcc}
F.~Lelli and G.~Maron, {Distributed Cooperative Laboratories: Networking,
  Instrumentation and Measurements}, Springer-Verlag New York Inc., New York,
  USA (2006) 269--277.

\bibitem{Grebosz2007251}
J.~Grebosz, Comp. Phys. Commun. 176 (2007) 251 -- 265.

\bibitem{gru}
{GRU/ViGRU documentation},
  \urlprefix\url{http://wiki.ganil.fr/gap/wiki/Documentation/Gru/Gru/}.

\bibitem{enx}
{ENX web site}, \urlprefix\url{http://enx.in2p3.fr/}.

\bibitem{debian}
{Debian web site}, \urlprefix\url{http://www.debian.org/}.

\bibitem{daqweb}
{AGATA DAQ web site}, \urlprefix\url{http://csngwinfo.in2p3.fr/}.

\bibitem{zab}
{Zabbix web site}, \urlprefix\url{http://www.zabbix.com/}.

\bibitem{xen}
{Xen web site}, \urlprefix\url{http://www.xen.org/}.

\bibitem{Venturelli2005}
R.~Venturelli and D.~Bazzacco, {LNL Annual Report 2004}, INFN-LNL, Legnaro,
  Italy (2005) 220.

\bibitem{Medina2004}
P.~M\'{e}dina, C.~Santos, and D.~Villaume, Proc. 21st IEEE Instr. Meas. Tech.
  Conf. 3 (2004) 1828--1832.

\bibitem{Schlarb2011_jass}
M.~Schlarb et~al., Eur. Phys. J. A  (2011), submitted.

\bibitem{Bruyneel2006764}
B.~Bruyneel, P.~Reiter, and G.~Pascovici, Nucl. Instr. Meth. A 569 (2006) 764
  -- 773.

\bibitem{Bruyneel2006774}
B.~Bruyneel, P.~Reiter, and G.~Pascovici, Nucl. Instr. Meth. A 569 (2006) 774
  -- 789.

\bibitem{Bruyneel2006}
B.~Bruyneel, {Characterization of Segmented Large Volume, High Purity Germanium
  Detectors}, Ph.D. thesis, Universität zu Köln, Cologne, Germany (2006),
  \urlprefix\url{http://kups.ub.uni-koeln.de/1858/}.

\bibitem{adl}
B.~Bruyneel, {Detector Simulation Software ADL},
  \urlprefix\url{http://www.ikp.uni-koeln.de/research/agata/index.php?show=download},
  unpublished.

\bibitem{Radeka1988}
V.~Radeka, Ann. Rev. Nucl. Part. Sci. 38 (1988) 217--277.

\bibitem{Mihailescu2000350}
L.~Mihailescu et~al., Nucl. Instr. Meth. A 447 (2000) 350 -- 360.

\bibitem{He2001250}
Z.~He, Nucl. Instr. Meth. A 463 (2001) 250 -- 267.

\bibitem{Birkenbach2011176}
B.~Birkenbach et~al., Nucl. Instr. Meth. A 640 (2011) 176 -- 184.

\bibitem{Bruyneel201192}
B.~Bruyneel, B.~Birkenbach, and P.~Reiter, Nucl. Instr. Meth. A 641 (2011) 92
  -- 100.

\bibitem{Bruyneel2010}
B.~Bruyneel et~al., LNL Annual Report 2010, INFN-LNL, Legnaro, Italy (2011)
  64--65.

\bibitem{OCC}
{Open CASCADE Technology, 3D modeling \& numerical simulation},
  \urlprefix\url{http://www.opencascade.org/}.

\bibitem{Geuzaine2009}
C.~Geuzaine and J.-F. Remacle, Int. J. Numer. Methods Eng. 79 (2009)
  1309--1331.

\bibitem{Kirk2006}
B.~Kirk et~al., Engineering with Computers 22 (2006) 237--254.

\bibitem{GSL}
M.~Galassi et~al., {GNU Scientific Library Reference Manual -- Third Edition
  (v1.12)}, Network Theory Limited, United Kingdom (2009).

\bibitem{Ljungvall2005PhD}
J.~Ljungvall, {Characterisation of the Neutron Wall and of Neutron Interactions
  in Germanium-Detector Systems}, Ph.D. thesis, Uppsala University, Uppsala,
  Sweden (2005),
  \urlprefix\url{http://urn.kb.se/resolve?urn=urn:nbn:se:uu:diva-5845}.

\bibitem{Schlarb2008}
M.~Schlarb, {Simulation and Real-Time Analysis of Pulse shapes from highly
  segmented Germanium detectors}, Ph.D. thesis, Technical University Munich,
  Munich, Germany (2008),
  \urlprefix\url{http://www.e12.physik.tu-muenchen.de/groups/agata/}.

\bibitem{Kroll2006691}
T.~Kr\"{o}ll and D.~Bazzacco, Nucl. Instr. Meth. A 565 (2006) 691 -- 703.

\bibitem{Olariu2007}
A.~Olariu, {Pulse Shape Analysis for the Gamma-ray Tracking Detector AGATA},
  Ph.D. thesis, Universit\'e Paris-Sud 11, Orsay France (2007).

\bibitem{Olariu2006}
A.~Olariu et~al., IEEE Trans. Nucl. Sci. 53 (2006) 1028--1031.

\bibitem{Recchia2009555}
F.~Recchia et~al., Nucl. Instr. Meth. A 604 (2009) 555 -- 562.

\bibitem{Soderstrom2011}
P.-A. S\"{o}derstr\"{o}m et~al., Nucl. Instr. Meth. A 638 (2011) 96--109.

\bibitem{Klupp2011}
S.~Klupp, {A Calibration Experiment for the AGATA Pulse Shape Analysis},
  Master's thesis, Technical University Munich, Munich, Germany (2011),
  \urlprefix\url{http://www.e12.physik.tu-muenchen.de/groups/agata/}.

\bibitem{Desesquelles2005}
P.~D\'{e}sesquelles et~al., 14th IEEE-NPSS Real Time Conf.  (2005) 100--102.

\bibitem{Desesquelles2009a}
P.~D\'{e}sesquelles et~al., Eur. Phys. J. A 40 (2009) 237--248.

\bibitem{Crespi2008PhD}
F.~C.~L. Crespi, {HPGe segmented detectors in $\gamma$-ray spectroscopy
  experiments with exotic beams}, Ph.D. thesis, Universit\'a degli Studi di
  Milano (2008), \urlprefix\url{http://hdl.handle.net/2434/152528}.

\bibitem{Crespi2007459}
F.~Crespi et~al., Nucl. Instr. Meth. A 570 (2007) 459 -- 466.

\bibitem{Desesquelles2009b}
P.~D\'{e}sesquelles et~al., Eur. Phys. J. A 40 (2009) 249--253.

\bibitem{Desesquelles2009542}
P.~D\'{e}sesquelles et~al., Nucl. Instr. Meth. B 267 (2009) 542 -- 547.

\bibitem{Desesquelles2009c}
P.~D\'{e}sesquelles et~al., J. Phys. G 36 (2009) 037001.

\bibitem{Recchia2008}
F.~Recchia, {In-beam test and imaging capabilities of the AGATA prototype
  detector}, Ph.D. thesis, Universit\`{a} degli Studi di Padova, Padova, Italy
  (2008), \urlprefix\url{http://npgroup.pd.infn.it/Tesi/PhD-thesisRecchia.pdf}.

\bibitem{vanderMarel1999538}
J.~van~der Marel and B.~Cederwall, Nucl. Instr. Meth. A 437 (1999) 538 -- 551.

\bibitem{Schmid199969}
G.~J. Schmid et~al., Nucl. Instr. Meth. A 430 (1999) 69 -- 83.

\bibitem{LopezMartens2004454}
A.~Lopez-Martens et~al., Nucl. Instr. Meth. A 533 (2004) 454 -- 466.

\bibitem{Bazzacco2004248}
D.~Bazzacco, Nucl. Phys. A 746 (2004) 248 -- 254, proceedings of the Sixth
  International Conference on Radioactive Nuclear Beams (RNB6).

\bibitem{fuzzycmeans}
G.~Suliman and D.~Bucurescu, Rom. Rep. Phys. 62 (2010) 27 -- 36.

\bibitem{Didierjean2010188}
F.~Didierjean, G.~Duch\^{e}ne, and A.~Lopez-Martens, Nucl. Instr. Meth. A 615
  (2010) 188 -- 200.

\bibitem{Ljungvall2005379}
J.~Ljungvall and J.~Nyberg, Nucl. Instr. Meth. A 550 (2005) 379 -- 391.

\bibitem{Ljungvall2005553}
J.~Ljungvall and J.~Nyberg, Nucl. Instr. Meth. A 546 (2005) 553 -- 573.

\bibitem{Atac2009554}
A.~Ata\c{c} et~al., Nucl. Instr. Meth. A 607 (2009) 554 -- 563.

\bibitem{Recchia200960}
F.~Recchia et~al., Nucl. Instr. Meth. A 604 (2009) 60 -- 63.

\bibitem{Soderstrom_phd_thesis_2011}
P.-A. Söderström, {Collective Structure of Neutron-Rich Rare-Earth Nuclei and
  Development of Instrumentation for Gamma-Ray Spectroscopy}, Ph.D. thesis,
  Uppsala University, Uppsala, Sweden (2011),
  \urlprefix\url{http://urn.kb.se/resolve?urn=urn:nbn:se:uu:diva-149772}.

\bibitem{Doncel2010614}
M.~Doncel et~al., Nucl. Instr. Meth. A 622 (2010) 614--618.

\bibitem{Brun199781}
R.~Brun and F.~Rademakers, Nucl. Instr. Meth. A 389 (1997) 81 -- 86.

\bibitem{ensdf}
J.~Tuli, {Evaluated Nuclear Structure Data File: A Manual for Preparation of
  Data Sets}, BNL-NCS-51655-01/02-Rev (2001),
  \urlprefix\url{http://www.nndc.bnl.gov/nndcscr/documents/ensdf/#ensdf}.

\bibitem{Radford1995297}
D.~C. Radford, Nucl. Instr. Meth. A 361 (1995) 297 -- 305.

\bibitem{Grid}
I.~Foster and C.~Kesselman, {The Grid: Blueprint for a New Computing
  Infrastructure}, Morgan Kaufmann Publishers, Inc., San Francisco, USA (1999).

\bibitem{Mendez_CLOSER_2011}
V.~M\'endez et~al., CLOSER 2011, Proceedings of the 1st International
  Conference on Cloud Computing and Services Science, Noordwijkerhout, The
  Netherlands, 507--511.

\bibitem{Kaci_Braga_2010}
M.~Kaci et~al., {IBERGRID 2010, Proceedings of the 4th Iberian Grid
  Infrastructure Conference}, Braga, Portugal, 482--484.

\bibitem{Kaci_AGATA_GCM}
M.~Kaci and V.~M\'endez, {The AGATA Grid Computing Model for Data Management
  and Data Processing},
  \urlprefix\url{http://ific.uv.es/grid/e-science/agata/gcm-dmdp.pdf},
  unpublished.

\bibitem{Kaci_INGRID_2011}
M.~Kaci et~al., {INGRID 2011, 6th Workshop for e-Science and e-Infrastructure},
  Santander, Spain.

\end{thebibliography}




%




\end{document}